\documentclass[aip,jcp,amsmath,amssymb,reprint,floatfix]{revtex4-1}
\pdfoutput=1
\usepackage{graphicx}
\usepackage{dcolumn}
\usepackage{bm}
\usepackage{natbib}
\usepackage{gensymb}
\usepackage[english]{babel}

\usepackage{graphicx}
\usepackage{alphalph}
\usepackage{wrapfig}
\usepackage{enumitem}
\usepackage{float}
\usepackage{xr}
\usepackage{zref-xr}

\usepackage{todonotes}          

\begin{document}

\title{Crystal Nucleation in Liquids: \\ Open Questions and Future Challenges in Molecular Dynamics Simulations}

\author{\it Gabriele C. Sosso}
\affiliation{Thomas Young Centre, London Centre for Nanotechnology and Department of Physics and Astronomy, University College London, Gower Street WC1E 6BT London, UK.}
\author{\it Ji Chen}
\affiliation{Thomas Young Centre, London Centre for Nanotechnology and Department of Physics and Astronomy, University College London, Gower Street WC1E 6BT London, UK.}
\author{\it Stephen J. Cox}
\affiliation{Thomas Young Centre, London Centre for Nanotechnology and Department of Physics and Astronomy, University College London, Gower Street WC1E 6BT London, UK.}
\altaffiliation{Current address: Chemical Sciences Division, Lawrence Berkeley National Laboratory, Berkeley, CA 94720, USA.}
\author{\it Martin Fitzner}
\affiliation{Thomas Young Centre, London Centre for Nanotechnology and Department of Physics and Astronomy, University College London, Gower Street WC1E 6BT London, UK.}
\author{\it Philipp Pedevilla}
\affiliation{Thomas Young Centre, London Centre for Nanotechnology and Department of Physics and Astronomy, University College London, Gower Street WC1E 6BT London, UK.}
\author{\it Andrea Zen}
\affiliation{Thomas Young Centre, London Centre for Nanotechnology and Department of Physics and Astronomy, University College London, Gower Street WC1E 6BT London, UK.}
\author{\it Angelos Michaelides}
\affiliation{Thomas Young Centre, London Centre for Nanotechnology and Department of Physics and Astronomy, University College London, Gower Street WC1E 6BT London, UK.}
\email{angelos.michaelides@ucl.ac.uk}



\begin{abstract}
The nucleation of crystals in liquids is one of nature's most ubiquitous phenomena, playing an important
role in areas such as climate change and the production of drugs.  As the early stages of nucleation involve exceedingly
small time and length scales, atomistic computer simulations can provide unique insight into the microscopic aspects of
crystallization.  In this review, we take stock of the numerous molecular dynamics simulations that in the last few decades
have unraveled crucial aspects of crystal nucleation in liquids.  We put into context the theoretical framework of
classical nucleation theory and the state of the art computational methods, by reviewing simulations of e.g. ice
nucleation or crystallization of molecules in solutions.  We shall see that molecular dynamics simulations have provided
key insight into diverse nucleation scenarios, ranging from colloidal particles to natural gas hydrates, and that in
doing so the general applicability of classical nucleation theory has been repeatedly called into question.  We have attempted
to identify the most pressing open questions in the field. 
We believe that by improving (i) existing interatomic potentials; and (ii) currently available enhanced sampling methods, the
community can move towards accurate investigations of realistic systems of practical interest, thus bringing
simulations a step closer to experiments.  

\end{abstract}



\maketitle

\clearpage

\tableofcontents

\clearpage

\section*{Abbreviations}

\begin{center}
\begin{table}[h!]
   {\renewcommand{\arraystretch}{1.2}
   \begin{tabular}{l l}
   BCC    & Body Centered Cubic \\
   cDFT   & Classical Density Functional Theory \\
   CNT    & Classical Nucleation Theory \\
   CNT    & Classical Nucleation Theory \\
   DFT    & Density Functional Theory \\
   DSC    & Differential Scanning Calorimetry \\
   FCC    & Face Centered Cubic \\
   FFS    & Forward Flux Sampling \\
   FTIRS  & Fourier Transform Infrared Spectroscopy \\
   HDL    & High Density Liquid \\
   LCH    & Labile Cluster Hypothesis \\
   LDL    & Low Density Liquid \\
   LSH    & Local Structure Hypothesis \\
   MD     & Molecular dynamics \\
   MetaD  & Metadynamics \\
   PNC    & Pre-Nucleation Cluster \\
   RHCP   & (Random)Hexagonal Close Packed \\
   SEM    & Scanning Electron Microscope \\
   sH     & Structure H \\
   sI     & Structure I \\
   sII    & Structure II \\   
   SMRT-TEM  & Single-Molecule-Real-Time - TEM \\
   TEM    & Transmission Electron Microscope \\
   TIS    & Transition Interface Sampling \\
   UMD    & Unbiased Molecular Dynamics \\
   US     & Umbrella Sampling \\
   XPS    & X-Ray Photoeletron Spectroscopy \\
   \end{tabular}
   }
   \label{ABB}
\end{table}
\end{center}

\clearpage
\section{Introduction}
\label{Introduction}

Crystal nucleation in liquids has countless practical consequences in science and technology, and it also
affects our everyday experience.  One obvious example is the formation of ice, which
influences global phenomena like climate change~\cite{bartels-rausch_chemistry:_2013,murray_heterogeneous_2010} as well as
processes happening at the nanoscale, like intracellular
freezing~\cite{mazur_cryobiology:_1970,lintunen_anatomical_2013}. On the other hand, controlling nucleation of
molecular crystals from solutions is of great importance to pharmaceuticals and particularly in
the context of drug design and production, as the early stages of crystallization impact on the crystal
polymorph obtained~\cite{erdemir_polymorph_2007,cox_selective_2007}.  Even the multibillion-dollar oil industry is
affected by the nucleation of hydrocarbon clathrates, which can form inside pipelines endangering 
extraction~\cite{sloan_fundamental_2003,hammerschmidt_formation_1934}. Finally, crystal nucleation is involved in
many processes spontaneously happening in living beings, from the growth of the beautiful Nautilus
shells~\cite{velazquez-castillo_nanoscale_2006} to the dreadful formation in our own brains of amyloid fibrils, which
are thought to be responsible for many neurodegenerative disorders like Alzheimers
disease~\cite{harper_atomic_1997,walsh_amyloid_1997}.

Each of the above scenarios start from a liquid below its melting
temperature. This \textit{supercooled
liquid}~\cite{Debenedetti:1996} is doomed, according to thermodynamics, to face a first-order phase transition
leading into a crystal~\cite{stevenson_ultimate_2011,fnote1}.
Before this can happen, however, 
a sufficiently large cluster of crystalline atoms (or molecules, or
particles) must form within the liquid, such that the free energy cost of
creating an interface between the liquid and the crystalline phase will be overcome by the free energy gain
of having a certain volume of crystal. This event stands at the heart of crystal nucleation, and how
the latter has been, is, and will be modeled by means of computer simulations is the subject of this review.  

The last few decades have witnessed an impressive body of experimental work devoted to crystal nucleation.  For
instance, thanks to novel techniques like e.g. Transmission Electron Microscopy at very low temperatures (cryo-TEM
microscopy), we are now able to peek in real-time into the early stages of
crystallization~\cite{habraken_ion-association_2013}. A substantial effort has also been made to understand which
materials, in the form of impurities within the liquid phase, can either promote or inhibit nucleation
events~\cite{murray_ice_2012}, a common scenario known as heterogeneous nucleation. However, our understanding of
crystal nucleation is far from being complete. This is because 
the molecular (or atomistic) details of the process are largely unknown due to the very small length scale
involved (nm), which is exceedingly challenging to probe in real time even by state of the art measurements. 
Hence the need for computer simulations and particularly molecular dynamics (MD), where the temporal evolution of the
liquid into the crystal is more or less faithfully reproduced.
Unfortunately,
crystal nucleation is a rare event, which can happen on timescales of e.g. seconds, far
beyond the reach of any conventional MD framework. In addition, a number of approximations within the computational models, the algorithms and the
theoretical framework used have been severely questioned for several decades. While the rush for computational methods
able to overcome this \textit{timescale problem} is more competitive than ever, we are almost always forced to base
our conclusions upon the ancient grounds of classical nucleation theory (CNT), a powerful theoretical tool that
nonetheless dates back 90 years to Volmer and Weber~\cite{ISI:000201367700009}.

Nonetheless, these are exciting times for the crystal nucleation community, as demonstrated by the many reviews covering several
aspects of this diverse
field~\cite{gebauer_pre-nucleation_2014,sear_quantitative_2014,vekilov_nucleation_vekilov_2010,xu_nucleation_2014,yi_molecular_2012,anwar_uncovering_2011,zahn_thermodynamics_2015}.
This particular review will focus almost exclusively on MD simulations of crystal nucleation of supercooled liquids and supersaturated solutions. We take
into account several systems, from colloidal liquids to natural gas hydrates, highlighting long standing issues as well
as recent advances.  While we will review a substantial fraction of the theoretical efforts in the field, mainly from the
last decade, our goal is
not to discuss in detail every contribution.  Instead, we try to pinpoint the most pressing
issues that still prevent us from furthering our understanding of nucleation.  

This paper is structured into three parts. In the first part we introduce the theoretical framework of CNT (Sec.~\ref{THEOF}),
the state of the art experimental techniques (Sec.~\ref{Experimental_Methods}) and 
the MD-based simulation methods (Sec.~\ref{SIMM}) that in the last few decades have provided insight into 
nucleation. In Sec.~\ref{MDSIM} we then put such computational approaches into context, describing
achievements as well as open questions concerning the molecular details of nucleation for different kinds of
systems, namely colloids (Sec.~\ref{COLL}), Lennard-Jones (LJ) liquids (Sec.~\ref{LJL}), atomic liquids (Sec.~\ref{BLJP}), 
water (Sec.~\ref{WAH}), nucleation from solution (Sec.~\ref{sec.MIS}) and natural gas hydrates (Sec.~\ref{sec:gas-hydrates}). 
In the third and last part of the paper (Sec.~\ref{Discussion}) we highlight future perspectives and open challenges in
the field.

\subsection{Theoretical Framework}
\label{THEOF}

\subsubsection{Classical Nucleation Theory}
\label{THEOF_1}

Almost every computer simulation of crystal nucleation in liquids invokes some elements~\cite{fnote_R1} of
classical nucleation theory (CNT). The latter has been discussed in great detail
elsewhere~\cite{Kelton2010279,kalikmanov_nucleation_2013,vehkamaki_classical_2006} and we include it here for the
sake of completeness and also to introduce various terms used throughout the review. 
Nonetheless, readers familiar with CNT can skip to Sec.~\ref{Experimental_Methods}.

CNT was formulated 90 years ago thanks to the contributions of Volmer and
Weber~\cite{ISI:000201367700009}, Farkas~\cite{FARKAS}~\cite{fnote8},
Becker and D\"{o}ring~\cite{ANDP:ANDP19354160806} and
Zeldovich~\cite{ISI:000202051200001}, on the basis of the pioneering ideas of none other than Gibbs
himself~\cite{gibbs_old}. CNT was created to describe the condensation of supersaturated vapors
into the liquid phase, but most of the concepts can also be applied to the crystallization of supercooled liquids 
and supersaturated solutions.
 According to CNT, clusters of crystalline atoms (or particles, or molecules) of any size are treated as
macroscopic objects, that is homogeneous chunks of crystalline phase separated from the surrounding liquid by a
vanishingly thin interface. This apparently trivial assumption is known as the capillarity approximation, which
encompasses most of the strengths and weaknesses of the theory. By embracing the capillarity approximation,
the interplay between the interfacial free energy $\gamma_{\mathcal{S}}$, and the free energy difference $\Delta
\mu_{\mathcal{V}}$ between the liquid and the crystal fully describes the thermodynamics of crystal nucleation. 
In three dimensions~\cite{fnote13}, the free
energy of formation $\Delta G_{\mathcal{N}}$ for a spherical crystalline nucleus of radius $r$ can thus be written as
the sum of a surface term and a volume term:

\begin{figure*}[t!]
\begin{center}
\includegraphics[width=15cm]{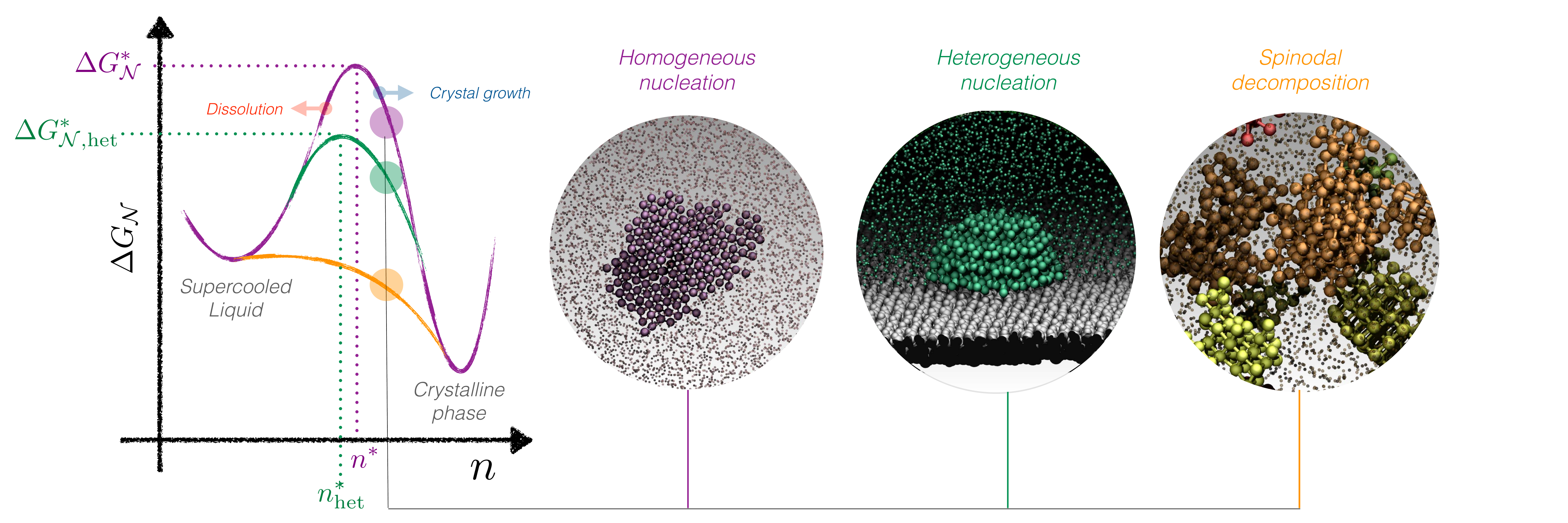}
\end{center}
\caption{Sketch of the free energy difference $\Delta G_{\mathcal{N}}$ as a function of the crystalline nucleus size
$n$. A free energy barrier for nucleation $\Delta G_{\mathcal{N}}^{\ast}$ must be overcome in order to proceed from  the
- metastable - supercooled liquid state to the thermodynamically stable crystalline phase via homogeneous nucleation
(purple).  Heterogeneous nucleation (green) can be characterized by a lower free energy barrier $\Delta
G_{\mathcal{N},\text{het}}^{\ast}$ and a smaller critical nucleus size $n^{\ast}_{\text{het}}$, while in the case of
spinodal decomposition (orange) the supercooled liquid is unstable with respect to the crystalline phase, and 
the transformation to the crystal proceeds in a barrierless fashion. 
The three snapshots depict a crystalline cluster nucleating within the supercooled
liquid phase (homogeneous nucleation) or thanks to the presence of a foreign impurity (heterogeneous nucleation), as
well as the simultaneous occurrence of multiple crystalline clusters in the unstable liquid. 
This scenario is often labeled as spinodal decomposition, albeit the existence of a genuine spinodal decomposition 
from the supercooled liquid to the crystalline phase has been debated (see text).}
\label{Theoretical_Framework_FIG_1} \end{figure*}

\begin{equation}
\Delta G_{\mathcal{N}} = \underbrace{4 \pi r^2 \gamma_{\mathcal{S}}}_{\text{Surface term}} - \underbrace{\frac{4\pi}{3} r^3 \Delta \mu_{\mathcal{V}}}_{\text{Volume term}} \ .
\label{cnt_1}
\end{equation}

\noindent This function, sketched in Fig.~\ref{Theoretical_Framework_FIG_1}, displays a maximum corresponding to the so
called critical nucleus size $n^{\ast}$

\begin{equation}
n^{\ast} = \frac{32 \pi \rho_{\mathcal{C}}}{3} \frac{\gamma_{\mathcal{S}} ^3}{\Delta \mu_{\mathcal{V}}^3} \ ,
\label{cnt_3}
\end{equation}

\noindent where $\rho_{\mathcal{C}}$ is the number density of the crystalline phase. The critical nucleus size
represents the number of atoms that must be included in the crystalline cluster for the free energy difference
$\Delta \mu_{\mathcal{V}}$ to match the free energy cost due to the formation of the solid liquid interface. Clusters of
crystalline atoms occur within the supercooled liquid by spontaneous, infrequent fluctuations, which eventually lead the
system to overcome the free energy barrier for nucleation

\begin{equation}
\Delta G_{\mathcal{N}}^{\ast} = \frac{16\pi}{3}\frac{\gamma_{\mathcal{S}} ^3}{\Delta \mu_{\mathcal{V}}^2} \ ,
\label{cnt_2}
\end{equation}

\noindent triggering the actual crystal growth (see Fig.~\ref{Theoretical_Framework_FIG_1}). 

The kinetics of crystal nucleation is typically addressed by assuming that no
correlation exists between successive events increasing or reducing the number of constituents of the crystalline
nucleus. In other words, the time evolution of the nucleus size is presumed to be a Markov process, in which atoms in the liquid either order
themselves one by one in a crystalline fashion or dissolve one by one into the liquid phase.  In addition, we state that
every crystalline nucleus lucky enough to overcome the critical size $n^{\ast}$ quickly grows to macroscopic dimensions on
a timescale much smaller than the long time required for that fortunate fluctuation to come about. If the above
mentioned conditions are met~\cite{fnote9}
the nucleation rate, i.e. the probability
per unit time per unit volume of forming a critical nucleus does not depend on time, leading to the following
formulation of the so called steady-state nucleation rate $\mathcal{J}$:

\begin{equation}
\mathcal{J}=\mathcal{J}_{0} \exp \left ( - \frac{\Delta G_{\mathcal{N}}^{\ast}}{k_B T}\right ) \ ,
\label{cnt_4}
\end{equation}

\noindent where $k_B$ is the Boltzmann constant and $\mathcal{J}_{0}$ is a prefactor which we discuss later.  The
steady-state nucleation rate is the central quantity in the description of crystallization kinetics, as much as the notion of
critical nucleus size captures most of the thermodynamics of nucleation.

All quantities specified up to now depend on pressure and most notably temperature. 
In most cases, the interfacial free energy
$\gamma_{\mathcal{S}}$ is assumed to be linearly dependent on temperature, while the free energy difference between the
liquid and the solid phase $\Delta \mu_{\mathcal{V}}$ is proportional to the supercooling
$\Delta T=T_\mathcal{M}-T$ (or the supersaturation). Several approximations exist to treat the
temperature dependence of $\gamma_{\mathcal{S}}$~\cite{baidakov_crystal_2012} and  $\Delta
\mu_{\mathcal{V}}$~\cite{hoffman_thermodynamic_1958}, which can vary substantially for different supercooled
liquids~\cite{thompson_approximation_1979}.  In any case, it follows from Eq.~\ref{cnt_2} that the free energy barrier
for nucleation $\Delta G_{\mathcal{N}}^{\ast}$ decreases with supercooling. In other words, the further we are from the
melting temperature $T_\mathcal{M}$, the larger the thermodynamic driving force for nucleation.

\begin{figure}[t!] \begin{center} \includegraphics[width=7cm]{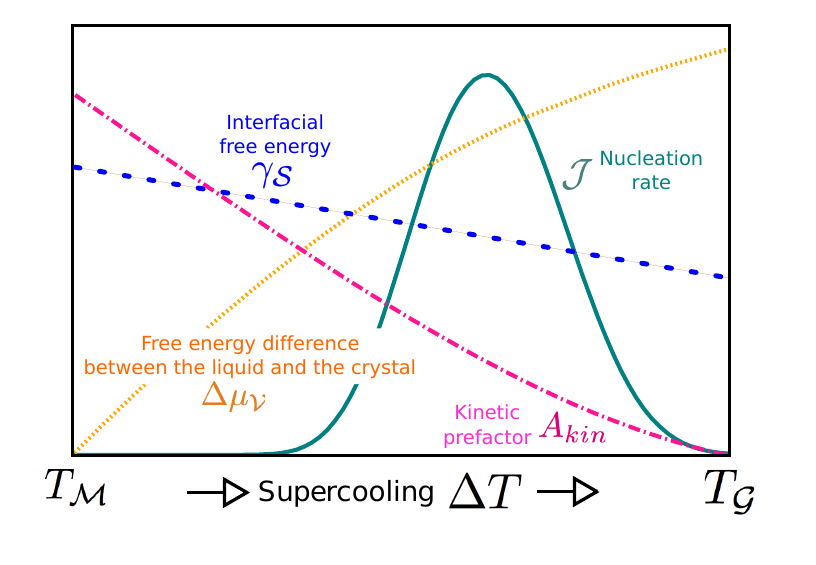} \end{center}
\caption{Illustration of how certain quantities from CNT vary as a function of supercooling $\Delta T$ for supercooled liquids. 
The free energy difference between the liquid and the solid phase $\Delta \mu_{\mathcal{V}}$, the
interfacial free energy $\gamma_{\mathcal{S}}$ and the kinetic prefactor $\mathcal{A}_{kin}$
are reported as a function of $\Delta T$ in a generic case of diffusion-limited nucleation, characterized by a maximum
in the steady state nucleation rate $\mathcal{J}$. $\Delta \mu_{\mathcal{V}}$ is zero at the melting temperature $T_{\mathcal{M}}$ and $\mathcal{A}_{kin}$
are vanishingly small at the glass transition temperature $T_{\mathcal{G}}$.}
\label{Theoretical_Framework_FIG_2} \end{figure}

Interestingly, in the case of supercooled liquids kinetics goes the other way, as the dynamics of the liquid slow down with supercooling, thus
hindering the occurrence of nucleation events.  In fact, while a conclusive expression for the prefactor
$\mathcal{J}_{0}$ is still lacking~\cite{auer_prediction_2001,schmelzer_determination_2010}, $\mathcal{J}_{0}$ it is
usually written within CNT as~\cite{kalikmanov_nucleation_2013}:

\begin{equation}
\mathcal{J}_{0}=\rho_{\mathcal{S}}\cdot\mathcal{Z}\cdot \mathcal{A}_{kin}
\label{kinp}
\end{equation}

\noindent where $\rho_{\mathcal{S}}$ is the number of possible nucleation sites per unit volume, $\mathcal{Z}$ is the
Zeldovich factor~\cite{kalikmanov_nucleation_2013,vehkamaki_technical_2007} (accounting for the fact that
several postcritical clusters may still shrink without growing into the crystalline
phase), and $\mathcal{A}_{kin}$ is a kinetic prefactor~\cite{auer_prediction_2001}. The latter should represent the
\textit{attachment rate}, that is the frequency with which the particles in the liquid phase reach the cluster
re-arranging themselves in a crystalline fashion. However, in a dense supercooled liquid $\mathcal{A}_{kin}$ also
quantifies the ease with which the system explores configurational space, effectively regulating the amplitude of the
fluctuations possibly leading to the formation of a crystalline nucleus.  In short, we can safely say that
$\mathcal{A}_{kin}$ involves the atomic or molecular mobility of the liquid phase, more often than not quantified in
terms of the self-diffusion coefficient $\mathcal{D}$~\cite{kalikmanov_nucleation_2013}, which obviously decreases with
supercooling.  Thus, for a supercooled liquid the competing trends of $\Delta G_{\mathcal{N}}^{\ast}$ and $\mathcal{A}_{kin}$ lead - in the case
of diffusion-limited nucleation~\cite{pcmbook} - to a maximum in the nucleation rate, as depicted in
Fig.~\ref{Theoretical_Framework_FIG_2}. The same arguments apply when dealing with e.g. solidification of metallic 
alloys~\cite{fredriksson_2012,citeulike:5777981}.
In the case of nucleation from solutions, $\gamma_{\mathcal{S}}$ and $\Delta \mu_{\mathcal{V}}$ depend mainly on supersaturation.
However, the dependence of the kinetic prefactor on supersaturation is much weaker than the temperature dependence of $\mathcal{A}_{kin}$ 
characteristic of supercooled liquids. As a result, there is usually no maximum in the nucleation rate as a function of supersaturation for nucleation
from solutions~\cite{KashchievBook2000}.

Although $\mathcal{A}_{kin}$ is supposed to play a minor role compared to the exponential term in Eq.~\ref{cnt_4}, the
kinetic prefactor has been repeatedly blamed for the quantitative
disagreement between experimental measurements and computed crystal nucleation rates~\cite{auer_prediction_2001,wette_nucleation_2007}.
Atomistic simulations could in
principle help to clarify the temperature dependence as well as the microscopic origin of $\mathcal{A}_{kin}$ and also
of the thermodynamic ingredients involved in the formulation of CNT. However, quantities like e.g.
$\gamma_{\mathcal{S}}$ are not only infamously difficult to converge within decent levels of
accuracy~\cite{angioletti-uberti_solid-liquid_2010,davidchack_ice_2012}, but can even be ill-defined in many situations.
For instance, it remains to be seen whether $\gamma_{\mathcal{S}}$, which in principle refers to a planar interface under
equilibrium conditions, can be safely defined when dealing with small crystalline clusters of irregular shapes. In fact,
the early stages of the nucleation process often involve crystalline nuclei whose size and morphology fluctuate on a
timescale shorter than the structural relaxation time of the surrounding liquid. On top of that, the dimensions of such
nuclei can be of the same order of the diffuse interface between the liquid and the solid phases, thus rendering the
notion of a well defined $\gamma_{\mathcal{S}}$ quite dangerous. 
As an example, Joswiak \textit{et al.}~\cite{joswiak_size-dependent_2013} have recently shown that for liquid water droplets
$\gamma_{\mathcal{S}}$ could strongly depends on the curvature of the droplet. The mismatch between the macroscopic interfacial
free energy and its curvature dependent value can spectacularly affects water droplets nucleation, as reported by atomistic simulations of
droplets characterized by radii of the order of $\sim$ 0.5-1.5 nm.
Some other quantities, like the size of the critical cluster, depend
in many cases rather strongly on the degree of supercooling.  This is the case of e.g. the critical nucleus size
$n^{\ast}$ that can easily span two orders of magnitude in just ten degrees of
supercooling~\cite{pereyra_temperature_2011,sanz_homogeneous_2013}.

\subsubsection{Two-step nucleation}
\label{THEOF_TS}

Given the old age of CNT, it is no surprise that substantial efforts have been devoted
to extend and/or improve its original theoretical framework. The most relevant modifications possibly concern the
issue of two-step nucleation. Many excellent works have reviewed this subject extensively (see e.g. 
Refs.~\citenum{vekilov_two-step_2010,de_yoreo_crystal_2013,gebauer_pre-nucleation_2014,zahn_thermodynamics_2015}),
so that in here we supply to the reader the essential concepts only.

\begin{figure*}[t!] 
\begin{center} 
\includegraphics[width=14cm]{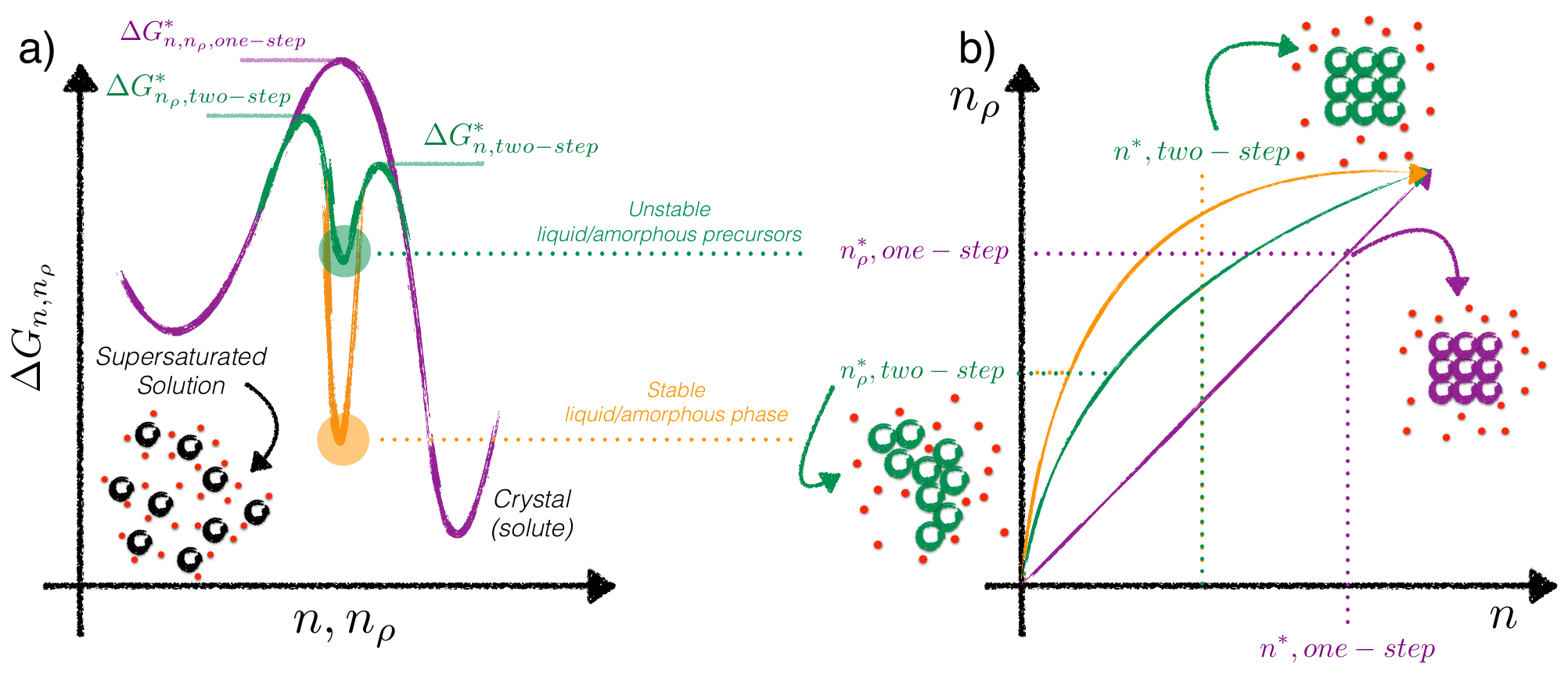} 
\end{center}
\caption{Schematic comparison of one-step versus two-step nucleation for a generic supersaturated solution.
a) Sketch of the free energy difference $\Delta G_{n,n_\rho}$ as a function of the number of solute molecules in the 
largest "connected" cluster (they can be ordered in a crystalline fashion or not) ($n_\rho$) and of the number of crystalline
molecules within the largest connected cluster ($n$). The one-step mechanism predicted by CNT (purple) is characterized by
a single free energy barrier for nucleation $\Delta G^*_{n,n_\rho,one-step}$. In contrast, the two-step nucleation requires
a free energy barrier $\Delta G^*_{n_\rho,two-step}$ to be overcome via a local density fluctuation of the solution, leading 
to a dense, but not crystalline-like precursor. The latter can be unstable (green) or stable (orange) with respect to the
liquid phase, being characterized by an higher (green) or lower (orange) free energy basin. 
Once this dense precursor has been obtained, the second step consists in climbing a second 
free energy barrier $\Delta G^*_{n,two-step}$ corresponding to the ordering of the solute molecules within the 
precursor from a disordered state to the crystalline phase. b) One-step (purple) and two-step (green and orange) nucleation 
mechanism visualized in the
density ($n_\rho$) - ordering ($n$) plane. The one-step mechanism proceeds along the diagonal, as both $n_\rho$ and $n$
increase at the same time, in such a way that a single free energy barrier has to be overcome. 
In this scenario, the supersaturated solution transforms continuously into the crystalline phase.
On the other hand, within a two-step nucleation scenario the system has to experience a 
favorable density fluctuation along ($n_\rho$) first, forming a disordered precursor which in a second step orders itself in 
a crystalline fashion, moving along the ($n$) coordinate and ultimately leading into the crystal.}
\label{Theoretical_Framework_FIG_3} 
\end{figure*}

In the original formulation of CNT, the system has to overcome a single free energy barrier, corresponding to a
crystalline nucleus of a certain critical size, as depicted in Fig.~\ref{Theoretical_Framework_FIG_3}.  When dealing
with crystal nucleation from the melt, it is rather common to consider the number of crystalline particles within the
largest connected cluster, $n$, as the natural reaction coordinate describing the whole nucleation process. In many
cases, the melt is dense enough so that local density fluctuations are indeed not particularly relevant, while the slow
degree of freedom is in fact the crystalline ordering of the particles within the liquid network. However, one can
easily imagine that in the case of e.g.  crystal nucleation of molecules in solutions the situation can be quite
different. Specifically, in a realistically supersaturated solution, a consistent fluctuation of the solute density
(concentration) could be required to just bring a number $n_\rho$ of solute molecules close enough to form a connected
cluster. Assuming that the molecules involved in such a density fluctuation will also order themselves in a crystalline
fashion on exactly the same timescale is rather counterintuitive.

In fact, the formation of crystals from molecules in
solution often happens according to a two-step nucleation mechanism that has no place in the original formulation of
CNT.  In the prototypical scenario depicted in Fig.~\ref{Theoretical_Framework_FIG_3}, a first free energy barrier
$\Delta G^*_{n_\rho,two-step}$ has to be overcome by means of a density fluctuation of the solute, such that a cluster
of connected molecules of size $n^*_\rho$ is formed. This object does not have any sort of crystalline order yet, and
according to the system under consideration can be either unstable or stable with respect to the supersaturated solution (see
Fig.~\ref{Theoretical_Framework_FIG_3}).  Subsequently, the system has to climb a second free energy barrier $\Delta G^*_{n,two-step}$ to order
the molecules within the dense cluster in a crystalline-like fashion.  A variety of different nucleation scenarios have
been loosely labeled as two-step, from crystal nucleation in colloids (see Sec.~\ref{COLL}) or Lennard-Jones liquids (see
Sec.~\ref{LJL}) to the formation of crystals of urea or NaCl (see Sec.~\ref{sec.MIS}), not to mention biomineralization (see e.g.
Refs.~\citenum{de_yoreo_crystal_2013,gebauer_pre-nucleation_2014}) and protein crystallization (see e.g. Refs.~\citenum{pan_nucleation_2005,vekilov_nucleation_2011}).

In all these cases, CNT as it is is simply not capable to deal with two-step nucleation. This is why in the last few
decades a number of extensions and/or modifications of CNT have been proposed and indeed successfully applied
in order to account for the existence of a two-step mechanism. In here we mention the phenomenological theory of
Pan \textit{et al.}~\cite{pan_nucleation_2005}, who wrote an expression for the nucleation rate assuming a free energy profile similar to the
one sketched in Fig.~\ref{Theoretical_Framework_FIG_3}, where dense metastable states are involved as intermediates toward the final crystalline structure.
The emergence of so-called pre-nucleation clusters (PNCs), i.e. stable states within supersaturated solutions which are known to play a very important role in the
crystallization of e.g. biominerals, has also been recently fit into the framework of CNT by Hu \textit{et al.}~\cite{hu_thermodynamics_2013}. 
The authors proposed a modified expression
for the excess free energy of the nucleus taking into account shape, size and free energy of the PNCs as well as the possibility for the PCNs to be either metastable or
stable with respect to the solution. A comprehensive review on the subject is offered by the work of Gebauer \textit{et al.}~\cite{gebauer_pre-nucleation_2014}.
It is worth noticing that these \textit{extensions} of CNT are mostly quite recent, as they have been triggered by overwhelming experimental evidence for two-step nucleation
mechanisms. 



\subsubsection{Heterogeneous Nucleation}
\label{THEOF_2}

CNT is also the tool of the trade for heterogeneous crystal nucleation, that is when nucleation occurs
thanks to the presence of a foreign phase (see Fig.~\ref{Theoretical_Framework_FIG_1}).  As a matter of fact, nucleation in liquids
happens heterogeneously more often than not, since in some cases the presence of foreign substances in contact with the
liquid can lower significantly the free energy barrier $\Delta G_{\mathcal{N}}^{\ast}$.  A typical example is given by
the formation of ice: as we shall see in Sec.~\ref{IceHON} and ~\ref{HIN}, it is surprisingly difficult to
freeze pure water, which invariably takes advantage of a diverse portfolio of impurities - from clay minerals to
bacterial fragments~\cite{murray_ice_2012} - in order to facilitate the formation of ice nuclei. 

Heterogeneous nucleation is customarily formulated within the CNT framework in terms of geometric
arguments~\cite{kalikmanov_nucleation_2013}.
Specifically:

\begin{equation}
\Delta G^{\ast}_{\mathcal{N} \text{(heterogeneous)}} =  \Delta G^{\ast}_{\mathcal{N} \text{(homogeneous)}} \cdot f(\theta)
\label{cnt_5}
\end{equation}

\noindent where $f(\theta)\leq1$ is the \textit{shape factor}, a quantity that accounts for the fact
that we have to balance three different interfacial free energies: $\gamma_{{\mathcal{S}}
\text{(crystal,liquid)}}$, $\gamma_{{\mathcal{S}} \text{(crystal,foreign phase)}}$ and $\gamma_{{\mathcal{S}}
\text{(liquid,foreign phase)}}$.  For instance, considering a supercooled liquid nucleating on top of an ideal
planar surface offered by the foreign phase, we obtain the so called Young's relation:

\begin{widetext}
\[
\gamma_{{\mathcal{S}} \text{(liquid,foreign phase)}} = \gamma_{{\mathcal{S}} \text{(crystal,foreign phase)}} + \gamma_{{\mathcal{S}} \text{(crystal,liquid)}} \cdot \cos\theta 
\label{CONT}
\]
\end{widetext}

\noindent where $\theta$ is the contact angle, i.e. a measure of the extent to which the crystalline nucleus \textit{wets} the
foreign surface.  Thus, the contact angle determines whether and how much it could be easier for a critical nucleus to
form in an heterogeneous fashion, as for $ 0\leq \theta < \pi$ the volume to surface energy ratio $\frac{\Delta
\mu_{\mathcal{V}}}{\gamma_{\mathcal{S}}}$ is larger for the spherical cap nucleating on the foreign surface compared to
the sphere nucleating in the liquid.  This simple formulation is clearly only a rough approximation of what happens in
reality. At first, the contact angle is basically a macroscopic quantity, of which the microscopic equivalent is in most
cases ill-defined on the typical length scales involved in the heterogeneous nucleation process~\cite{fnote10}. 
In
addition, in most cases the nucleus will not be shaped like a spherical cap, and to make things more complicated,
typically many different nucleation sites with different morphologies can exist on the same impurity.
Finally, the kinetic prefactor $\mathcal{A}_{kin}$ in heterogeneous nucleation becomes even more obscure,
as it is plausible that the foreign phase will affect the dynamical properties of the supercooled
liquid.

\subsubsection{Nucleation at Strong Supercooling}
\label{THEOF_3}

Moving towards strong supercooling, several things can happen to the supercooled liquid phase. Whether or not one can
avoid the glass transition largely depends on the specific liquid under consideration and on the cooling rate (see e.g.
Ref.~\citenum{cardinaux_interplay_2007}).  Assuming we are able to cool the system sufficiently slowly, hence
avoiding both the glass transition and crystal nucleation, one can in principle enter a supercooled regime where the
liquid becomes unstable with respect to the crystalline phase. This region of the phase diagram is known as the
\textit{spinodal region}, where the tiniest perturbation of e.g. the local density or the degree of ordering leads the
system toward the crystalline phase without paying anything in terms of free energy (see
Fig.~\ref{Theoretical_Framework_FIG_1}).  In fact, below a certain critical temperature $T_{\mathcal{SP}}$, the free
energy barrier for nucleation is zero, and the liquid transforms spontaneously into the crystal on very short timescales. The same picture
holds for molecules in solution, as nicely discussed by e.g. Gebauer \textit{et al.}~\cite{gebauer_pre-nucleation_2014},
and it cannot, by definition, be described by conventional CNT, according to which a small $\Delta
G_{\mathcal{N}}^{\ast}$ persists even at the strongest supercoolings~\cite{fnote11}.

While spinodal regimes have been observed in a variety of scenarios~\cite{spinobook} the existence of a proper spinodal
decomposition from the supercooled liquid to the crystalline phase has been debated (see e.g.
Ref.~\citenum{bartell_supercooled_2007}).  Enhanced sampling MD simulations~\cite{trudu_freezing_2006} which we discuss
in Sec.~\ref{LJL} have suggested that barrierless crystal nucleation is possible at very strong supercooling, while
other works claim that this is not the case (see e.g. Ref.~\citenum{peng_temperature-dependent_2008}).  In here,
we just note that at strong supercooling - not
necessarily within the presumed spinodal regime - a number of assumptions upon which CNT rely become, if not
erroneous, ill-defined. The list is long, and in fact a number of nucleation theories~\cite{kalikmanov_nucleation_2013}
able to at least take into account the emergence of a spinodal decomposition exist, although they have been mostly
formulated for condensation problems. In any case the capillarity approximation is most likely to fail at
strong supercoolings, since the size of the critical nucleus becomes exceedingly small, down to loosing its meaning in
the event of a proper spinodal decomposition. On top of that, we shall see for instance in Sec.~\ref{LJL} that the shape
of the crystalline clusters is anything but spherical at strong supercooling, and that at the same time the kinetic
prefactor assumes a role of great importance. As a matter of fact, nucleation at strong supercooling may very well be
dominated by $\mathcal{A}_{kin}$, as the mobility of the supercooled liquid is what really matters when the free energy
barrier for nucleation approaches vanishingly small values.  We care about strong supercooling because this is the
regime in which most computational studies have been performed.  Large values of $\Delta T$ imply high
nucleation rates and smaller critical nuclei, although as much as we move away from $T_{\mathcal{M}}$ we progressively
invalidate most of the assumptions of CNT. 

At this point, having highlighted some of the substantial approximations of CNT~\cite{fnote12},
and especially in light of its old age, the reader might be
waiting for us to introduce the much more elegant, accurate and comprehensive theories that experiments and simulations
surly embrace nowadays. Sadly, this is not the case. Countless flavors of nucleation theories exist.  Many of them, like
e.g. Dynamical Nucleation Theory~\cite{schenter_dynamical_1999}, Mean-field Kinetic Nucleation
Theory~\cite{kalikmanov_mean-field_2006}, and Coupled Flux
Theory~\cite{russell_linked_1968,peters_coupling_2011,wei_coupled-flux_2000,kelton_time-dependent_2000}, are mainly limited to condensation problems, and some
others have only rarely been applied to e.g. crystallization in glasses~\cite{Kelton2010279}, such as the Diffuse Interface
Theory~\cite{granasy_diffuse_1993,granasy_diffuse_1995}.  Several improvements upon CNT have been proposed, targeting
specific aspect like the shape of the crystalline nuclei ~\cite{prestipino_systematic_2012} or the finite size of the
non-sharp crystal-liquid interface~\cite{joswiak_size-dependent_2013}. Nucleation theories largely unrelated to CNT can
also be found, like classical Density Functional Theory (cDFT)~\cite{kahl_classical_2009,lowen_critical_2007,neuhaus_density_2014,lutsko_recent_2010}
(\textit{classical}, not to be misinterpreted with the celebrated quantum mechanical framework of Hohenberg and
Kohn~\cite{citeulike:1747293}).  A fairly complete inventory of nucleation theories, together with an excellent review
of nucleation in condensed matter, can be found elsewhere~\cite{kelton_nucleation_2010}. In here, we do not
enter into the details of any of these approaches, as indeed none of them has been consistently used to model
crystal nucleation in liquids. This is because CNT, while having many shortcomings, is a simple yet powerful
theory, able to capture at least qualitatively the thermodynamics and kinetics of nucleation for very different systems,
from liquid metals to organic crystals. It has been easily extended to include heterogeneous nucleation, and it is
fairly easy to take into consideration multicomponent systems like binary mixtures as
well~\cite{kelton_nucleation_2010,kalikmanov_nucleation_2013}.

\subsection{Experimental Methods}
\label{Experimental_Methods}

Several different experimental approaches have been employed
to understand the thermodynamics and the kinetics of crystal nucleation in liquids.
While this review discusses almost exclusively theory and simulations, we present in this section a concise overview of the state of the art 
experimental techniques, in order to highlight their capabilities as well as their limitations.

\begin{figure*}[t!]
\begin{center}
\includegraphics[width=15cm]{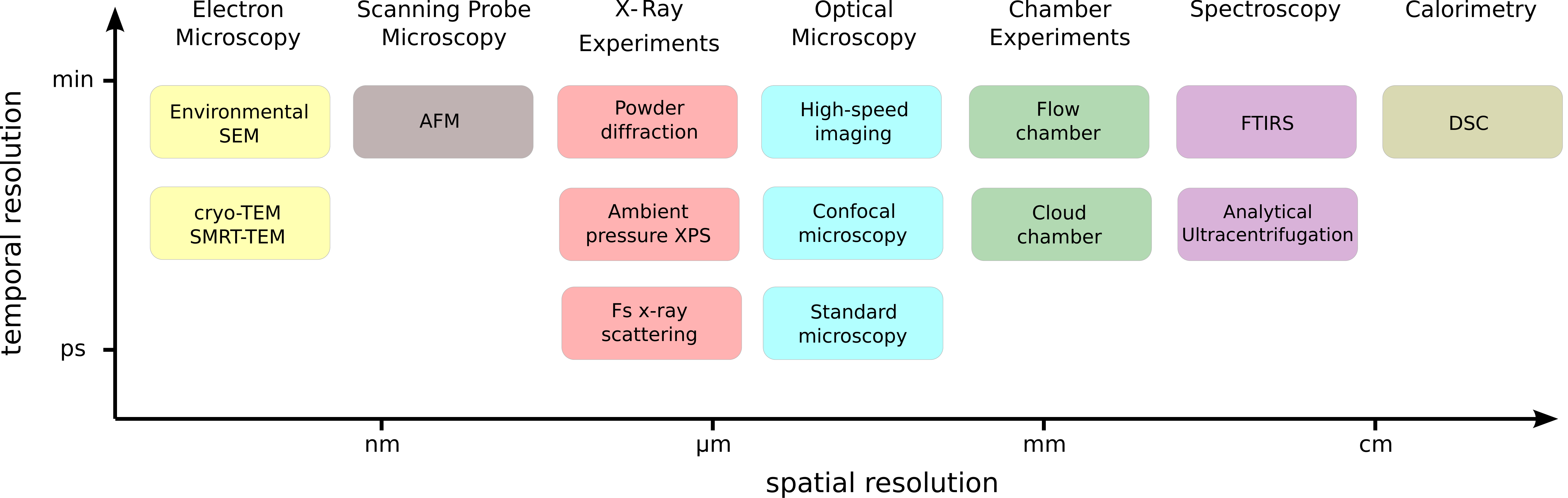}
\end{center}
\caption{Overview of some of the experimental methods that have been applied to characterize nucleation. 
A schematic of the spatial and temporal resolution typical of each 
approach is reported on the x- and y-axis, respectively.}
\label{fig.experiments.overview}
\end{figure*}

A schematic synopsis focusing on both spatial and temporal resolutions is sketched in
Fig.~\ref{fig.experiments.overview}, while an inventory of notable applications is reported in
Table~\ref{tab.experiments.overview}.
As we have said nucleation is a dynamical process usually happening on very small time and length scales (ns and nm respectively).
Thus, obtaining the necessary spatial and temporal resolutions is a tough technical challenge.

Indeed, true microscopic~\cite{fnote3} insight has rarely been achieved.
For instance, colloids offer a playground where simple microscopy can image the
particles involved in the nucleation events, which in turn happen on such long timescales (seconds) that a full
characterization in time of the process has been achieved~\cite{pusey_phase_1986,zhang_experimental_2014}. 
Specifically, confocal microscopy has lead to 3D
imaging of colloidal systems, unraveling invaluable information about e.g. the critical
nucleus size~\cite{gasser_real-space_2001,dinsmore_three-dimensional_2001}. 

In a similar fashion Sleutel \textit{et al.} achieved molecular resolution of the formation of two-dimensional glucose-isomerase crystals
by means of atomic force microscopy~\cite{sleutel_observing_2014}.
This particular investigation features
actual movies showing crystal growth as well as the dissolution of pre-critical clusters, also providing
information about the influence of the substrate. In addition, cryo-TEM 
techniques have recently delivered 2D snapshots of nucleation events at very low temperatures. In selected cases, where 
the timescales involved are again on the order of seconds, dynamical details have been obtained, as e.g. in the case of
CaCO$_3$~\cite{pouget_initial_2009,nielsen_situ_2014}, metal phosphate~\cite{chung2009multiphase} or magnetite~\cite{baumgartner_nucleation_2013}. 

However, more often than not crystal nucleation in liquids takes place within time windows too small (ns) to allow for a 
sequence of snapshots to be taken with high spatial-resolution instruments. In these cases, microscopic insights
cannot be obtained, and much more \textit{macroscopic} measurements have to be performed.
In this context, several experimental approaches aim at examining a 
large number of independent nucleation events for a whole set of rather small configurations of the system, basically
performing an ensemble average. For example in \textit{droplet experiments}, nucleation is
characterized as a function of time
or temperature. Freezing is identified for each nucleation event within the ensemble of available configurations 
by techniques such as femtosecond X-Ray scattering~\cite{sellberg_ultrafast_2014,laksmono_anomalous_2015},
optical microscopy~\cite{pusey_phase_1986,zhang_experimental_2014,campbell_is_2015,li_investigating_2012} 
or powder X-Ray diffraction~\cite{pusey_structure_1989,zhu_crystallization_1997,ehre_water_2010}. 
From these data the nucleation rate is often reconstructed by either measuring metastable zone widths~\cite{ildefonso2012nucleation,kadam2012new,kubota2008new,kashchiev2010effect,sangwal2011recent,peters2011supersaturation,sangwal2009novel,kashchiev2010dependence} 
or induction times~\cite{kashchiev1991induction,lindenberg2009effect,roelands2004development,roelands2006analysis,jiang2010crystal,teychene2008nucleation}, 
(several examples are listed in e.g. Refs.~\citenum{kashchiev2003review,davey2013nucleation,vetter2013modeling,kulkarni2013crystal,kubota2012effect}),
providing in this way a solid connection to theoretical frameworks such as CNT (see Sec.~\ref{THEOF}).

An essential technical detail within this class of measurements is that the volume available for each nucleation event
has to be as small as possible, in order to reduce the occurrence of multiple nucleation events within the same
configuration. High throughput devices such as the
lab-on-a-chip~\cite{stan_microfluidic_2009} can significantly improve the statistics of the nucleation events, thus
enhancing the capabilities of these approaches.

Another line of action focuses on the study of large, macroscopic systems. Freezing is detected
by e.g. differential scanning calorimetry~\cite{bogoeva-gaceva_nucleation_2001,davies_studies_2009,charoenrein_use_1989,marcolli_efficiency_2007,pinti_ice_2012,rasmussen_dsc:_1976}, Fourier transform infrared spectroscopy (FTIRS)~\cite{ochshorn_towards_2006,manka_freezing_2012,bhabhe_freezing_2013,yang_effect_2008,fujiwara_paracetamol_2002}, analytical 
ultracentrifugation~\cite{gebauer_stable_2008} or
some flavor of chamber experiments~\cite{rogers_continuous-flow_2001,demott_african_2003,tobo_impacts_2012,lihavainen_homogeneous_2001,konstantinov2000analysis,finnegan2003new,murray_heterogeneous_2010,hiranuma_ice_2015}. In this case, the frozen fraction of the overall system and/or the nucleation
temperatures can be obtained, and in some cases nucleation rates have been extracted (see
Table~\ref{tab.experiments.overview}).

Finally, experimental methods that can detect nucleation and the formation of the crystal (predominantly by means of optical microscopy)
but do not provide any microscopic detail have helped to shed light on issues such as the role of the solvent or impurities.
This is usually possible by examining the amount
of crystalline phase obtained along with its structure.

Even though there are a large number of powerful experimental techniques and new ones emerging (e.g. ultrafast X-ray~\cite{sellberg_ultrafast_2014}),
it is still incredibly challenging to obtain microscopic level insight into nucleation from experiments. As we shall see now MD simulations provide
a powerful complement to experiment.

\newpage

\begin{table*}[t!]
\centering													
\begin{tabular}{l|l}
\cline{1-2}
Methods                        & Examples          																														\\ \hline
Confocal Scanning Microscopy   & colloids~\cite{gasser_real-space_2001,dinsmore_three-dimensional_2001}, oogenesis in \textit{Xenopus}~\cite{gard_organization_1991}															\\
AFM                            & glucose-isomerase~\cite{sleutel_observing_2014} 																						\\
SMRT-TEM, HREM                 & organic crystals~\cite{harano_heterogeneous_2012,nakamura_movies_2013}, metal phosphate~\cite{chung2009multiphase}  																\\
cryo-TEM                       & CaCO$_3$~\cite{pouget_initial_2009,nielsen_situ_2014}, magnetide~\cite{baumgartner_nucleation_2013}, mcm41~\cite{regev_nucleation_1996}\\
fs X-Ray Scattering            & ice~\cite{sellberg_ultrafast_2014,laksmono_anomalous_2015}              																\\
high speed VIS or IR imaging   & ice~\cite{bauerecker_monitoring_2008} 																									\\
Analytical Ultracentrifugation & CaCO$_3$~\cite{gebauer_stable_2008}             																						\\
Powder Diffraction             & colloids~\cite{pusey_structure_1989,zhu_crystallization_1997}, ice~\cite{ehre_water_2010}               								\\
FTIRS                          & ice~\cite{ochshorn_towards_2006,manka_freezing_2012,bhabhe_freezing_2013}, glycine~\cite{yang_effect_2008}, paracetamol~\cite{fujiwara_paracetamol_2002,nagy2008determination}\\
Optical Microscopy             & colloids~\cite{pusey_phase_1986,zhang_experimental_2014}, ice~\cite{campbell_is_2015,li_investigating_2012}      						\\
Ambient Pressure XPS           & ice~\cite{bluhm_premelting_2002,ketteler_nature_2007}               																	\\ 
DSC                            & glass fibers~\cite{bogoeva-gaceva_nucleation_2001}, hydrates~\cite{davies_studies_2009}, ice~\cite{charoenrein_use_1989,marcolli_efficiency_2007,pinti_ice_2012}, metal-alloy~\cite{rasmussen_dsc:_1976} \\ 
Environmental SEM              & CaP~\cite{barrere_nano-scale_2004}, ice~\cite{zimmermann_ice_2008}																		\\
Flow Chamber                   & ice~\cite{rogers_continuous-flow_2001,demott_african_2003,tobo_impacts_2012}, $n$-penthanol~\cite{lihavainen_homogeneous_2001}			\\
Cloud Chamber                  & ice~\cite{konstantinov2000analysis,finnegan2003new,murray_heterogeneous_2010,hiranuma_ice_2015}                  						\\ \hline \end{tabular}
\caption{Selection of experimental approaches that have been employed to study nucleation phenomena, along with some examples of systems examined.}
\label{tab.experiments.overview}
\end{table*}

\subsection{Molecular Dynamics Simulations}
\label{SIMM}
\subsubsection{Brute Force Simulations}
\label{Brute_force}

When dealing with crystal nucleation in liquids, atomistic simulations should provide a detailed picture of the
formation of the critical nucleus. 
The simplest way to achieve this is by so-called
brute force MD simulations, which involve cooling the system to below the freezing temperature and then following its time
evolution until nucleation is observed.
Brute force simulations are the
antagonist of enhanced sampling simulations, where specific computational techniques are used to alter the
dynamics of the system so as to observe nucleation on a much smaller timescale.  Monte Carlo (MC) techniques,
although typically coupled with enhanced sampling techniques, can be used to recover $\Delta
G_{\mathcal{N}}^{\ast}$~\cite{auer_numerical_2004,schilling_crystallization_2011,punnathanam_crystal_2006}, but the
calculation of $\mathcal{A}_{kin}$ requires other methods, like Kinetic Monte Carlo (KMC)~\cite{auer_prediction_2001}.
The natural choice to simulate nucleation event is instead in MD simulations, which provide directly
the temporal evolution of the system.  

MD simulations aimed at investigating nucleation are usually performed in the isothermal-isobaric ensemble NPT, where P (usually
ambient pressure) and $T<T_m$ are kept constant by means of a barostat and a thermostat respectively. Such computational
tweaking is a double-edged sword. 
In fact, nucleation and most notably crystal growth are exothermic processes~\cite{fnote2}, and within the length scale
probed by conventional atomistic simulations (1-10$^4$ nm) is necessary to keep the system at constant temperature.  On
the other hand, in this way dynamical and structural effects in both the liquid and the crystalline phases due to the
heat developed during the nucleation events are basically
neglected~\cite{zhang_microcanonical_2015,perez_molecular_2011,wedekind_finite-size_2006}. Although the actual
extent of these effects is not yet clear, forcing the sampling of the canonical ensemble is expected to be especially
dangerous when dealing with very small systems affected by substantial finite size effects.
More importantly, thermostats
and barostats affect the dynamics of the system. Small coupling constants and clever approaches (e.g. stochastic
thermostats~\cite{bussi_canonical_2007}) can be employed to limit the effects of the thermostats, but in general care
must be taken. The same reasoning applies for P and barostats as well. A density change of the system is usually
associated with nucleation~\cite{oxtoby_crystal_2003}, the crystalline phase being more (or less, in the case of e.g.
water) dense than the liquid parent phase. 

\begin{figure}[t!] \vspace{0.0cm} \begin{center} \includegraphics[width=8cm]{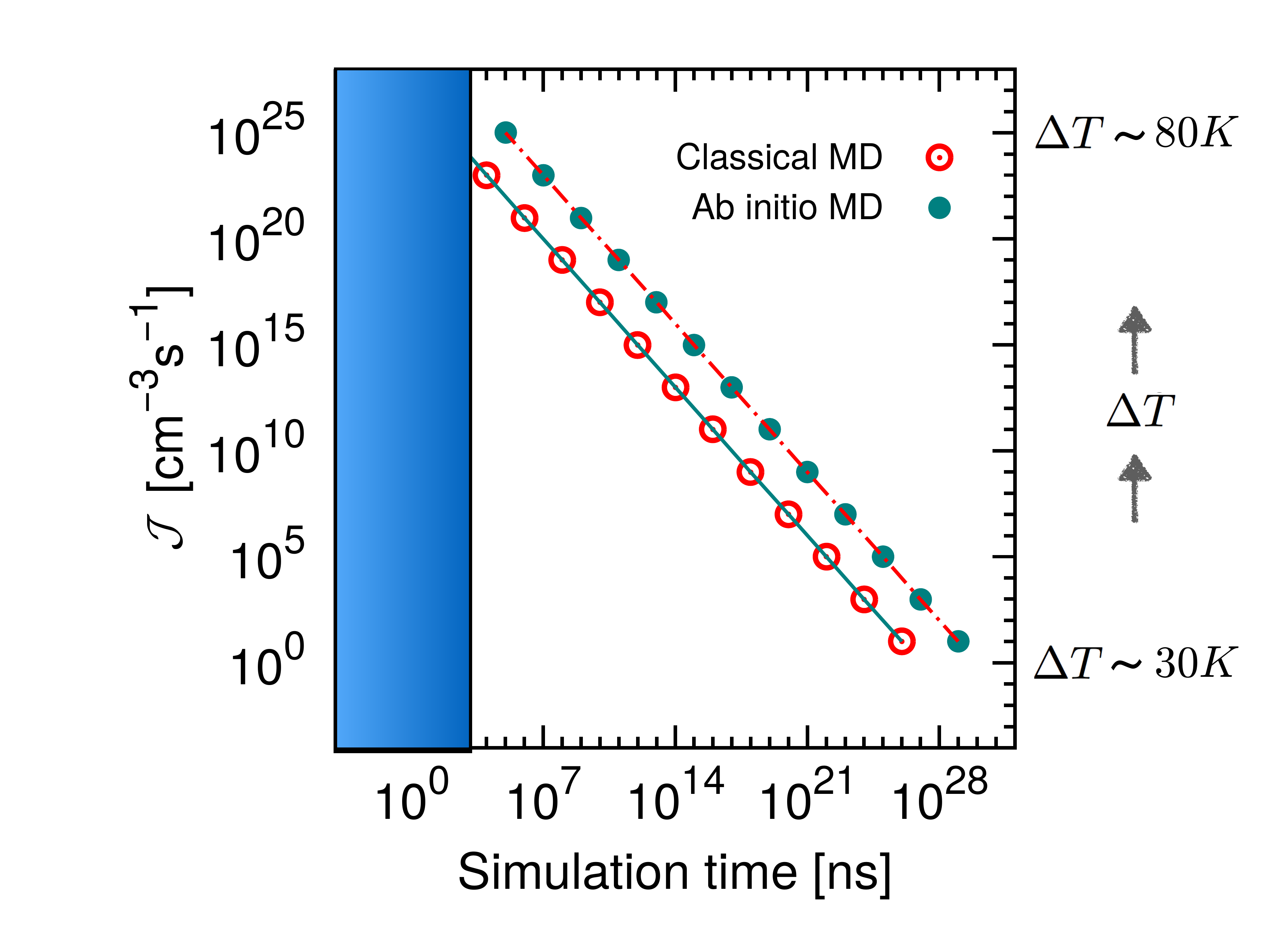} \end{center}
\caption{Nucleation rate $\mathcal{J}$ as a function of the simulation time needed within a MD simulation to observe a single nucleation event.
The blue shaded region highlights the approximate simulation times currently affordable by classical MD simulations; clearly
only very fast nucleation processes can be simulated with brute force MD. For homogeneous ice nucleation, $\mathcal{J}=10^0$
and $\mathcal{J}=10^{25}$ can typically be observed for $\Delta T=30 K$ and $\Delta T=80 K$ respectively. 10$^5$ and 10$^2$ molecules have been considered
in the derivation of classical and \textit{ab initio} simulation times respectively, together with the number density of a generic supercooled liquid
$\rho_{\mathcal{L}}=0.01 \ \text{molecules}\cdot$\AA$^{-3}$}
\label{FIG_Brute_force_2} \end{figure}

Three conditions must be fulfilled to extract $\mathcal{J}$ from brute force MD simulations:
\begin{enumerate} \item{The system must be allowed to evolve in time until spontaneous fluctuations lead to a nucleation
event.} \item{The system size must be significantly larger than the critical nucleus.} \item{A
significant statistics of nucleation events must be collected.} \end{enumerate} Each one of these conditions is 
surprisingly difficult to fulfill.  The most daunting obstacle is probably the first one because of the so-called \textit{timescale
problem}~\cite{nielaba_bridging_2002,abrams_enhanced_2013}. In most cases, nucleation is a rare event, meaning that it
usually happens on a very long timescale; precisely how long depends strongly on $\Delta T$.
A rough estimate of the number of simulation steps required to observe a nucleation event within a molecular dynamics run
is reported in Fig.~\ref{FIG_Brute_force_2}. Under the fairly optimistic assumption that classical MD simulations can
cope with up to $\sim 10^5$ molecules on a timescale of nano/micro-seconds, there is only a very narrow sets of
conditions for which brute force classical MD simulations could be used to investigate nucleation, usually only
at strong supercooling.  Timescales typical of first principles simulations, also reported in
Fig.~\ref{FIG_Brute_force_2} assuming up to $\sim 10^2$ molecules, indicate that unbiased \textit{ab initio}
simulations of nucleation events are unfeasible.

The second important condition is the size of the system.
The number of atoms (or molecules) in the system defines the timescale accessible to the simulation, and thus the severity of
the timescale problem. The reason we need large simulation boxes, i.e. significantly larger than the size of the
critical nucleus, is because periodic boundary conditions will strongly affect nucleation (and growth) if even the
precritical nuclei are allowed to interact with themselves. This usually leads to unrealistically high nucleation
rates. This issue worsens at mild supercooling, where the
critical nucleus size rapidly increases towards dimensions not accessible by MD simulations.

Third, it is not sufficient to collect information on just one nucleation event.
Nucleation is a stochastic event following a Poisson distribution (at least ideally, see
Sec.~\ref{THEOF}), and so to obtain the nucleation rate, one needs to pile up decent
statistics. 

Taking these issues into consideration, various approaches for obtaining $\mathcal{J}$ have emerged.
One approach, known as the Yasuoka-Matsumoto method~\cite{yasuoka_molecular_1998}, involves
simulating a very large system, so that different nucleation events can be observed within a single run.
In this case large simulation boxes are needed in order to collect sufficient statistics and to avoid spurious interactions between different nuclei.
Another family of methods involves running many different simulations using much smaller systems, which is usually computationally
cheaper.
Once a collection of nucleation events
has been obtained, several ways to extract $\mathcal{J}$ can be employed. The simplest ones (Mean
Lifetime~\cite{skripov_metastable_1974}, Survival Probability~\cite{horst_determination_2003,yi_molecular_2013} methods)
involve the fitting of the nucleation times to Poisson statistics. A more in-depth technique, the
so-called Mean First-Passage method~\cite{wedekind_new_2007} allows for a detailed analysis of the nuclei
population, but requires a probability distribution in terms of nucleus size.

The literature offers a notable number of works in which brute force MD simulations have been successfully applied. Most of
them rely on one way to circumvent the above mentioned issues, particularly the time scale problem.
As we shall see in Sec.~\ref{MDSIM}, in order to simulate nucleation events we almost always have to either choose a very simple system, 
or to increase
the level of approximation sometimes dramatically, for instance by coarse-graining the interatomic potential used.

\subsubsection{Enhanced Sampling Simulations}
\label{ESM}

In the previous section we introduced the timescale problem, the main reason 
brute force MD simulations are generally not feasible when studying crystal nucleation.
Enhanced sampling methods alter how the system
explores its configurational space, so that nucleation events can be observed within a reasonable amount of
computational time. Broadly speaking, one can distinguish between free energy methods and path sampling methods, both of which
have been extensively discussed elsewhere (see e.g. Refs.~\citenum{van_erp_dynamical_2012,abrams_enhanced_2013,dellago_transition_2009,schlick_molecular_2009}).
Thus, only the briefest of introductions is needed here.

Of the many enhanced sampling methods, only an handful have
been successfully used to compute crystal nucleation rates. This is because we need information about both
the thermodynamics of the system (the free energy barrier for nucleation $\Delta G_{\mathcal{N}}^{\ast}$) and the kinetics of the
nucleation process (the kinetic prefactor $\mathcal{A}_{kin}$). 
When dealing with
crystal nucleation in supercooled liquids, free energy based methods are rather common, such as umbrella
sampling (US)~\cite{torrie_monte_1974,torrie_nonphysical_1977,kumar_weighted_1992} or
metadynamics~\cite{laio_escaping_2002,laio_metadynamics:_2008,barducci_well-tempered_2008}. In both cases, and indeed in
almost all enhanced sampling methods currently available, the free energy surface of the actual system is coarse
grained by means of one or more order parameters, or collective variables. The choice of the order parameter is
not trivial and can have dramatic consequences. An external bias is then
applied to the system, leading to a modified sampling of the configurational space that allows for the reconstruction of
the free energy profile with respect to the chosen order parameter, and thus for the computation of the free energy
barrier. This approach has been successful in a number of cases.
However, there is a price to be paid: by introducing an extra term into the system Hamiltonian, the actual
dynamics of the system is to some extent hampered, and much of the insight into the nucleation mechanism is
lost. Besides, $\Delta G_{\mathcal{N}}^{\ast}$ is only half of the story. In
order to obtain $\mathcal{A}_{kin}$, one needs complementary methods, usually aimed at estimating the probability for the system
on top of the nucleation barrier - in the space of the selected order parameter - to get back to the liquid phase or
to evolve into the crystal. Most frequently such methods are based on some flavor of transition state
theory~\cite{eyring_activated_1935,wigner_transition_1938,anderson_statistical_1973,hanggi_reaction-rate_1990}, like the Bennet-Chandler
formulation~\cite{chandler_statistical_1978,charles_h._bennett_molecular_1977}, and require a massive set of MD or
KMC simulations to be performed.  

On the other hand, the ever growing family of path sampling methods can provide direct access to the kinetics of the
nucleation process. These approaches rely again on the definition of an order parameter, but instead of applying an
external bias potential, an importance sampling is performed so as to enhance the naturally occurring fluctuations of the
system. 
Within the majority of the path sampling approaches used nowadays, including 
Transition Interface Sampling~\cite{erp_novel_2003,moroni_investigating_2004,juraszek_efficient_2013} (TIS) and Forward Flux
Sampling~\cite{allen_simulating_2006,allen_forward_2009} (FFS),  
the ensemble of paths connecting the liquid and the crystal is divided into a series of interfaces according to
different values of the order parameter. By sampling the probability with which the system crosses each one of these
interfaces, a cumulative probability directly related to the nucleation rate can be extracted. Other path sampling
techniques such as Transition Path Sampling~\cite{dellago_efficient_1998,bolhuis_sampling_1998} (TPS) rely instead on the sampling of the
full ensemble of the reactive trajectories.
In both cases, by means of additional
simulations involving e.g. committor analysis distribution~\cite{geissler_kinetic_1999} and thermodynamic
integration~\cite{kirkwood_statistical_1935}, one can subsequently extract the size of the critical nucleus and the free
energy difference between the solid and the liquid phases respectively.
Many different path sampling methods are available, but to our knowledge only TPS, TIS 
and most prominently FFS have allowed for estimates of crystal nucleation rates.
Under certain conditions, path sampling methods do not alter the
dynamics of the system, allowing for invaluable insight into the nucleation mechanism.  However, they are
particularly sensitive to the slow dynamics of strongly supercooled systems, which hinder the sampling of the paths and
makes them exceedingly expensive computationally.
While the last few decades have taught us that enhanced sampling techniques are effective in tackling crystal
nucleation of colloids (see Sec.~\ref{COLL}), Lennard-Jones melts (see Sec.~\ref{LJL}) or other atomic liquids (see Sec.~\ref{BLJP}), only 
recently have these techniques been applied
to more complex systems. 
One exceedingly challenging scenario for
simulations of nucleation is provided by the formation of crystals from solutions characterized by very low solute
concentration. While this occurrence is often encountered in real systems of practical interest, it is clearly extremely
difficult for MD simulations, even if aided by conventional enhanced sampling techniques, to deal with just a few solute
molecules dissolved within 10$^{3-6}$ solvent molecules. In these cases, the diffusion of the solute plays a role of
great relevance, and the interaction between solvent and solute can enter the nucleation mechanism itself. Thus, obtaining
information about the thermodynamics, let alone the kinetics of nucleation at very low solute concentration is presently a formidable task. However,
efforts have been devoted to further our understanding of e.g. solute migration and solute-nuclei association, as
demonstrated by the pioneering works of Gavezzotti \textit{et al.}~\cite{gavezzotti_crystal_1997,gavezzotti_molecular_1999} 
and more recently by Kawska \textit{et al.}~\cite{kawska_atomistic_2006,kawska_atomistic_2008}. In the latter work the
authors illustrate an approach the relies on the modeling of subsequent growth step, where solute particles (often
ions) are progressively added to the (crystalline or not) cluster. After each one of these growth steps, a structural
optimization of the cluster and the solvent by means of MD simulations is performed. While this method cannot offer
quantitative results in terms of the thermodynamics and/or the kinetics of nucleation, it can in principle provide
valuable insight into the very early stages of crystal nucleation when dealing with solutions characterized by very low
solute concentration.

On a final note, we mention seeded MD simulations. This technique relies on simulations in which a crystalline nucleus
of a certain size is inserted into the system at the beginning of the simulation.  While useful information about
critical nucleus size can be obtained in this
way~\cite{cacciuto_onset_2004,browning_nucleation_2008,kalikka_nucleus-driven_2012,sanz_homogeneous_2013}, the method
does not usually allow for a direct calculation of the nucleation rate.  However, seeded MD simulations are one of the
very few methods by which we can currently investigate solute precipitate nucleation (see e.g. Knott \textit{et
al.}~\cite{knott_homogeneous_2012}). There, the exceedingly slow attachment rate of the solute often prevents both free
energy as well as transition path sampling enhanced sampling methods from being applied effectively.

As we shall see in the next few sections, the daunting computational cost, together with the delicate choice of order parameter and
the underlying framework of CNT, still make enhanced sampling simulations of crystal nucleation in liquids
an intimidating challenge.

\section{Selected Systems}
\label{MDSIM}
We have chosen to review different classes of systems, which we shall present in order of increasing complexity.  
We start in Secs.~\ref{COLL} and~\ref{LJL} with colloids and Lennard-Jones liquids respectively. These systems are described
by simple interatomic potentials that allow large scale MD simulations, and thus with them many aspects of CNT
can be investigated and nucleation rates calculated.  In some cases, the latter can can be directly compared to experimental
results. As such, colloids and Lennard-Jones liquids represent a sort of benchmark for MD simulations of crystal nucleation in liquids, although we shall see that our understanding of crystal nucleation is far from satisfactory even
within these relatively easy playgrounds.
In Sec.~\ref{BLJP} we discuss selected atomic liquids of technological interest such as liquid metals,
supercooled liquid silicon and phase change materials for which nucleation happens on exceedingly small timescales.
As the first example of a molecular system, we then focus of the most important liquid of them all, water. 
We review the body of computational work devoted
to unravel the homogeneous (Sec.~\ref{IceHON}) as well as the heterogeneous (Sec.~\ref{HIN}) formation of ice, offering an
historical perspective guiding the reader through the many advances that have furthered our understanding of ice
nucleation in the last decades.

Following this, we present an overview of nucleation from solution
(Sec.~\ref{sec.MIS}),
where simulations have to deal with solute and solvent. We take into account systems of great practical relevance such as urea
molecular crystals, highlighting the complexity of the nucleation mechanism which is very
different from what CNT predicts. 
Finally, Sec.~\ref{sec:gas-hydrates} is devoted to 
the formation of gas hydrates. 

As a rule of thumb, increasing the complexity of the system raises more questions about the validity of the assumptions
underpinning CNT.
The reader will surely notice that simulations
have revealed many drawbacks of CNT along the way, and that reaching decent agreement for the nucleation rate
$\mathcal{J}$ between experiments and simulations still remains a formidable task.

\subsection{Colloids}
\label{COLL}


Hard sphere model systems take a special place in nucleation studies.  One reason for this is the simplicity of the
interatomic potential customarily used to model them: the only interaction a hard sphere particle experiences comes from
elastic collisions with other particles.  Because there is no attractive force between particles, a hard sphere system
is entirely driven by entropy.  As a consequence, the phase diagram is very simple, and can be entirely described
with one single parameter, the volume fraction $\Phi$.  Only two different phases are possible: a fluid and a crystal.
At a volume fraction $\Phi < 0.494$ the system is in its fluid state, at $0.492 < \Phi < 0.545$ it will be a mix between
fluid and crystalline states, and at $\Phi > 0.545$ the thermodynamically most stable phase is the crystal.  The
transformation from fluid to crystal happens via a first order phase transition \cite{hoover_melting_1968}.
Despite their simplicity, systems behaving like hard spheres can be prepared experimentally.  Colloids made of polymers
are commonly used for this purpose, the most prominent example being  poly-methylmetharylate (PMMA) spheres coated with
a layer of poly-12-hydroxystearic acid.  After synthesizing the spheres, they are dissolved in a mixture of cis-decaline
and tetraline, which enables the use of a wide range of powerful optical techniques in order to, for example,
investigate nucleation~\cite{antl_preparation_1986, phan_phase_1996}.  The possibility of using these large hard spheres
in nucleation experiments has two major advantages: First, a particle size larger than the wavelength used in microscopy
experiments makes it possible to track the particle trajectories in real space.  Second, the timescale of the nucleation
process happens on the second scale, which allows the experimentalists to follow the complete nucleation process in
detail.  Compared to other systems it is therefore possible to directly observe the critical nucleus by e.g. confocal
microscopy (see Sec.~\ref{Experimental_Methods}) which is of
crucial importance for understanding nucleation.  These qualities of hard sphere systems make them 
ideal candidates to advance our understanding of nucleation.  As such it is
not surprising that the freezing of hard spheres is better characterized than any other nucleation scenario,
and in fact a number of excellent reviews in this field 
already exist~\cite{palberg_colloidal_1997,palberg_crystallization_1999,anderson_insights_2002, sear_nucleation:_2007, gasser_crystallization_2009, palberg_crystallization_2014}.
Our aim here is therefore not to give a detailed overview of the field, but to highlight some of the milestones and key
discoveries and connect them to other nucleation studies.  In order to keep the discussion reasonably brief, we limit
the latter to neutral and perfectly spherical hard sphere systems. However, we note that a sizable amount of work has
been devoted to a diverse range of colloidal systems, such as non-spherical particles~\cite{Bettina_hard_cuboids_2004,
haji_2009_hard_tetrahedra, damasceno_2012_HS_self_assembly, agarwal_2011_mesophase_behaviour, ni2011crystal,
thapar_2014_rotator_phase_nucleation}, charged particles and mixtures
of different colloidal particles~\cite{auer2002crystallization, wette2005crystallization, punnathanam2006crystal,
wette2007nucleation, williams2008crystallization, peters2009competing} just to name a few.

\begin{figure}
\centering
\includegraphics[width=0.50 \textwidth]{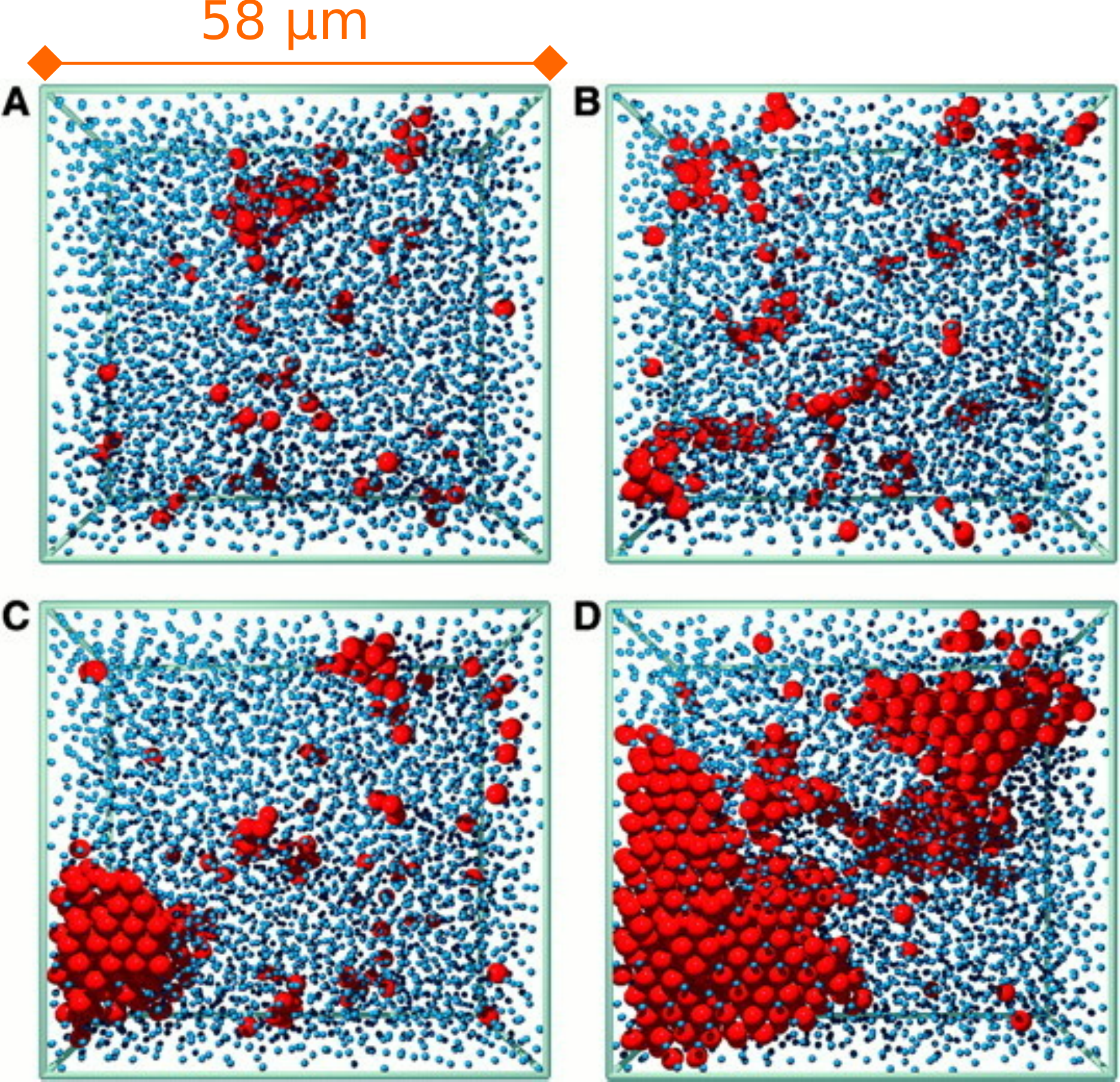}
\caption{Crystallization of PMMA with $\Phi=0.45$ observed by confocal microscopy.  Red (large) and blue (small) spheres show crystal-
and liquid-like particles respectively.
The size of the observed volume is 58 $\mu m$
by 55 $\mu m$ by 20 $\mu m$, containing about 4000 particles.  After shear melting the
sample, snapshots were taken after 20 min (A), 43 min (B), 66 min (C) and 89 min (D).  The time series shows how an
aspherical nucleus forms and grows over time.  Reprinted with permission from Ref.~\citenum{gasser_real-space_2001}. Copyright 2001, The American Association for the Advancement of Science.}
\label{fig: nucleation}
\end{figure}

Readers interested in the state of the art around 2000 are referred to other reviews~\cite{palberg_colloidal_1997,
palberg_crystallization_1999}.  In the early 2000's two major advances in the field were made, one on the
theoretical side, the other experimentally. In 2001 Auer and Frenkel~\cite{auer_prediction_2001} computed absolute nucleation rates of a
hard sphere system using KMC simulations. They did so by calculating $P_{crit}$, the
probability of forming a critical nucleus spontaneously, and $\mathcal{A}_{kin}$, the kinetic prefactor separately and
without any parameters.  This made a direct comparison between simulations and experiments possible.  The
outcome was surprising and worrisome.  They found that experimental and theoretical nucleation rates
disagreed by several orders of magnitude.  This was surprising, because simulations did really well in describing all
sorts of properties of hard spheres before.  It was worrisome because only very few sound
approximations were made by Auer and Frenkel to obtain their nucleation rates.  Their theoretical approach seemed to be as
good as it gets.  The authors' suggestion, that the problem lay in experiments, or more precisely in the interpretation
of experiments, showed a possible way to resolve the discrepancy.  In the same year, Gasser \emph{et
al.}~\cite{gasser_real-space_2001} conducted ground-breaking experiments.  The authors imaged the
nucleation of a colloidal suspension in real space using confocal microscopy.  Four snapshots of their system containing
approximately 4000 particles are shown in Fig.~\ref{fig: nucleation}.
This was a big step, because previously nucleation had been investigated indirectly, using for example the structure factor
obtained from light scattering experiments.  In their study, they were able to directly measure the size of a critical
nucleus for the first time. Achieving sufficient temporal and spatial resolution at the same time so far is only possible for colloidal systems 
(for more details about experimental techniques see Sec.~\ref{Experimental_Methods}). 
They found that the nucleus was rather aspherical with a rough surface, both of these
effects are completely neglected in CNT. Note that the aspherical nucleus also appears in e.g. LJ systems (see sec 2.2.1). In addition, a random hybrid close-packed structure (RHCP) stacking for the hard spheres was observed, in good
agreement with Auer and Frenkel~\cite{auer_prediction_2001}.  This is interesting, because slightly different systems
such as soft spheres and Lennard Jones particles seem to favor BCC stacking. However,
Gasser's
study did not resolve the discrepancy between experimental and simulated nucleation
rates, as their results were in agreement with earlier small-angle light scattering experiments~\cite{schatzel_density_1993}.  

Much of the subsequent work focused on trying to resolve this discrepancy between experiments
and simulation.  
A step forward was made in 2006 and 2007~\cite{schope_two-step_2006, schope_effect_2007}. Sch\"ope \emph{et
al.} found experimental evidence supporting a two step crystallization (see Sec.~\ref{THEOF_TS}) process in hard sphere systems.  Other systems
such as proteins or molecules in solution (see Sec.~\ref{sec.MIS}) 
were well known at that time to crystallize via a more complex mechanism than that assumed by CNT.  Even for
hard sphere systems, two step nucleation processes were reported before 2006~\cite{harland_observation_1995,
cheng_colloidal_2001, francis_bragg_2002}; the occurrence of that mechanism was attributed to details of the
polydispersity of the hard spheres however.  The new insight Sch\"ope \emph{et al.} provided in 2006 and 2007 was that
the two step nucleation process is general, and as such does not depend on either polydispersity or volume fraction.
In 2010, simulations performed by Schilling \emph{et al.}~\cite{schilling_precursor-mediated_2010} supported
these experimental findings.  Using unbiased MC simulations the authors were able to reproduce the evolution of the
structure factor from previous experiments.  Not even the simplest model system seemed to follow the traditional picture
assumed in CNT.  Could this two step mechanism explain, why the computational rates~\cite{auer_prediction_2001} disagree
with experiments?  At first this seems like a tempting explanation, because Auer and Frenkel had to introduce order
parameters to calculate absolute nucleation rates.  That however automatically presupposes a
reaction pathway, which might not necessarily match the nucleation pathway taken in experiments.
Filion \emph{et al.}~\cite{filion_crystal_2010} showed in the same year however, that very different computational
approaches (brute force MD, US and FFS which we described earlier (see Sec.~\ref{ESM})) led to the same nucleation rates, all in agreement with Auer and
Frenkel.  They therefore concluded that the discrepancy between simulations and experiments did not lie in the computational
approach employed by Auer and Frenkel.  They offered two possible explanations, one being that hydrodynamic effects,
completely neglected in the simulations, might play a role; the other possible difficulties in interpreting
the experiments.  Schilling  \emph{et al.}~\cite{schilling_crystallization_2011} tried to address one of the key issues
when comparing experiments with simulations: uncertainties and error estimation.  Whilst the determination of the most
characteristic quantity in hard sphere systems, the volume fractions, is straightforward for simulations, experimentalists are
confronted with a more difficult task there.  The typical error in determining the volume fraction experimentally is
about $\pm 0.004$, which translates into an uncertainty in the nucleation rate of about an order of magnitude. On taking these considerations into
account, the authors concluded that the discrepancy can be explained by statistical errors and uncertainties.

Does this mean the last 10 years of research tried to explain a discrepancy which actually is not there?
Filion \emph{et al.}~\cite{filion_simulation_2011} rightfully pointed out, that whilst the rates between experiments and
simulations coincide at high volume fraction, they still clearly disagree in the low volume fraction regime.  No simple
rescaling justified by statistical uncertainty could possibly resolve that.  In their paper, they also addressed a
different issue.  In a computational study in 2010, Kawasaki and Tanaka~\cite{kawasaki_formation_2010} obtained, by means of 
Brownian Dynamics~\cite{fnote14},
nucleation rates in good agreement with experiments, contrary to the nucleation rates computed by Auer and Frenkel using brute force MD.  
It should be
noted that they did not use a pure hard sphere potential, but a Weeks-Chandler-Andersen potential instead.  Was the
approximation of a hard sphere system, something that can never be fully realized in experiments, the problem all the
time?  What Filion \emph{et al.} showed is that different computational approaches (brute force MD, US and FSS) all lead to the same
nucleation rates, all of them in disagreement with what Kawasaki and Tanaka found.  Through a detailed evaluation of
their approach and that of Kawasaki and Tanaka they agreed that their rates are more reliable.  The discrepancy
was back on the table, where it still remains, and is as large as ever.  

For a detailed comparison between experimental and computational
rates, the reader is referred to Ref.~\citenum{palberg_crystallization_2014}.  What we want to leave the
reader with here is that still today, the disagreement between simulations and experiments in the simplest system
persists. It is worth mentioning, that this fundamental disagreement between simulations and experiments is not unique to colloids.
Other systems such as water (Sec.~\ref{IceHON} and~\ref{HIN}) and molecules in solution (Sec.~\ref{sec.MIS}) 
also show discrepancies of several orders of magnitude in nucleation rates. 
This long standing debate is of great relevance to all investigations dealing with
systems modeled via any flavor of hard sphere potential. A notable example in this context is the crystallization
of proteins, which are usually treated as hard spheres. While basically neglecting most of the complexity of
these systems, this substantial approximation has allowed for a number of computational
studies~\cite{shiryayev_crystal_2004,rosenbaum_protein_1996,piazza_interactions_2000,lomakin_liquid-solid_2003,liu_toward_2010,liu_self-assembly-induced_2009,george_predicting_1994,doye_controlling_2007,dixit_crystal_2000,dixit_comparison_2001,chang_determination_2004} that, although outside
the scope of this review, certainly contributed to furthering our understanding of the self-assembly of biological
particles. 
\subsection{Lennard-Jones Liquids}
\label{LJL}

Having discussed hard spheres, the first step towards more realistic systems involves the inclusion of attractive interactions. 
The Lennard-Jones liquid is a widely studied model system that achieves this. It can be seen as
the natural extension of the hard-sphere model, to which it becomes equivalent when the strength of the
attractive interactions goes to zero.
LJ liquids were first reported in 1924~\cite{jones_determination_1924}, and since then
they have been the subject of countless computational studies. 
LJ potentials allow for exceedingly fast MD simulations, and 
a wide range of thermodynamic
information is available for them, such as the
phase diagram~\cite{baidakov_crystal_2010, hoef_free_2000, de_wette_crystallization_1969,
khrapak_accurate_2011, luo_nonequilibrium_2004} and the interfacial free energy~\cite{morris_anisotropic_2003,
davidchack_direct_2003, broughton_molecular_1986}. 

The stable structure of the LJ system up to $T_{\mathcal{M}}$
is a face centered cubic (FCC) crystal, slightly less stable in free energy is an 
hexagonal close packed (HCP) structure which in turn is significantly more stable then a third body centered cubic (BCC) 
phase~\cite{bolhuis_entropy_1997,
desgranges_controlling_2007}. 
With his study of liquid argon in 1964, Rahman reported what is probably the first LJ MD simulation~\cite{rahman_correlations_1964}. 
His findings showed good agreement with experimental data for the pair distribution function and the self
diffusion coefficient, thus demonstrating that LJ potentials 
can properly describe noble elements in their liquid form at ambient pressure. This conclusion was validated later by
Verlet~\cite{verlet_computer_1967} and McGinty~\cite{mcginty_molecular_1973}. As far as we know, 
nucleation of LJ liquids was investigated for the first time in 1969 by de
Wette~\cite{de_wette_crystallization_1969} and in 1976 by
Mandell \textit{et al.}~\cite{mandell_crystal_1976} for two-dimensional and three-dimensional systems, respectively. 

\vspace{0.5cm}
\noindent \textbf{\textit{Non-spherical Nuclei}} \\
Early simulations~\cite{wolde_numerical_1996,wolde_numerical_1999} investigating the condensation of LJ vapors into liquid 
already indicated a substantial discrepancy with CNT rates. It is worth 
noticing that the order parameter for crystal-like particles presented by
ten Wolde \textit{et al.}~\cite{wolde_numerical_1996} fostered a considerable amount of later work devoted to improve the order parameters 
customarily used to 
describe crystal nucleation from the liquid phase (see e.g. Ref.~\citenum{lechner_accurate_2008}).
In 2008, Kalikmanov \textit{et
al.}~\cite{kalikmanov_argon_2008} compared CNT and cDFT (see Sec.~\ref{THEOF}) simulations with
condensation data of argon. It turned out that CNT spectacularly failed to reproduce experimental condensation
rates, underestimating them by up to 26 orders of magnitude. 
This disagreement triggered a number of computational studies 
aimed at clarifying the assumption of the sphericity of the critical nucleus within the freezing of LJ liquids.
By embedding pre-existing spherical clusters into supercooled LJ liquids, Bai and
Li~\cite{bai_test_2005,bai_calculation_2006} found values of the critical nucleus size in excellent agreement with CNT
within a broad range of temperatures. However, these results have been disputed by e.g.  the umbrella sampling
simulations of Wang \textit{et al.}~\cite{wang_homogeneous_2007} and the path sampling investigation of Moroni \textit{et
al.}~\cite{moroni_interplay_2005}. In both cases the nuclei became less spherical with increasing $\Delta T$.  In
addition, Moroni \textit{et al.} pointed out that the critical nucleus size is determined by a non trivial interplay
between the shape, the size and the degree of crystallinity of the cluster. Such a scenario is clearly much more
complex than the usual CNT picture as it violates the capillarity approximation (see~\ref{THEOF_1}).  
Non-spherical nuclei have also been observed by Trudu \textit{et
al.}~\cite{trudu_freezing_2006}, who extended the conventional CNT formula to account for ellipsoidal nuclei. Such a
tweak gave much better estimations of both the critical nucleus size and the nucleation barrier. 
Recall that the shape of the critical nuclei can be observed experimentally in very few cases (see Sec.~\ref{Experimental_Methods} and~\ref{COLL}).

However, at very strong supercooling things fell apart because of the emergence of spinodal effects (see
Sec.~\ref{THEOF}). Note that CNT fails at strong supercooling even without the occurrence of spinodal
effects, as the time lag (transient time) needed for the structural relaxation into the steady-state regime results in a
time dependent nucleation rate~\cite{sear_quantitative_2014}. For instance, Huitema \textit{et
al.}~\cite{huitema_thermodynamics_2000} have shown that incorporating the time dependence into the kinetic prefactor
yields an improved estimate of nucleation rates. In fact, by embedding 
extensions to the original CNT framework one can recover in some cases a reasonable agreement between simulations and
experiments even at strong supercooling. As an example, Peng \textit{et al.}~\cite{peng_parameter-free_2010}  have also
shown that including enthalpy-based terms in the formulation of the temperature dependence of $\gamma_{\mathcal{S}}$
substantially improves the outcomes of CNT.

\vspace{0.5cm}
\noindent \textbf{\textit{Polymorphism}} \\
Another aspect that has been thoroughly addressed within crystal nucleation of LJ liquids is the structure of the
crystalline clusters involved. The mean-field theory approach of Klein and
Leyvraz~\cite{klein_crystalline_1986} suggests a decrease of the nucleus density as well as an increase of BCC character
when moving towards the spinodal region. These findings were
confirmed by the umbrella sampling approach of ten Wolde \textit{et al.}~\cite{ten_wolde_numerical_1995,wolde_numerical_1996,wolde_homogeneous_1999} 
who reported a BCC shell surrounding FCC cores.
Furthermore, Wang \textit{et al.}~\cite{wang_homogeneous_2007} have shown
that the distinction between the crystalline clusters and the surrounding liquid phase falls off as a function of
$\Delta T$. In fact, the free energy barrier for nucleation, computed by means of umbrella sampling (see
Sec.~\ref{ESM}) simulations, turned out to be of the order of $k_B T$ at $\Delta T=52$\%.  In addition, the
nuclei undergo substantial structural changes towards non-symmetric shapes, a finding validated by the metadynamics
simulations of Trudu \textit{et al.}~\cite{trudu_freezing_2006}. The same authors investigated the
nucleation mechanism close to the critical temperature $T_{\mathcal{S}\mathcal{P}}$
for spinodal decomposition (see Sec.~\ref{THEOF_3}), where the free energy basin corresponding
to the liquid phase turned out to be ill-defined, i.e. already overlapping with the free energy basin 
of the crystal.
Such a finding, basically suggested that below $T_{\mathcal{S}\mathcal{P}}$ there is no free energy
barrier for nucleation, indicating that the liquid is unstable rather than metastable and that the crystallization
mechanism has changed from nucleation towards the more collective process of spinodal decomposition (see
Sec.~\ref{THEOF_3} and Fig.~\ref{Theoretical_Framework_FIG_1}).

\begin{figure}[h!]
\begin{center}
\includegraphics[width=7cm]{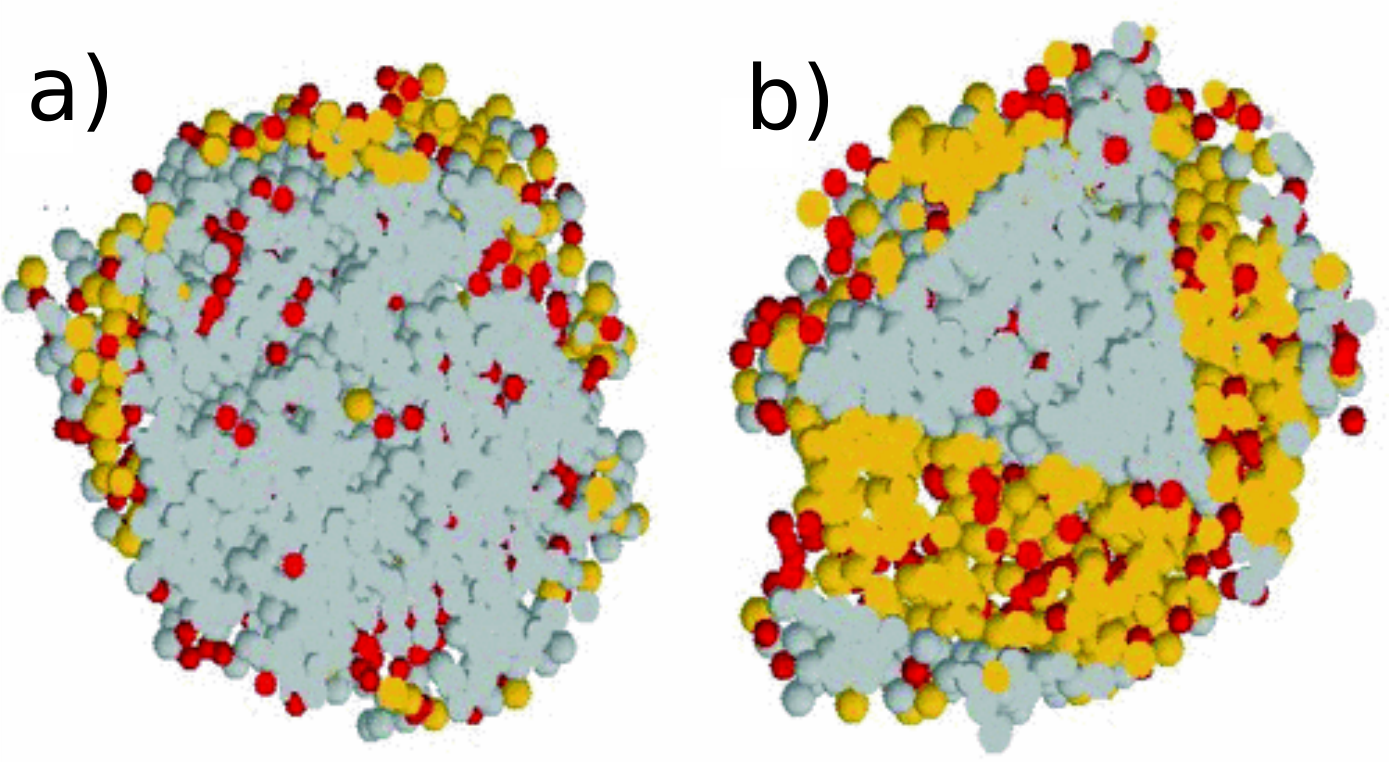}
\end{center}
\caption{Cross section of post-critical crystalline clusters of ~5000 LJ particles for 
a) $\Delta T=10$\% and b) $\Delta T=22$\%.
FCC-, HCP- and BCC-like particles are depicted in gray, yellow and red respectively. 
At $\Delta T=22$\% substantial HCP domains form within the crystallite,
while at $\Delta T=10$\% HCP particles can be found almost exclusively on the
surface of the FCC core.
Reprinted with permission from Ref.~\citenum{desgranges_controlling_2007}. Copyright 2007, American Physical Society.}
\label{LJPOLYY}
\end{figure}

Insights into the interplay between nucleation and polymorphism have been provided by the simulations of e.g. ten Wolde
\textit{et al.}~\cite{ten_wolde_numerical_1995,wolde_numerical_1996,wolde_homogeneous_1999}, suggesting that within the
early stages of the nucleation process the crystalline clusters are BCC-like, turning into FCC crystalline kernels
surrounded by BCC shells later on.  These findings were validated by Desgranges \textit{et
al.}~\cite{desgranges_molecular_2006} and Wang \textit{et al.}~\cite{wang_homogeneous_2007}.

More recently, Wang \textit{et al.}~\cite{wang_density_2013} performed a cDFT study in order to
determine the difference between the free energy barrier for nucleation required for the creation of a FCC or a BCC
critical nucleus.  In addition the difficulty for nucleation of the three different crystal orientations for FCC was
ranked (100) $>$ (110) $>$ (111). These studies confirm the presence of a two-step mechanism (see Sec.~\ref{THEOF_TS}) and the
validity of Ostwald's step rule~\cite{fnote4} for the LJ model. As we will see
later (e.g. homogeneous ice nucleation, Sec.~\ref{IceHON}), nucleation via metastable phases has also
been observed for more complicated liquids.  Important contributions regarding polymorph control during crystallization
have been made by Desgranges and
Delhommelle~\cite{desgranges_controlling_2007,desgranges_molecular_2006,delhommelle_crystal_2011} who investigated
nucleation under different thermodynamic conditions. While keeping the temperature constant and altering the pressure,
they were able to influence the amount of BCC particles. This reached up to a point where the nucleus was almost purely
BCC-like. By calculating the BCC-liquid line in the phase diagram it was shown that these nucleation events had occurred
in the BCC existence-domain. Additionally the transformation from HCP to FCC during crystal growth, well after the
critical nucleus size has been reached, was studied by changing temperature under constant pressure conditions. As
depicted in Fig.~\ref{LJPOLYY}, at $\Delta T=10$\%
a small number of HCP atoms were observed surrounding the FCC core, while at $\Delta T=22$\% much larger HCP domains 
formed within the crystallite, suggesting that the conversion from HCP to FCC is hindered at higher temperatures.
On a final note, we underline that many findings related to polymorphism are often quite dependent on the choice of the
order parameters employed. This issue is not limited to LJ systems, and it is especially important when dealing with
 similarly dense liquid and crystalline phases (e.g. metallic liquids), where 
order parameters usually struggle to distinguish properly the different crystalline
phases from the liquid~\cite{delhommelle_crystal_2011}.  In particular, it remains to be seen whether the fractional BCC, FCC and HCP
content of the LJ nuclei we have discussed will stand the test of the last generation of order parameters.

\vspace{0.5cm}
\noindent \textbf{\textit{Heterogeneous Nucleation}} \\
Heterogeneous crystal nucleation has also been investigated for a variety of LJ systems.
For instance, Wang \textit{et al.}~\cite{wang_homogeneous_2007} calculated by means of umbrella sampling simulations (see Sec.~\ref{ESM}) 
the free energy barrier for heterogeneous nucleation of a LJ
liquid on top of an ideal impurity, represented by a single layer of LJ particles arranged in an honeycomb lattice.
By explicitly varying the lattice spacing $a_{Sub}$ of the substrate, 
they calculated
$\Delta G_{\mathcal{N}}^{\ast}$ as a function of $a_{Sub}-a_{Equi}$, $a_{Equi}$ being the lattice spacing of the equilibrium crystalline
phase~\cite{fnote5}. 
It turned out that $\Delta G_{\mathcal{N}}^{\ast}$ displays a minimum for $a_{Sub}-a_{Equi}=0$, while for large $a_{Sub}-a_{Equi}$
nucleation proceeds within the bulk of the supercooled liquid phase.
These findings support the early argument of the zero lattice mismatch introduced by Turnbull and Vonnegut~\cite{turnbull_nucleation_1952} 
to justify the striking effectiveness of AgI crystals in promoting heterogeneous ice nucleation. In fact, in several situations one can
define a disregistry or lattice mismatch $\delta$ as 

\begin{equation} 
\label{eqn.mismatch} 
\delta = \frac{a_{Sub}-a_{Equi}}{a_{Equi}} \ .
\end{equation}

\begin{figure}[h!]
\begin{center}
\includegraphics[width=9cm]{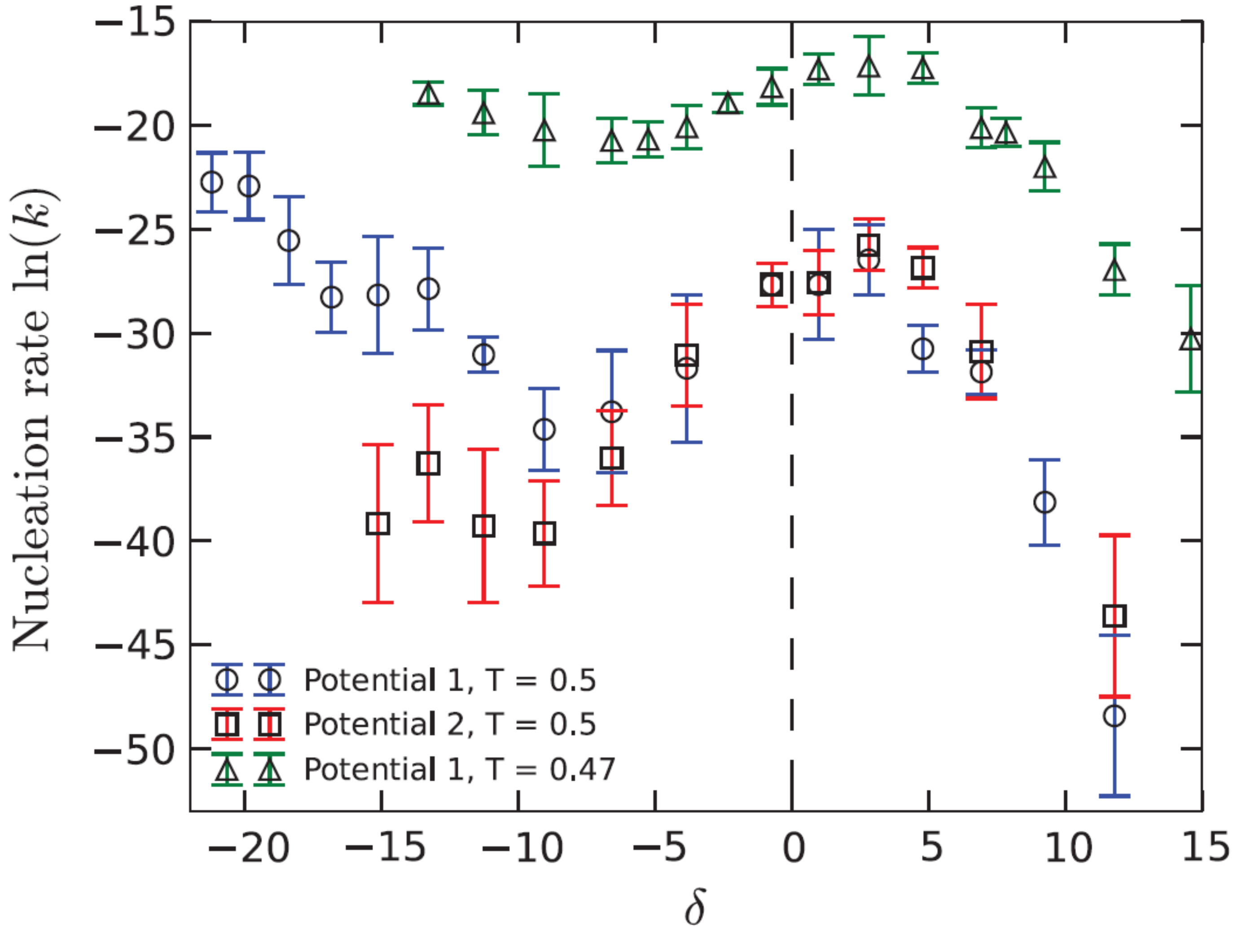}
\end{center}
\caption{Nucleation rates computed with the FFS method for a rigid hexagonal surface of LJ atoms in contact with a LJ
liquid. Potential 1 and 2 describe the interaction between substrate and liquid and differ only slightly by the $\sigma$
they use. Error bars are standard deviations from 5 FFS runs. These results show that the maximum in the nucleation rate
occurs at non-zero values of the lattice mismatch $\delta$. 
Reprinted with permission from Ref.~\citenum{mithen_computer_2014}. Copyright 2014, AIP Publishing LLC.}
\label{fig.lj.ffs_rates} \end{figure}

\noindent Values of $\delta$ close or even better equal to zero have often been celebrated as the main ingredient that makes a crystalline impurity particularly
effective in promoting heterogeneous nucleation. However, the universality of this concept has been severely questioned in the last few decades, as we
shall see in Sec.~\ref{HIN} for heterogeneous ice nucleation. 
Nonetheless, it seems that the zero lattice mismatch argument can hold for certain simple cases, as demonstrated by e.g. Mithen and Sear~\cite{mithen_computer_2014}, 
who studied heterogeneous nucleation of LJ liquids on the (111) and (100) faces of a FCC crystal by means of FFS simulations (see Sec.~\ref{ESM}).
They reported a maximum in the heterogeneous nucleation rate for a small, albeit non-zero, value of $\delta$ (see Fig.~\ref{fig.lj.ffs_rates}).
The difference between this study and Wang \textit{et al.}~\cite{wang_homogeneous_2007} is simply because many more values of 
$\delta$ were taken into account by Mithen and Sear~\cite{mithen_computer_2014}, thus allowing the
maximum of $\mathcal{J}$ to be determined more precisely.
On a different note, Dellago \textit{et al.}~\cite{jungblut_heterogeneous_2011} also investigated by TIS (see Sec.~\ref{ESM})
simulations the heterogeneous crystal nucleation of LJ supercooled liquids on very small crystalline impurities. They found that even tiny crystalline clusters
of just $\sim$10 LJ particles can actively promote nucleation, and that the morphology of the substrate
 can play a role as well. Specifically, while FCC-like clusters
were rather effective in enhancing nucleation rates, no substantial promotion was observed for icosahedrally ordered seeds.

MC simulations performed by Page and Sear~\cite{page_crystallization_2009} have demonstrated that confinement effects can be of great relevance as well.
They computed heterogeneous nucleation rates for a LJ liquid walled in two flat crystalline planes characterized by a certain angle $\theta_{Sub}$.
A maximum of $\mathcal{J}$ was found for a specific value of $\theta_{Sub}$, boosting the nucleation rate by several orders of magnitude with respect to the
promoting effect of a flat crystalline surface. In addition, different values of $\theta_{Sub}$ led to the formation of different crystalline polymorphs.

Finally, Zhang \textit{et al.}~\cite{zhang_wall-induced_2014} recently probed the influence of structured and structurless LJ potential
walls or nucleation rates. Both types of wall were found to increase the temperature at which nucleation occurs. However, this effect
became negligible when moving towards vanishingly small liquid-wall interaction strengths.
We shall see in Sec.~\ref{HIN} that the interplay between the morphology of the substrate and the strength of the liquid-substrate interaction can lead to a
diverse range of nucleation behavior.

\vspace{0.5cm}
\noindent \textbf{\textit{Finite Size Effects}} \\
MD simulations of LJ liquids are computationally cheap, making them the perfect
candidates to examine how finite-size effects impact upon crystal nucleation.
The seminal work of Honeycutt and Andersen~\cite{honeycutt_small_1986} took into account up to 1300 LJ particles
at $\frac{k_BT}{\epsilon_{LJ}}=0.45$, which turned out to be too few to completely rule out the effect of periodic 
boundary conditions. In fact, the authors suggested that extra care had to be taken because of the diffuseness of the interface
between the supercooled liquid phase and the crystalline nucleus, which can induce an artificial long range order in the
system leading to a nonphysically fast nucleation rate. These findings are particularly relevant as the critical nucleus size
at this $\Delta T$ is of the order of just a few tens of particles, representing a tiny fraction of the whole system.
Only a few years later Swope and Andersen~\cite{swope_106-particle_1990} investigated the same effects by taking into 
account up to $10^6$ LJ particles subjected to the same strong supercooling probed by Honeycutt and Andersen~\cite{honeycutt_small_1986}. 
According to their large scale MD simulations, 15,000 particles seem to be sufficient to avoid
finite-size effects. This outcome must be carefully pondered, as nowadays the vast majority of simulations dealing with
crystallization of realistic systems cannot obviously afford to take into account system sizes three order of magnitude larger than
the size of the critical nucleus. 
Consistent with Honeycutt and Andersen~\cite{honeycutt_small_1986}, 
Huitema \textit{et al.}~\cite{huitema_thermodynamics_2000} examined
the nucleation of a LJ liquid in a wide range of temperatures (70-140 K).
While nonphysical, instantaneous crystallization was observed for systems of the order of
$\sim$500 particles, simulation boxes containing about 10,000 particles seem to be free from finite-size effects.

It is also worth pointing out that Peng \textit{et
al.}~\cite{peng_parameter-free_2010} have recently described a novel class of finite-size effects unrelated to periodic
boundary conditions. In fact, they have shown that the equilibrium density of critical 
nuclei $\mathcal{P}_{Equi}$~\cite{fnote6}
can effectively influence the absolute value of nucleation rates.
Specifically, at very strong supercooling the critical nuclei will on average form very shortly after the transient time, while at
mild $\Delta T$ the stochastic nature of nucleation will lead to a consistent scatter of the nucleation times.
In other words, in this latter scenario either exceedingly large systems must be taken into account, or a sizable number of
independent simulations must be performed in order to deal with the long tails of the distribution of nucleation times.

\subsection{Atomic Liquids}
\label{BLJP}

Various interatomic potentials have been developed to deal with atomic
liquids. Examples include
the Sutton-Chen potentials~\cite{doi:10.1080/09500839008206493} for several metals and the Tosi-Fumi potential~\cite{FUMI196431} 
for molten salts like NaCl.
Terms accounting for the directionality of covalent bonds have been included in
e.g. the Stillinger-Weber potential~\cite{PhysRevB.31.5262} for Si, the bond order potentials
of Tersoff~\cite{PhysRevB.37.6991,PhysRevB.39.5566} for Si, GaAs and Ge or the reactive potential of 
Brenner~\cite{0953-8984-14-4-312} for carbon-based systems. Another class of
interatomic potential is based on the concept of local electronic density, and includes for instance the
Finnis-Sinclair potentials~\cite{doi:10.1080/09500839008206493,doi:10.1080/01418618708204464} for metallic systems,
the whole family of the Embedded Atoms Method (EAM) potentials~\cite{Daw1993251} and
the Glue potential~\cite{0295-5075-26-8-005,PhysRevLett.57.719} for Au and Al.

\begin{figure*}[t!]
\begin{center}
\includegraphics[width=14cm]{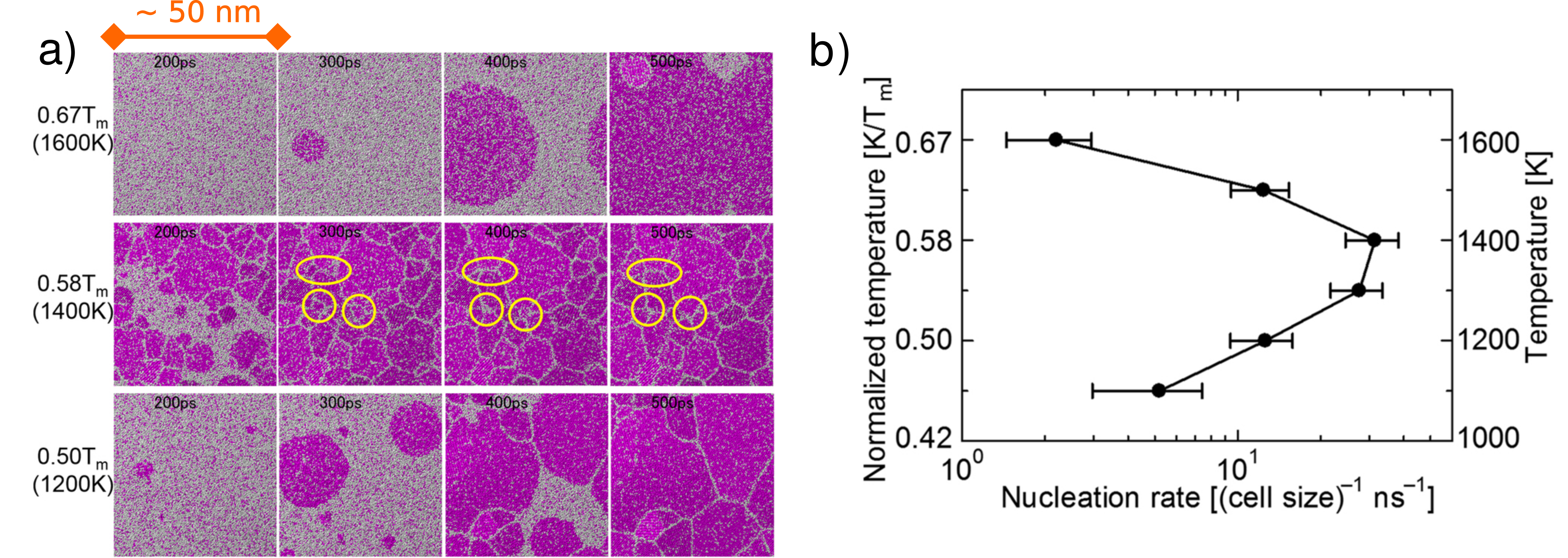}
\end{center}
\caption{Crystal nucleation of supercooled Fe by means of large scale MD simulations.
a) Snapshots of trajectories at different temperatures.
Crystalline (BCC) atoms are depicted in purple. Yellow circles highlight small crystalline grains doomed to be incorporated into
the bigger ones later on because of grain coarsening.
b) Nucleation rate as a function of temperature. 
Reprinted with permission from Ref.~\citenum{shibuta_million-atom_2014}. Copyright 2015, Nature Publishing Group.}
\label{LJHUGE}
\end{figure*}

Many of these potentials are still incredibly cheap in terms of computer time, thus allowing for large scale, unbiased
MD simulations. Recently, massively parallel MD runs succeeded in nucleating supercooled liquid Al~\cite{hou_formation_2015}
and Fe~\cite{shibuta_homogeneous_2015} using an EAM and a Finnis-Sinclair potential respectively. 
As up to $10^6$ atoms were taken into account, actual grain boundaries were observed,
providing unprecedented insight in to the crystal growth process.
The nucleation of BCC Fe crystallites and the evolution of the resulting grain boundaries at different temperatures
can be appreciated in Fig.~\ref{LJHUGE}a. The sizable dimension of the simulation boxes ($\sim$50 nm) allowed
nucleation events to be observed within hundreds of ps, and grain coarsening, i.e. the process by which small crystallites end up 
incorporated into the bigger ones, is also clearly visible. Mere visual inspection of the nucleation
trajectories depicted in Fig~\ref{LJHUGE}a suggests different nucleation regimes as a function of temperature.
In fact, the same authors have calculated a temperature profile for the nucleation rate, shown in Fig~\ref{LJHUGE}b,
which demonstrates the emergence of a maximum of $\mathcal{J}$ characteristic of diffusion limited nucleation (see Sec.~\ref{THEOF_1}).

A field that has greatly benefited from MD simulations is the crystallization of metal clusters, as nicely reviewed by
Aguado \textit{et al.}~\cite{aguado_melting_2011}. For instance, it is possible to probe the interplay
between the size of the clusters and the cooling rate upon crystal nucleation and growth.  In this context,
Shibuta~\cite{shibuta_molecular_2012} reported three different outcomes for supercooled liquid Mo nanoparticles
- modeled by means of a Finnis-Sinclair potential - namely the formation of a BCC single crystal, a glassy state or a
polycrystalline phase. In some cases, nucleation rates obtained from simulations were consistent with CNT, as
in the case of Ni nanodroplets~\cite{yakubovich_kinetics_2013} where nucleation events were again observed by means of
brute force MD simulations using the Sutton-Chen potential.  The influence of the redox potential on the nucleation
process has also been investigated. Milek and Zahn~\cite{milek_molecular_2014} employed an enhanced flavor of the EAM
potential to study the nucleation of Ag nanoparticles from solution. They established that the outcome of nucleation
events is strongly influenced by the strength of the redox potential, able to foster either a rather regular FCC phase
or a multi-twinned polycrystal.  Similar to what was done for LJ liquids, the effects of confinement were assessed for Au
nanodomains modeled via the Glue potential by Pan and Shou~\cite{pan_single_2015}. According to their findings, smaller
domains facilitate crystal nucleation. L\"u and Chen~\cite{lu_surface_2012} have instead investigated surface
layering-induced crystallization of Ni-Si nanodroplets, using a modified EAM potential. It seems that for this
particular system atoms proximal to the free surface of the droplet assume a crystalline-like ordering on very short
timescales, thus triggering crystallization in the inner regions of the system. No such effect has been reported instead
in the case of surface-induced crystallization in supercooled tetrahedral liquids like Si and Ge, investigated by Li
\textit{et al.}~\cite{li_surface-induced_2009} via FFS simulations employing both Tersoff or Stillinger-Weber
potentials. The presence of the free surface facilitates crystal nucleation for this class of systems as well, but
surface layering was not observed. Instead, the authors claimed that the surface reduces the free energy barrier for
nucleation as it introduces a pressure-dependent term in the volume free energy change expected for the formation of the
crystalline clusters. 
The situation is quite different for surface induced ice nucleation, at least according to the
coarse-grained mW model of Molinero and Moore~\cite{molinero_water_2009}. In fact, Haji-Akbari \textit{et
al.}~\cite{haji-akbari_suppression_2014} have recently investigated ice
nucleation in free-standing films of supercooled mW water using both FFS and US, finding that in these systems
crystallization is inhibited in the proximity of the vapor-liquid interface. Very recently, Gianetti \textit{et
al.}~\cite{gianetti_computational_2016} extended
the investigation of Haji-Akbari \textit{et al.}~\cite{haji-akbari_suppression_2014} to the 
crystallization of a whole family of modified Stillinger-Weber liquids
with different degrees of \textit{tetrahedrality} $\lambda$, locating a crossover from surface-enhanced to bulk-dominated
crystallization in free-standing films as a function of $\lambda$. Another seminal study by Li~\cite{li_ice_2013}, again using FFS, focused
on homogeneous ice nucleation within supercooled mW water nano-droplets, where nucleation rates turned out to be
strongly size dependent and in general consistently smaller (by several orders of magnitude) then the bulk case. 
FFS was also applied by Li \textit{et al.}~\cite{li_nucleation_2009} to examine homogeneous nucleation of supercooled
Si. FFS has also been successful in predicting homogeneous crystal nucleation rates in molten NaCl, modeled via a
Tosi-Fumi potential by Valeriani \textit{et al.}~\cite{valeriani_rate_2005}. Large discrepancies between their results
and experimental nucleation rates can be appreciated when CNT is used to extrapolate the calculations to the milder
supercooling probed by the actual measurements. Given that the authors obtained consistent results using two
different enhanced sampling methods, this study hints again at the many pitfalls of CNT. 

\begin{figure}[t!]
\begin{center}
\includegraphics[width=7cm]{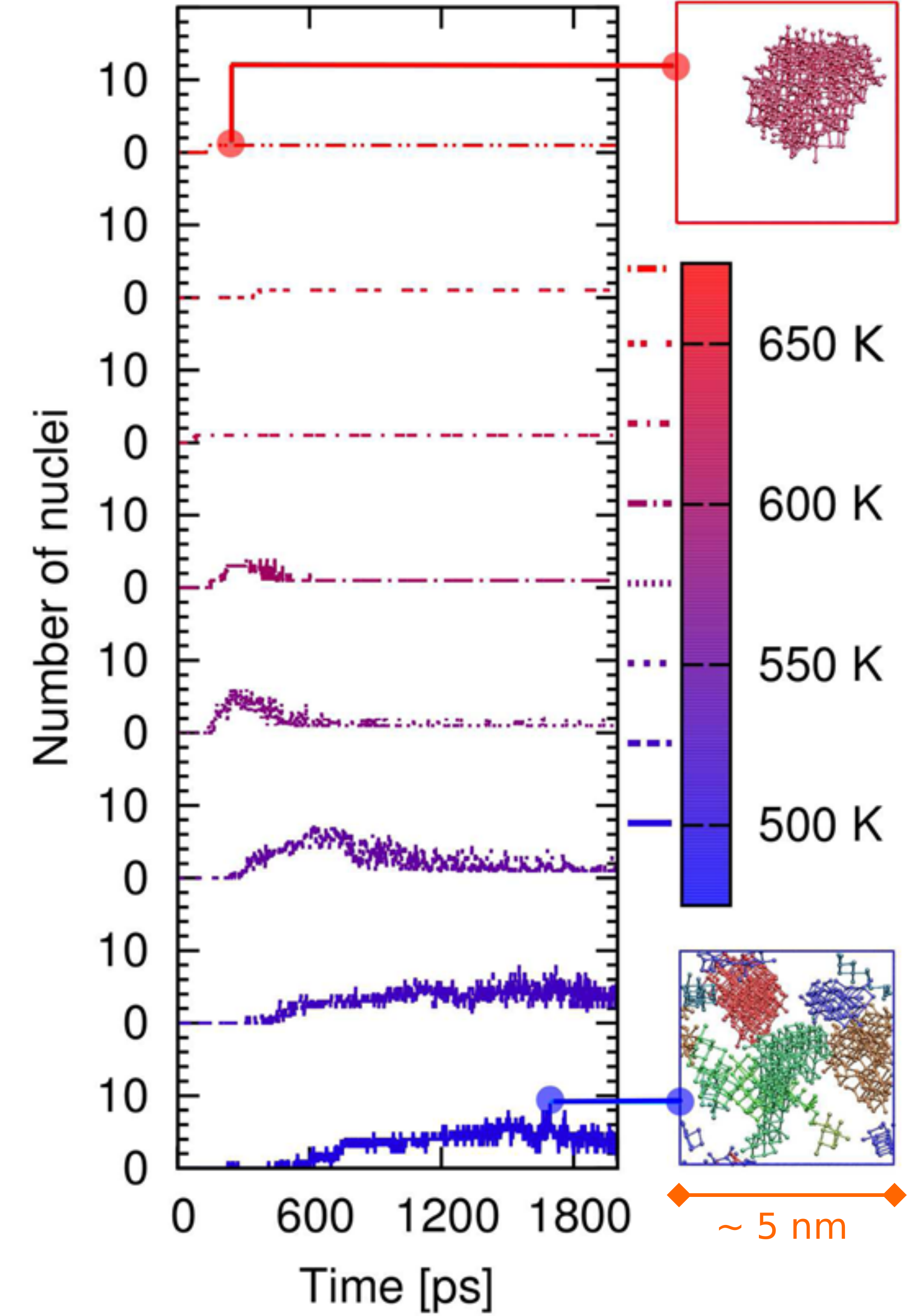}
\end{center}
\caption{Fast crystallization of supercooled GeTe by means of MD simulations with neural network derived potentials.
The number of crystalline
nuclei larger than 29 atoms at different temperatures in the supercooled liquid phase is reported as a function of time
(notice the exceedingly small timescale at strong supercooling).  Two snapshots at the highest and lowest temperatures
showing only the crystalline atoms are also reported. At high temperature, a single nucleus is present, while several
nuclei (each one depicted in a different color) appear at low temperature. 
The number of nuclei first increases and then decreases due to coalescence.
Reprinted with permission from Ref.~\citenum{sosso_fast_2013}. Copyright 2013, American Chemical Society.}
\label{SPCM} \end{figure}

\vspace{0.5cm}
\noindent \textbf{\textit{Phase Change Materials}} \\
A unique example of a class of materials for which nucleation can be effectively addressed by brute force MD simulations
is given by the so called phase change materials~\cite{raoux_phase_2010,lencer_design_2011}.  Phase change materials are
systems of great technological interest as they are widely employed in optical memories (e.g. DVD-RW) and in a
promising class of non volatile memories known as Phase Change Memory (PCM)~\cite{wuttig_phase-change_2007}, based
on the fast and reversible transition from the amorphous to the crystalline phase.
While crystal nucleation in amorphous systems, especially metallic and covalent glasses is
outside the scope of this review, we refer the reader to the excellent
work of Kelton and Greer~\cite{Kelton2010279} for a detailed introduction.
Phase change materials used in optical and electronic devices are typically tellurium
based chalcogenide alloys (see Ref.~\citenum{lencer_design_2011}). The family of the pseudobinary compounds
(GeTe)$_{x}$(Sb$_{2}$Te$_{3}$)$_{y}$ represents a prototypical system. Although both the structure and the dynamics of
these systems is far from trivial, nucleation from the melt takes place within the ns timescale for a wide range
of supercooling~\cite{raoux_phase_2010,lencer_design_2011,wuttig_phase-change_2007}. 
Thus with phase change materials we have a great opportunity to
investigate nucleation in a complex system by means of brute force MD
simulations. We note that the crystallization of these systems has been extensively characterized by different experimental
techniques (particularly TEM and AFM, see Sec.~\ref{Experimental_Methods}; the crystallization kinetics has also 
been recently investigated by means of ultrafast-heating 
calorimetry~\cite{orava_characterization_2012} and ultrafast X-ray imagining~\cite{zalden_how_2015}),
but because of the exceedingly high nucleation
rates, it is difficult to extract information about the early stages of the nucleation process. Thus, in this scenario
simulations could play an important role.
Unfortunately, Phase Change Materials require \textit{ab initio} methods or sophisticated  
interatomic potentials with first principles accuracy.
In fact, several attempts have been made to study nucleation in phase change materials by \textit{ab
initio} MD in very small systems~\cite{hegedus_microscopic_2008,lee_textitabinitio_2011}. While these
studies provided useful insights into the nucleation mechanism, severe finite size effects prevented the
full characterization of the crystallization process. The limited
length and timescale typical of first principles calculations has recently been outstripped in the case of the
prototypical phase change material GeTe by the capabilities of a neural network interatomic
potential~\cite{sosso_neural_2012}. Those potentials allow for a computational speedup of several orders of magnitude compared
to conventional \textit{ab initio} methods, while retaining an accuracy close to that of the
latter~\cite{behler_generalized_2007}. While nucleation rates have not been calculated yet, detailed investigations of
homogeneous and heterogeneous nucleation have already been reported~\cite{sosso_fast_2013,sosso_heterogeneous_2015}.
For instance, as shown in Fig.~\ref{SPCM}, a single crystalline nucleus formed in a 4000-atom model of 
supercooled liquid GeTe in the 625-675 K temperature regime. Within a few hundred picoseconds, 
several nuclei appear below 600 K, suggesting that the free energy
barrier for nucleation is vanishingly small for this class of materials just above the glass transition 
temperature. This is because of the \textit{fragility}~\cite{debenedetti_supercooled_2001} of the supercooled liquid, which displays a 
substantial atomic mobility even at large supercoolings~\cite{sosso_breakdown_2012}. Thus, in this particular case the kinetic prefactor
$\mathcal{A}_{kin}$ (see Eq.~\ref{kinp}) is not hindered that much by the strong supercooling,
while the free energy difference
between the liquid and the crystal $\Delta \mu_{\mathcal{V}}$ (see Eq.~\ref{cnt_1}) skyrockets as expected, leading
to the exceedingly high nucleation rates characteristic of these materials.

In conclusion, while MD simulations have by no means exhausted the field of crystal nucleation of atomic liquids,
they have certainly provided insight into a number of interesting systems and paved the way for the study
of more complex systems, as we shall see in the following sections.

\subsection{Water}
\label{WAH}
\subsubsection{Homogeneous Nucleation}
\label{IceHON}

Ice nucleation impacts many different areas, ranging from aviation~\cite{potapczuk2013aircraft,ye2013anti} to biological cells~\cite{padayachee2009cryopreservation} and Earth's climate~\cite{baker1997cloud, carslaw2002cosmic}.
It is therefore not surprising that a considerable body of work has been carried out to understand this fundamental process.
We cannot cover it all here, instead we give a general overview of the field, starting with
a discussion of nucleation rates.
This allows us to directly compare experiments and simulations and to identify strengths and weaknesses of different approaches.
We then discuss insights into the nucleation mechanism.

\subsubsection*{Nucleation rates}
\begin{figure*}[t!]
\begin{centering}
\centerline{\includegraphics[width=14cm]{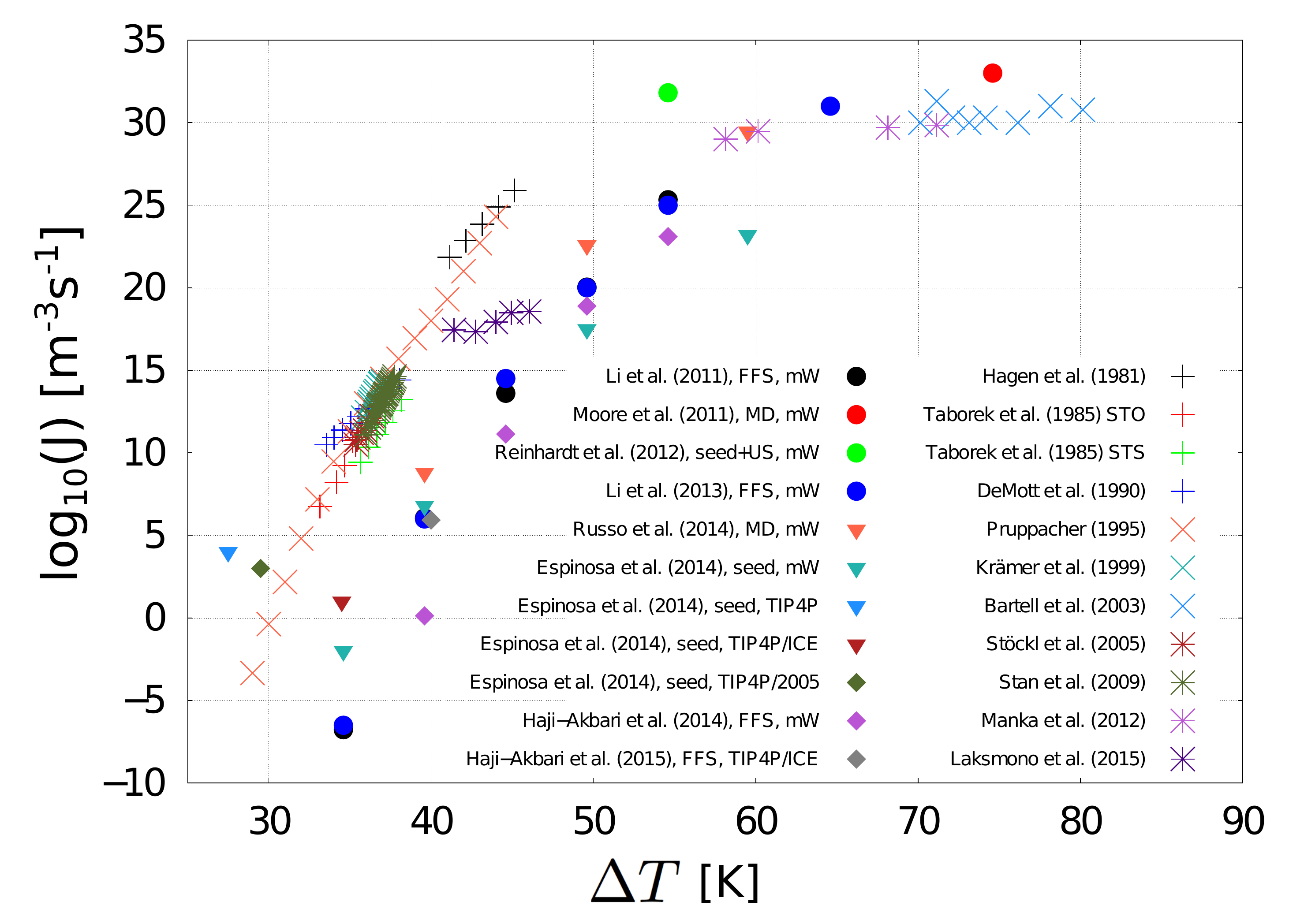}}
\par\end{centering}
\protect\caption{Compilation of homogeneous nucleation rates for water, obtained by experiments and simulations.
The x-axis shows the
supercooling with respect to the melting point of different water models or 273.15 K for experiment.  The y-axis
shows the logarithm of the nucleation rate in $m^{-3}s^{-1}$.  Rates obtained with computational approaches are shown as
filled symbols, experimental rates are shown otherwise.  For all computational studies, the method as well as the water
force field used are specified.  The nucleation study of Sanz \emph{et al.}~\cite{sanz_homogeneous_2013} is not included
in the graph, because this study was conducted at a small supercooling (20 K) which results in a very low estimated
nucleation rate far outside this plot (would correspond to -83 on y-axis).  Taborek \emph{et al.} performed measurements with
different setups, namely using sorbitan tristearate (STS) and sorbitan trioleate (STO) as surfactants, with different
droplet volumes (specified in the legend in $\mu m$, e.g. STS [surfactant] 6 [$\mu m$]).
Data for the graph was taken from Refs.~\citenum{li_homogeneous_2011, moore_structural_2011, reinhardt_free_2012, 
li_ice_2013, russo_new_2014, espinosa_homogeneous_2014, haji-akbari_direct_2015, haji-akbari_suppression_2014, hagen_homogeneous_1981, 
taborek1985nucleation, demott1990freezing, pruppacher1995new, kramer1999homogeneous, bartell2003nucleation, stockel2005rates, 
stan2009microfluidic, manka_freezing_2012}.}
\label{fig: rates}
\end{figure*}

An important goal for both experiments and simulations is to extract nucleation rates.
Experimental nucleation rates have been measured over a broad range of temperatures, most often with
micrometer sized water droplets so as to avoid heterogeneous nucleation. In Fig.~\ref{fig: rates} we bring together
nucleation rates obtained from various experiments. Along with this we also report computed nucleation rates. 

Accessing nucleation rates from MD simulations became feasible only in the past few years due to advances in force 
fields (such as the coarse-grained mW~\cite{molinero_water_2009} potential) and enhanced sampling techniques described earlier (see Sec.~\ref{ESM}).
These methods have therefore been widely used, not only for homogeneous but also heterogeneous nucleation studies (see Sec.~\ref{HIN}).
From the comparison of experimental and computational nucleation rates reported in Fig.~\ref{fig: rates} 
few things are apparent.
First, nucleation rates vary hugely with supercooling, by a factor of more than 10$^{35}$.
Second, nucleation rates differ substantially (approximately 10 orders of magnitude) between simulations (filled symbols) and experiments (crossed symbols) at relatively small supercooling ($\approx$ 30-50 K).
At larger supercoolings the agreement appears to be slightly better, even though very few simulations have been reported at very strong supercooling.
The third striking feature is that whilst experimental results agree well with each other (within 1-2 orders of magnitude) the computational rates
differ much more from each other by a factor of approximately 10$^{10}$.

What is the cause of disagreement between different computational approaches?
Part of the reason is certainly that different water models lead to different rates, see for example Espinosa et al.~\cite{espinosa_homogeneous_2014}.
But even if the same water model is employed, rates do not agree with each other very well.
A neat example is offered by nucleation rates obtained using the mW model. An early study by Moore \textit{et
al.}~\cite{moore_ice_2010} succeeded in calculating the Avrami exponent~\cite{avrami_kinetics_1939,avrami_kinetics_1940}
for the crystallization kinetics of ice from brute force MD simulations at very strong supercooling, obtaining results
remarkably similar to experiment~\cite{hage_crystallization_1994,hage_kinetics_1995}.  However, mW
nucleation rates turned out to be far less encouraging.  In fact, Li \emph{et al.}~\cite{li_homogeneous_2011} and
Reinhardt and Doye~\cite{reinhardt_free_2012} both performed simulations using the mW model, obtaining nucleation rates
that differed by around 5 orders of magnitude.  The only major difference was the enhanced sampling technique employed,
FFS by Li \emph{et al.} and US in Reinhardt and Doye.  The statistical uncertainty of both approaches (1-2 orders of
magnitude) is much smaller than the 5 orders of magnitude discrepancy between the two studies.  It was also shown that
both methods agree very well with each other for e.g. colloids~\cite{filion2010crystal} (see Sec.~\ref{COLL}).  The use
of different computational approaches therefore also seems to be an unlikely source of the disagreement.  What the cause
is remains elusive.

Because we cannot cover all of the work shown in Fig.~\ref{fig: rates} in detail here, we now discuss just two studies.
First Sanz et al.~\cite{sanz_homogeneous_2013}, which agrees best with the experimental rates.  The authors used the
TIP4P/2005 as well as the TIP4P/Ice water models in combination with the \emph{seeding technique} (for more details the
reader is referred to the original paper).  Seeding involves considerably more assumptions than for example US or FFS.
In particular, the approach assumes a CNT-like free energy profile - albeit it does not usually employ the macroscopic
interfacial free energy. Furthermore, the temperature dependence of key quantities such as $\gamma_{\mathcal{S}}$ and
$\Delta\mu_{\mathcal{V}}$ (see Sec.~\ref{THEOF_1}) is approximated.  Nevertheless, the agreement between their
nucleation rates and experiment seemingly outperforms other approaches.  In a more recent paper, Espinosa \textit{et
al.}~\cite{espinosa_homogeneous_2014} obtained nucleation rates for a few other water models as well. However, it should
be noted that the good agreement between the nucleation rates reported in
Refs.~\citenum{sanz_homogeneous_2013,espinosa_homogeneous_2014} and the experimental data could originate
from error cancellation. In fact, while the rather conservative definition of crystalline nucleus adopted in these works
will lead to small nucleation barriers (and thus to higher nucleation rates), the TIP4P family of water models is
characterized by small thermodynamic driving forces to nucleation~\cite{haji-akbari_direct_2015}, which in turn results in smaller nucleation rates.

The second work we briefly discuss is the very recent study (2015) of Haji-Akbari and
Debenedetti~\cite{haji-akbari_direct_2015}.  The authors directly calculated the nucleation rate at 230 K of an all-atom
model of water (TIP4P/ICE) using a novel FFS sampling approach~\cite{haji-akbari_direct_2015}.  This was a \textit{tour
de force} but strikingly, their rates differed from experiment by around 11 orders of magnitude.  The authors point that
this might be as close as one can actually get to experiment with current classical water models.  This is because of
the extreme sensitivity of nucleation rates to thermodynamic properties such as $\gamma_{\mathcal{S}}$ and
$\Delta\mu_{\mathcal{V}}$, which according to CNT enter exponentially (Sec.~\ref{THEOF_1}) in the definition of
$\mathcal{J}$.  For instance, an uncertainty of only 6-7\% for $\gamma_{\mathcal{S}}$ at 235 K leads to an error of
about 9 orders of magnitude in $\mathcal{J}$~\cite{li_homogeneous_2011}.  Experimental estimates for $\gamma$ range from
25 to 35 mN/m~\cite{granasy2002interfacial}, computational estimates from about 20~\cite{malkin_Isd_2012} to 35
mN/m~\cite{angell1973anomalous}. As another example, Haji-Akbari and Debenedetti~\cite{haji-akbari_direct_2015} have explicitly quantified the extent
to which the TIP4P/Ice model underestimates the free energy difference $\Delta\mu_{\mathcal{V}}$ between the crystalline and
liquid phase. It turned out that the mismatch between $\Delta\mu_{\mathcal{V}} (TIP4P/Ice)$ and
$\Delta\mu_{\mathcal{V}}(Experimental)$ alone leads to an overestimation of the free energy barrier for nucleation of about
60\%, which translates in nucleation rates up to 9 orders of magnitude larger. In fact, taking into account such a
discrepancy brings the results of Haji-Akbari and Debenedetti within the confidence interval of the experimental data.
Thus, it is clear that we simply do not know some key quantities accurately enough to expect perfect
agreement between simulations and experiments.

Besides issues of modeling water/ice accurately, finite size effects can be expected to also play a role (as they do
with Lennard-Jones systems (Sec.~\ref{LJL}) and molecules in solution (Sec.~\ref{sec.MIS})).  Only recently was this
issue addressed explicitly for ice nucleation by English and Tse~\cite{english2015massively} in unbiased simulations
with the mW model.  They were able to simulate systems containing nearly 10 million waters on a microsecond timescale,
and found that larger systems favor crystallization precursor formation compared to smaller ones.  Interestingly,
lifetimes of the precursors were found to be less sensitive to system size.  A quantitative understanding of finite size
effects on the nucleation rates remains elusive nevertheless.

In summary it can be said that in terms of accurate nucleation rates, experiments are still clearly superior to simulation.
However, the advantage of simulations is that the nucleation mechanism can also be obtained, which
 at present is not accessible from experiments, although femtosecond X-ray laser spectroscopy might be able to 
partially overcome this limitation in the near future~\cite{sellberg_ultrafast_2014}.

%
\subsubsection*{Nucleation mechanism}

In 2002 Matsumoto \emph{et al.}~\cite{matsumoto2002molecular} were the first to report a nucleation event in an unbiased simulation based on an all-atom model of water (TIP4P).
Their landmark paper opened the doors to the study of ice nucleation at an atomistic level.
They found that nucleation took place once a sufficient number of long-lived hydrogen bonds were formed with a nucleus of ice.
Recent evidence suggests, that most likely their nucleation trajectory was driven by finite size effects~\cite{sanz_homogeneous_2013}.
Together with the simulations of Vrbka \textit{et al.}~\cite{vrbka_homogeneous_2006}, also affected by severe finite size effects~\cite{sanz_homogeneous_2013}, the work of Matsumoto remains to date the only seemingly unbiased MD simulation observing homogeneous ice nucleation with an all atom force field.

\begin{figure*}
\includegraphics[width=14cm]{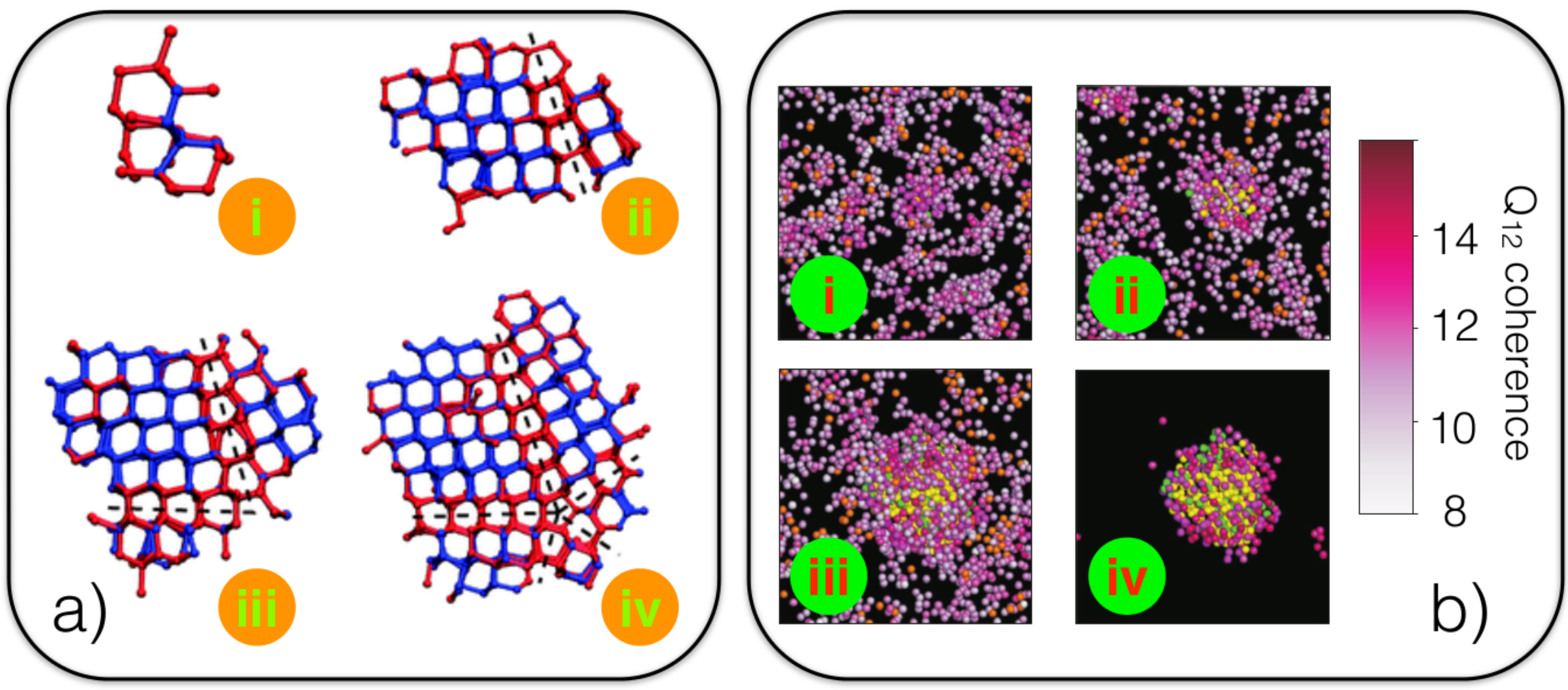}
\protect\caption{a) Formation of a topological defect with 5-fold symmetry during homogeneous ice nucleation.  The snapshots (i-iv)
show the time evolution of the defect structure, indicated by black dashed lines. I$_c$ and I$_h$ water molecules are
shown in blue and red respectively. Reprinted with permission from Ref.~\citenum{li_homogeneous_2011} (Copyright 2011, Royal Society of Chemistry), in which Li \emph{et al.} performed
FFS simulations of models containing about 4000 mW water molecules. b) Nucleation of an ice cluster forming
homogeneously from $I_0$-rich pre-critical nuclei. Water molecules belonging to I$_c$, I$_h$, a clathrate-like phase and $I_0$ are depicted in yellow,
green, orange and magenta respectively. i-ii: a critical nucleus forms in a $I_0$-rich region.  iii: the crystalline cluster evolves in a post critical nucleus, formed
by a I$_c$-rich core surrounded by a I$_0$-rich shell. iv: the same post critical nucleus depicted in iii, but only
particles with 12 or more connections (among ice-like particles) are shown. The colormap refers to the order parameter $Q_{12}$ specified in 
Ref.~\citenum{russo_new_2014}, from which this picture has been reprinted with permission (Copyright 2014, Nature Publishing Group). The unbiased MD simulations on which the analysis is based feature
10,000 mW molecules.}
\label{fig: ice-nucleation}
\end{figure*}

What really enabled the community to investigate ice formation at a molecular level was the development of the 
coarse grained mW potential for water~\cite{molinero_water_2009} in the early 2010's.
Using unbiased MD simulations based on the mW force field Moore and Molinero in 2011~\cite{moore_structural_2011} provided evidence that in the supercooled regime around $T_h$ the fraction of four fold coordinated water molecules increases sharply prior to a nucleation event.
In a separate work the same authors suggested~\cite{moore_is_2011} that at very strong supercooling the critical nucleus is mostly made of
cubic ice, which subsequently evolves into a mixture of stacking disordered cubic and hexagonal ice layers.
In the same year, Li \emph{et al.}~\cite{li_homogeneous_2011} identified another structural motif that might play a role in ice nucleation.
They consistently observed a topological defect structure in growing ice nuclei in their FFS simulations based on the mW representation of water.
This defect, depicted in Fig.~\ref{fig: ice-nucleation}a, can be described as a twin boundary with 5-fold symmetry, 
and it has also been observed~\cite{li_nucleation_2009} in nucleation simulations of tetrahedral liquids simulated via the Stillinger-Weber potential, upon 
which the mW coarse-grained model is built.


\begin{figure*}[t]
\begin{centering}
\centerline{\includegraphics[width=0.75 \textwidth]{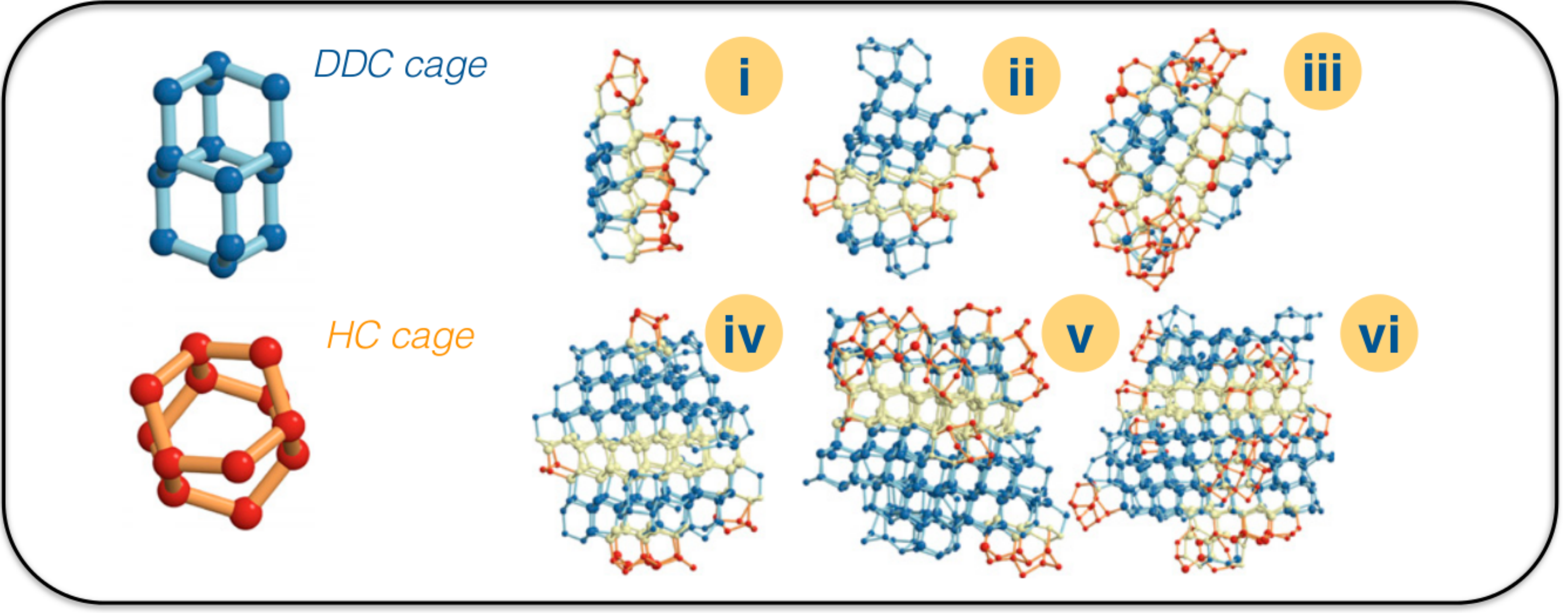}}
\par\end{centering}
\protect\caption{On the left, a typical double-diamond cage (DDC, blue) and an hexagonal cage (HC, red), the building blocks of $I_c$ and $I_h$ respectively.
i-vi: temporal evolution of an ice nucleus from the early stages of nucleations (i-ii) up to post-critical dimensions (v-vi), as observed in the
FFS simulations of Haji-Akbari and Debenedetti~\cite{haji-akbari_direct_2015}. About 4,000 water molecules, modeled via the TIP4P/Ice potential~\cite{abascal_potential_2005}
have been considered in the NPT ensemble at $\Delta T\sim $ 40 K. One can clearly notice the abundance of DDC cages throughout the whole temporal
evolution. In contrast, HC-rich nuclei have only a marginal probability to cross the nucleation barrier (see text). 
Reprinted with permission from Ref.~\citenum{haji-akbari_direct_2015}. Copyright 2015, National Academy of Sciences.}
\label{fig: ice-nucleationDB} \end{figure*}

In 2012 another big leap in understanding the nucleation mechanism of ice from a structural point of view was made by
combining experimental and computational techniques~\cite{malkin_Isd_2012}.  Malkin \emph{et al.} showed, that ice
forming homogeneously is stacking disordered (the corresponding ice structure was called I$_{sd}$), meaning that it is
made out of cubic and hexagonal ice layers stacked in a random fashion.

In 2014 two studies substantiated the potential relevance of precursor structures prior to ice formation.  Palmer
\emph{et al.} provided evidence for a liquid-liquid transition in supercooled water in a molecular model of water
(ST2)~\cite{palmer_metastable_2014}.  In their study, the authors sampled the energy landscape of supercooled water and
found two metastable liquid basins corresponding to low (LDL) and high density (HDL) water.  The appealing idea behind
the transition from HDL to LDL prior to ice nucleation is that LDL is structurally closer to ice than HDL.  Note that
the existence of two metastable liquid basins was not a general finding, the mW model does not have a basin for LDL for
example~\cite{moore_structural_2011}. Indeed the presence of this liquid-liquid phase transition is a highly debated issue~\cite{chandler2014illusions,palmer2014response}.

Another conceptually similar idea is ice formation via ice 0, $I_0$, proposed by Russo \emph{et
al.}~\cite{russo_new_2014}.  Instead of a liquid-liquid phase transition which transforms water into another liquid
state prior to nucleation, the authors propose a new ice polymorph ($I_0$) to bridge the gap between supercooled water
and ice.  $I_0$ is a metastable ice polymorph and is structurally similar to the supercooled liquid.  It has a low
interfacial energy with both, liquid water and ice $I_c/I_h$.  Russo \emph{et al.} therefore proposed $I_0$ to bridge
liquid water to crystalline $I_c/I_h$.  And indeed, the authors found $I_0$ at the surface of growing ice nuclei in MD
simulations, we show part of a nucleation trajectory in Fig.~\ref{fig: ice-nucleation}b.  Furthermore, they showed
that the shape of the nucleation barrier is much better described by a core-shell-like model ($I_c/I_h$ core surrounded
by $I_0$) compared to the classical nucleation model.  This is important, because it suggests that models which are
solely based on CNT assumptions might not be appropriate to describe homogeneous ice nucleation.

However, the emergence of $I_0$ has not yet been reported by any other nucleation study, including the
recent work of Haji-Akbari and Debenedetti~\cite{haji-akbari_direct_2015} we have previously mentioned in the context of
nucleation rates. In there, the authors performed a topological analysis of the nuclei, validated by the substantial
statistics provided by the FFS simulations. As depicted in Fig.~\ref{fig: ice-nucleationDB}, the majority of nuclei that reach the
critical nucleus size contain a large amount of double-diamond cages (DDC, the building blocks of $I_c$), while nuclei
rich in hexagonal cages (HC, the building blocks of $I_h$) have an exceedingly low probability to overcome the free
energy barrier for nucleation. In addition, even postcritical nuclei have an high content of DDC cages, while HC cages
do not show any preference to appear within the core of the postcritical nuclei. This evidence is consistent with the findings
reported in Ref.~\cite{moore_structural_2011} and in contrast with the widely invoked scenario in which a kernel of
thermodynamically stable polymorph (in this case $I_h$) is surrounded by a shell of a less stable crystalline structure (in this case
$I_c$).

In he past few years the understanding of homogeneous ice nucleation has improved dramatically.  We now have a good
understanding of the structure of ice that forms through homogeneous nucleation, stacking disordered ice.  Furthermore
there is very good agreement (within two orders of magnitude) between experimental nucleation rates at a certain
temperature range.  Computational methods face the problem of being very sensitive to some key thermodynamic properties,
the nucleation rates they predict are therefore less accurate.  On the other hand they allow us to study conditions
which are very challenging to probe experimentally, and also provide insight into the molecular mechanisms involved in
the crystallization process.

\subsubsection{Heterogeneous Ice Nucleation}
\label{HIN}

As mentioned in the previous section, homogeneous ice nucleation becomes extremely slow at moderate supercooling.
This seems at odds with our everyday experience -- we do not, for example, have to wait for temperatures to reach
$-30$\,\degree C before we have to use a deicer on our car windows. 
In fact, the formation of ice in nature happens almost
exclusively heterogeneously, thanks to the presence of foreign particles. These ice nucleating agents facilitate
the formation of ice by lowering the free energy barrier for nucleation (see Fig.~\ref{Theoretical_Framework_FIG_1}).
Indeed, the work of Sanz \textit{et al.}\cite{SanzValeriani2013sjc}, in which homogeneous ice
nucleation was studied using seeded MD (see Sec.~\ref{Brute_force}) found rates so low at
temperatures above $\Delta T=20$\,K that they concluded that all ice nucleation above this
temperature must occur heterogeneously. 
Homogeneous nucleation is still of great importance in atmospheric processes and climate modeling, 
as in certain conditions both heterogeneous and homogeneous nucleation are feasible routes toward 
the formation of ice in clouds, as reported in e.g. Ref.~\citenum{herbert_sensitivity_2015}.

\begin{figure*}[t!]
\begin{center}
\includegraphics[width=15cm]{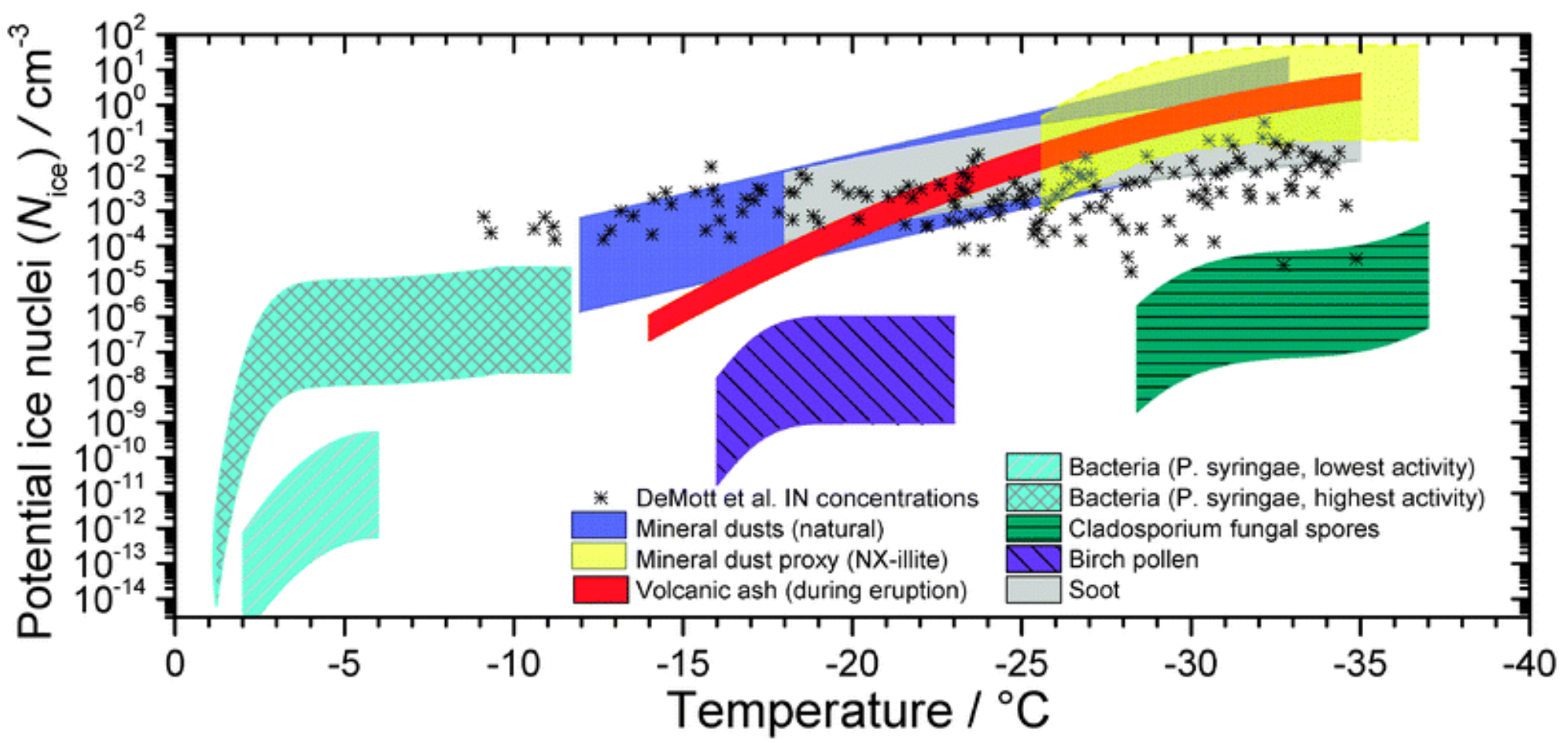}
\end{center}
\caption{The \textit{potential immersion mode ice nuclei concentrations} $N_{ice}$, a measure of the efficiency of a given substance to boost heterogeneous ice nucleation, 
is reported as a function of temperature for a range of atmospheric aerosol species. Note the wide range of nucleating capability for materials as diverse as soot or
bacterial fragments over a very broad range of temperatures. Reprinted with permission from Ref.~\citenum{murray-review}. Copyright 2012, Royal Society of Chemistry.}
\label{FIG_MUR}
\end{figure*}

In addition to the challenges (both computational and experimental) faced when investigating homogeneous ice nucleation,
one also has to consider the structure of water-surface interface and how this impacts on the nucleation rate.
Generally, the experimental data for the rates and characterization of the interfacial structure come from two different
communities: climate scientists have provided us with much information on how various particles, often dust particles or
biological matter such as pollen, affect ice nucleation (as depicted in Fig.~\ref{FIG_MUR}), while surface
scientists have invested a lot of effort in trying to understand, at the molecular level, how water interacts with and
assembles itself at surfaces (see e.g. Ref.~\citenum{NM:review}). This means that there is a huge gap in our
understanding, as the surfaces of the particles used to obtain rates are often not characterized, whereas surface
science experiments are generally carried out at pristine, often metallic, surfaces under ultra-high vacuum conditions.
We will see in this section that computational studies have gone some way to bridging this gap, although there is still
much work to be done should we wish to quantitatively predict a material's ice nucleating efficacy.

\begin{figure*}[t]
  \centering
  \includegraphics[width=0.7\linewidth]{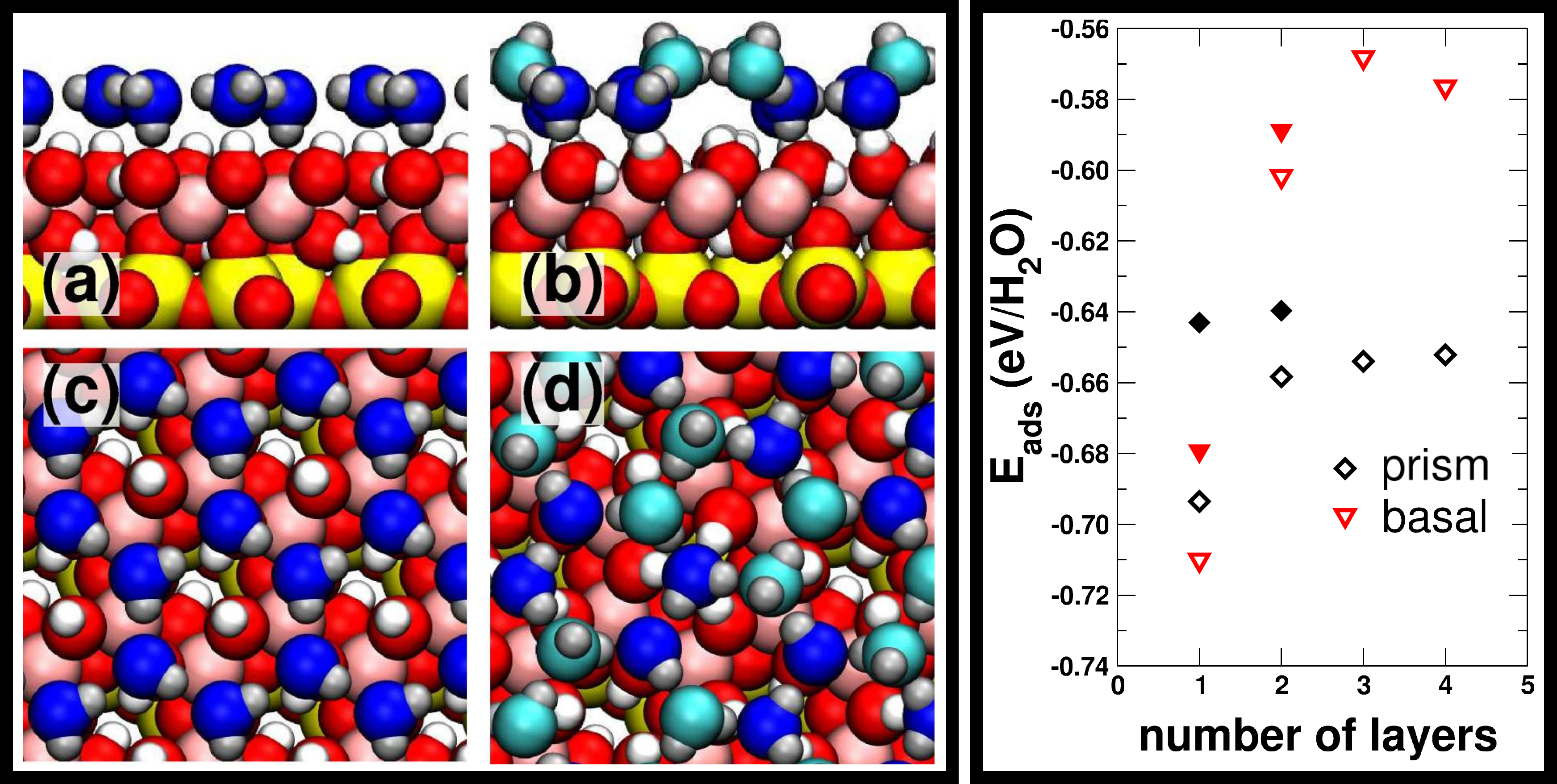}
  \caption{The amphoteric nature of kaolinite is important to its ice
    nucleating ability. The left panel shows ice-like contact layers
    at the kaolinite surface, with (a) the basal face of ice adsorbed
    on kaolinite, and (b) with its prism face adsorbed. The
    structure in (c) is the same as (a) viewed from above, likewise
    for (d) and (b). The right panel shows the adsorption energy of
    ice on kaolinite when bound either through its basal face (red
    data) or prism face (blue data), for a varying number of layers of
    ice. (The empty symbols were obtained with a classical force field
    and the filled symbols with DFT.) When only the contact layer is
    present, the basal face structure is more stable than the prism
    face, but as soon as more layers are present, the prism face
    becomes more stable. This can be understood by the ability of the
    prism face to donate hydrogen bonds to the surface, and to the
    water molecules above, due to the `dangling' hydrogen bonds seen
    in (b) and (d). Figure reprinted with permission from
    Ref.~\protect\citenum{FD:kaolinite}. Copyright 2013, Royal Society of Chemistry.}
  \label{fig:hetice:kao}
\end{figure*}

\vspace{0.5cm}
\noindent \textbf{\textit{Water on crystalline surfaces}} \\
From a computational perspective, it is the surface science experiments that lend themselves most readily to modeling.
In fact, even relatively expensive computational methods such as Density Functional Theory (DFT), which have not
featured much in this article, have proven indispensable in furthering our understanding of how water behaves at
surfaces, especially when used in conjunction with experiment (see e.g. Refs.~\citenum{feibelman:2010:prl}, and~\citenum{NM:review} for an overview). 
As such, early computational studies focused on understanding how the
surface affected the first few layers of water, especially with respect to the concept of lattice mismatch
(see Sec.~\ref{LJL}),
where a surface that has a structure commensurate with ice acts as a template for the crystal. Nutt and Stone
\cite{nutt:2002,Nutt2004,sadtchenko2002} investigated the adsorption structures of water at a model hexagonal surface
and at BaF$_{2}$ (111), using interaction potentials derived from \textit{ab initio} calculations. Although the surfaces
under investigation had structures that matched the basal face of ice well, Nutt and Stone found disordered structures
of water to be more favorable than ice-like overlayers. Using DFT, Hu and Michaelides investigated the adsorption of
water on the (001) face of the clay mineral kaolinite \cite{xiaoliang2007, xiaoliang2008}, a known ice nucleating agent
in the atmosphere. (The (001) surface of kaolinite exposes a pseudo-hexagonal arrangement of OH groups that were
proposed to be the cause of its good ice nucleating ability \cite{PK97}.) While they found that a stable ice-like layer
could form at the surface, the amphoteric nature of the kaolinite surface, depicted in Fig.~\ref{fig:hetice:kao} 
meant that all water molecules could
participate in four hydrogen bonds, making further growth on top of the ice-like layer unfavorable. Croteau \textit{et al.}
\cite{patey2008,patey2009} investigated adsorption of water on kaolinite using the CLAYFF+SPC/E potentials
\cite{cygan:clayff,berendsen:spce} and grand canonical Monte Carlo (GCMC). While some hexagonal patches of water were
seen in the contact layer, the overall structure was mostly disordered, and the hexagonal structures that did form were
strained relative to those found in ice. Also using GCMC, Cox \textit{et al.} \cite{pccp-gcmc} investigated the role of lattice
mismatch using model hexagonal surfaces and TIP4P water \cite{jorgensen:926}. It was found that for atomically flat
surfaces, a nominally zero lattice mismatch produced disordered contact layers comprised of smaller sized rings (i.e.
pentagons and squares), and hexagonal ice-like layers were only observed for surfaces with larger lattice constants.

Prior to \emph{ca.} 2010, the above types of study were state-of-the-art for simulations of heterogeneous ice
nucleation. While they provide evidence that properties such as lattice match alone are insufficient to explain a
material's ice nucleating ability, as ice nucleation itself was not directly observed, only inferences could be drawn
about how certain properties may actually affect ice nucleation. Yan and Patey \cite{patey2011} investigated the
effects of electric fields on ice nucleation using brute force molecular dynamics (the electric fields were externally
applied and were not due to an explicit surface). They found that the electric field need only act over a small range (e.g.
10\,\AA) and that the ice that formed near the `surface' was ferroelectric cubic ice, although the rest of the ice that
formed above was not. Cox \textit{et al.} performed simulations of heterogeneous ice nucleation \cite{FD:kaolinite} in
which both the atomistic nature of the water and the surface was simulated explicitly, using TIP4P/2005 water
\cite{vega:tip4p-2005} and CLAYFF \cite{cygan:clayff} to describe kaolinite. Despite the fact that the simulations were
affected by finite size effects, the simulations revealed that the amphoteric nature of the kaolinite
\cite{xiaoliang2007,xiaoliang2008} was important to ice nucleation. In the liquid, a strongly bound contact layer was
observed, and that for ice nucleation to occur, significant rearrangement in the above water layers was required. It was
found that ice nucleated with its prism face, rather than its basal face, bound to the kaolinite, which was unexpected
based on the theory that the pseudo-hexagonal arrangement of OH groups at the surface was responsible for templating the
basal face. Cox \textit{et al.} rationalized the formation of the prism of ice at the kaolinite due to its ability to donate
hydrogen bonds both to the surface and to the water molecules above (see Fig.~\ref{fig:hetice:kao}), 
whereas the basal face maximizes hydrogen bonding
to the surface only \cite{xiaoliang2007,xiaoliang2008}. More recent simulation studies, employing rigid or constrained models
of kaolinite have also found the amphoteric
nature of the kaolinite surface to be important \cite{ZielkePatey2015sjc}. However, the heterogeneous nucleation mechanism of
water on clays is yet to be validated by unconstrained simulations unaffected by substantial finite size effects.

\vspace{0.5cm}
\noindent \textbf{\textit{Hydrophobicity and Surface Morphology}} \\
As in the case of simulations of homogeneous ice nucleation, the use of the coarse grained mW potential
\cite{molinero_water_2009} has seen the emergence of computational studies that actually quantify the ice nucleating
efficiency of different surfaces. Recently, Lupi \textit{et al.} \cite{molinero:het-jacs} investigated ice nucleation at
carbonaceous surfaces (both smooth graphitic and rough amorphous surfaces) using cooling ramps to measure
non-equilibrium freezing temperatures $\Delta T_{\text{f}} \equiv T_{\text{f}} - T^{\text{homo}}_{\text{f}}$, where
$T_{\text{f}}$ is the temperature at which ice nucleates in the presence of a surface, and $T^{\text{homo}}_{\text{f}} =
201 \pm 1$\,K is the temperature at which homogeneous ice nucleation is observed. It was found that the rough amorphous
surface did not enhance ice nucleation ($\Delta T_{\text{f}} = 0$\,K) whereas the smooth graphitic surfaces promoted ice
nucleation ($\Delta T_{\text{f}} = 11$--$13$\,K). This was attributed to the fact that the smooth graphitic surface
induced a \emph{layering} in the density profile of water above the surface, whereas the rough amorphous surface did
not. (Lupi and Molinero quantified the extent of layering as $\mathcal{L}=\int_0^{z_{\text{bulk}}}
[\frac{\rho(z)}{\rho_{0}} -1 ]{^2} \mathrm{d}z$, where $\rho(z)$ is the density of water at a height $z$ above the
surface, and $\rho_{0} \equiv \rho(z_{\text{bulk}})$ where $z_{\text{bulk}}$ is a height where the density profile is
bulk-like.) In a subsequent paper using the same methodology, Lupi and Molinero \cite{molinero:het-jpca} investigated
how the hydrophilicity of graphitic surfaces affected ice nucleation. The hydrophilicity of the surface was modified in
two different ways: first, by uniformly modifying the water-surface interaction strength; and second, by introducing
hydrophilic species at the surface. It was found that the two ways produced qualitatively different results - uniformly
modifying the interaction potential lead to enhanced ice nucleation, whereas increasing the density of hydrophilic
species was detrimental to ice nucleation (although the surfaces still enhanced nucleation relative to homogeneous
nucleation). It was concluded that the hydrophilicity is not a good indicator of the ice nucleating ability of graphitic
surfaces. As for the difference between increasing the hydrophilicity by uniform modification of the interaction
potential and by introducing hydrophilic species, Lupi and Molinero again saw that the extent of layering in water's
density profile above the surface correlated well with the ice nucleating efficacy. The general applicability of the
layering mechanism, however, was left as an open question.

Cox \textit{et al.} \cite{CoxMichaelides2015sjc,CoxMichaelides2015sjc-2} addressed the question of the general applicability of
the layering mechanism by investigating ice nucleation rates over a wider range of hydrophilicities (by uniformly
changing the interaction strength) on two surfaces with different morphologies: (1) the (111) surface of a face centered
cubic LJ crystal (FCC-111) that provided distinct adsorption sites for the water molecules; and (2) a graphitic surface,
similar to that of Lupi \textit{et al.} \cite{molinero:het-jacs}. While it was found that the layering mechanism (albeit with a
slight modification to the definition used by Lupi \textit{et al.}) could describe the ice nucleating behavior of the graphitic
surface, at the FCC-111 surface no beneficial effects of layering were observed. This was attributed to fact that the
FCC-111 surface also affected the structure of the water molecules in the second layer above the surface, in a manner
detrimental to ice nucleation. It was concluded that layering of water above the surface can be beneficial to ice
nucleation, but only if the surface presents a relatively smooth potential energy surface to the water molecules.

\begin{figure*}[t!]
\begin{center}
\includegraphics[width=15cm]{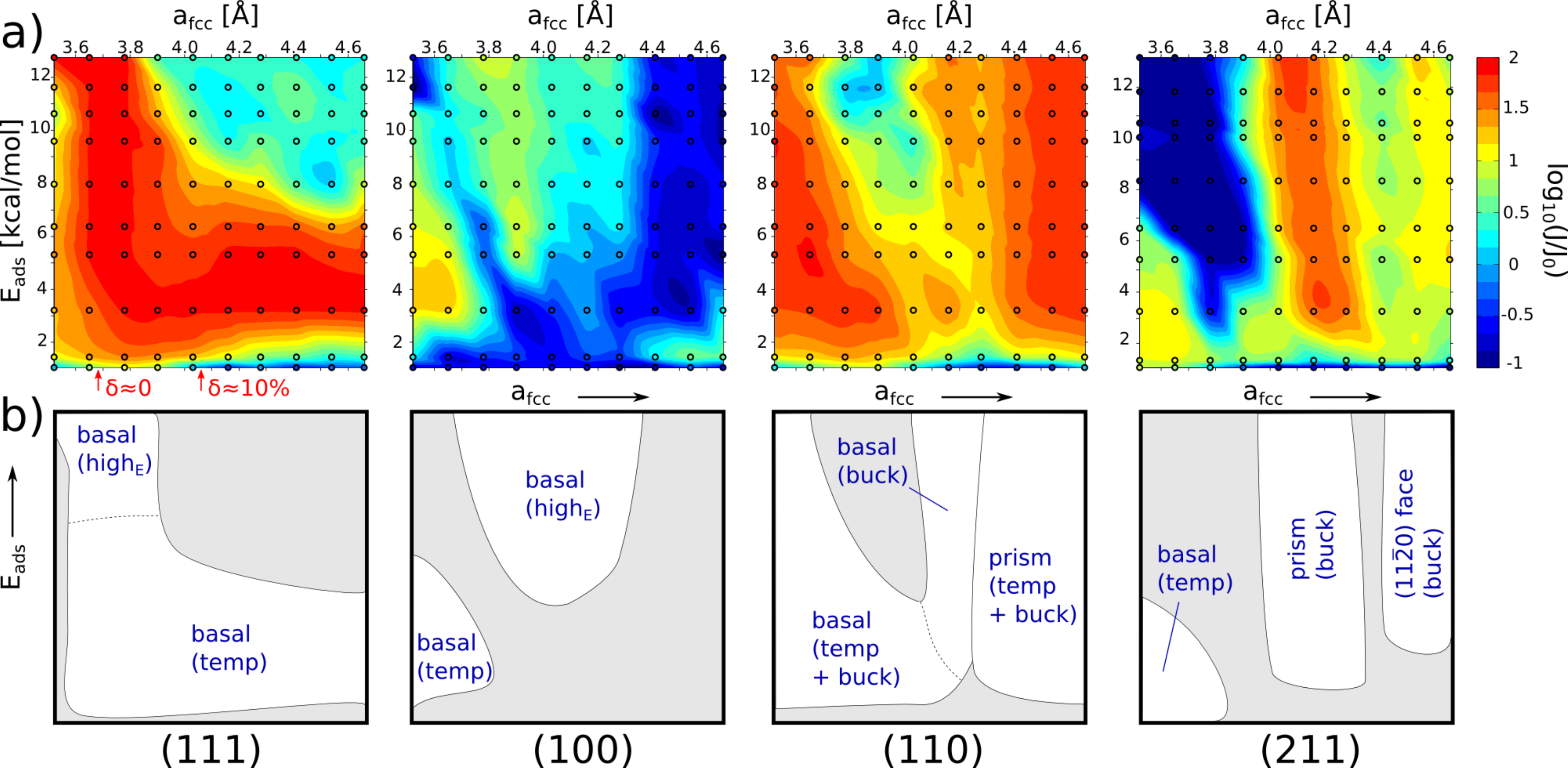}
\end{center}
\caption{
Interplay between surface morphology and water-surface interaction on the heterogeneous ice nucleation rate.
a) Heat maps representing the values of ice nucleation rates
on top of four different FCC surfaces ((111), (100), (110), (211)), plotted as a function of the adsorption
energy E$_{ads}$ and the lattice
parameter
$a_{FCC}$. The lattice mismatch $\delta$ with respect to ice on (111) is indicated below the
graph. The values
of the nucleation rate $\mathcal{J}$ are reported as log$_{10}$ ($J/J_0$), where $J_0$ refers
to the homogeneous
nucleation rate at the same temperature.
b) Sketches of the different regions (white areas) in the ($E_{ads}$,$a_{FCC}$ ) space in which a significant enhancement
of the nucleation rate is observed.
Each region is labeled according to the face of Ih nucleating and growing
on top of the surface
(basal, prismatic or (11-20)), together with an indication of what it is
that enhances the nucleation. "temp", "buck", and "highE" refer to the in-plane template of
the first overlayer,
the ice-like buckling of the contact layer, and the nucleation for high
adsorption energies on
compact surfaces.
Reprinted with permission from Ref.~\citenum{FitznerMichaelides2015sjc}. Copyright 2015, American Chemical Society.}
\label{figure2}
\end{figure*}

The studies at the carbonaceous and FCC-111 surfaces
\cite{molinero:het-jacs,molinero:het-jpca,CoxMichaelides2015sjc,CoxMichaelides2015sjc-2} discussed above hinted 
that the heterogeneous nucleation mechanism could be very different at different types of surfaces. 
While there is experimental evidence that e.g. different carbon nanomaterials are capable of boosting ice nucleation (see e.g. 
Ref.~\citenum{whale_ice_2015}), most experiments can only quantify the ice nucleating ability of the substrates 
(see Sec.~\ref{Experimental_Methods}). However, the structure of the water-substrate interface and any insight into the
morphology of the nuclei are typically not available, making simulations essential to complement the experimental picture.
In this respect, Zhang \textit{et al.}~\cite{zhang_impact_2014} have assessed that the (regular) pattering of a generic
crystalline surface at the nano scale can strongly affect ice formation.
More generally, the interplay between hydrophobicity and surface morphology has been recently elucidated by Fitzner \textit{et
al.}~\cite{FitznerMichaelides2015sjc}. Brute force MD simulations of heterogeneous ice nucleation were performed for the mW water model on top
of several crystalline faces of a generic FCC crystal, taking into account different values of
$\epsilon_{\mathcal{W}\mathcal{S}}$ as well as different values of the lattice parameter.
The latter is involved in the rather dated~\cite{turnbull_nucleation_1952} concept of the zero lattice mismatch, which
we introduced in Sec.~\ref{LJL} (see Eq.~\ref{eqn.mismatch}) and that has been often quoted as the
main requirement of an effective ice nucleating agent.
However, a surprisingly non-trivial interplay between hydrophobicity and morphology was observed, as depicted in
Fig.~\ref{figure2}. Clearly neither the layering nor the lattice mismatch alone
are enough to explain such a diverse scenario. In fact, the authors proposed three additional microscopic factors that can
effectively aid heterogeneous ice nucleation on crystalline surfaces: (i) An in-plane templating of the first water overlayer on top of the
crystalline surface; (ii) A first overlayer buckled in
an ice-like fashion; and (iii) Enhanced nucleation in regions of the liquid beyond the first two overlayers, possibly aided by dynamical effects and/or
structural templating effects of the substrate extending past the surface-water interface.
In addition, it turned out that different lattice parameters can lead to the nucleation and growth of up to three different faces of
ice (basal, prismatic, secondary prismatic (\{11$\bar{2}$0\})) on top of the very same surface, adding a layer of
complexity to the nucleation scenario. Insights into the interplay between hydrophobicity and morphology have also very recently been
obtained by Bi \textit{et al.}~\cite{bi_heterogeneous_2016}, who investigated heterogeneous ice nucleation on top of graphitic surfaces by means
of FFS simulations using the mW model. Among their findings, 
the authors suggested that the efficiency of ice nucleating agents can be a function not only of surface chemistry 
and surface crystallinity, but of the elasticity of the substrate as well.

\vspace{0.5cm}
\noindent \textbf{\textit{Computational Methods and Models}} \\
Enhanced sampling techniques have also been used to investigate heterogeneous ice nucleation. Reinhardt and Doye
\cite{ReinhardtDoye2014sjc} used umbrella sampling with the mW model to investigate nucleation at a smooth planar
interface and at an ice-like surface. They found that the flat planar interface did not help nucleate ice, and that
homogeneous nucleation was the preferred pathway. One explanation given for this was that, as the density of liquid
water is higher than that of ice, an attractive surface favors the liquid phase. It was also noted that the mW potential
imposes an energetic penalty for non-tetrahedral triplets and by removing neighbors at the surface, this energetic
penalty is decreased, and this reduction in tetrahedrality favors the liquid phase. Cabriolu and Li recently studied ice
nucleation at graphitic surfaces using forward flux sampling \cite{cabriolu_ice_2015}, again with the mW model. Under
the assumption that $\Delta \mu_{\mathcal{V},\text{water,ice}}$ depends linearly on $\Delta T$ and that
$\gamma_{\mathcal{S}}$ does not depend on $\Delta T$, Cabriolu \textit{et al.} have also extracted the values
of the contact angle at different temperatures, which, along with the free energy barrier, turn out to be consistent
with CNT for heterogeneous nucleation (see Sec.~\ref{THEOF_2}). Although intriguing, the generality of this finding to surfaces that include
strong and localized chemical interactions remains an open question.

We have seen that for both homogeneous and heterogeneous nucleation, using the coarse grained mW model has greatly
enhanced our ability to perform quantitative, systematic simulation studies of ice nucleation. We must face the fact,
however, that this approach will only further our understanding of heterogeneous ice nucleation so far. As discussed 
for kaolinite \cite{xiaoliang2007,xiaoliang2008,FD:kaolinite,ZielkePatey2015sjc}, an explicit
treatment of the hydrogen bonds is essential in describing heterogeneous ice nucleation. 
In addition to this, the mW model (as well as the majority of the fully atomistic water models) 
cannot take into account surface charge effects.  Surfaces can polarize water molecules in the
proximity of the substrate, alter their protonation state and even play a role in determining the equilibrium structure of
the liquid at the interface. In light of recent studies~\cite{aber_strong_2005,patey2011}, it seems that these effects can heavily affect
nucleation rates of many different systems.
How then, do we proceed? The
answer is not clear. As discussed, enhanced sampling techniques such as as umbrella sampling
\cite{ReinhardtDoye2014sjc} and forward flux sampling \cite{cabriolu_ice_2015} have been applied to heterogeneous ice
nucleation with the mW model, and we have seen the latter applied successfully to homogeneous nucleation with an
all-atom model of water \cite{haji-akbari_direct_2015}; the computational cost, however, was huge. Although the presence
of an ice nucleating agent should help reduce this cost, the parameter space that we wish to study is large and to
systematically study how the various properties of a surface affect ice nucleation requires the investigation of many
different surfaces.

There is another computational issue that also requires attention. Simulating heterogeneous ice nucleation under
\textit{realistic conditions} does not just mean mild supercooling -- we also need realistic models of the surfaces that we
wish to study! Most studies of kaolinite have only considered the planar
interface, even though in nature kaolinite crystals have many step-edges and defects. Ice nucleation at AgI has also
recently been studied \cite{ZielkePatey2015sjc-2,FrauxDoye2014sjc}, although bulk truncated structures for the exposed
crystal faces were used. In the case of AgI (0001) this is problematic, as the wurtzite structure of the crystal means
that this basal face is polar and likely to undergo reconstruction~\cite{tasker_stability_1979}. 
Furthermore, AgI is photosensitive and it has been shown experimentally that exposure to light
enhances its ice nucleating efficacy \cite{anderson1976supersaturation}, suggesting structural motifs at the surface
very different from those expected from the bulk crystal structure are important. The development of computational
techniques to determine surface structures, along with accurate force fields to describe the interaction with water,
will be essential if we are to fully understand heterogeneous ice nucleation.

\subsection{Nucleation from Solution} 
\label{sec.MIS}

Understanding crystal nucleation from solution is a problem of great practical interest, influencing for instance
pharmaceutical, chemical and food processing companies.  Being able to obtain a microscopic
description of nucleation and growth would allow the selection of specific crystalline polymorphs, which in turn
can have an enormous impact on the final product~\cite{price_predicting_2014}. An (in)famous case illustrating the 
importance of this issue is the drug 
Ritonavir~\cite{bauer_ritonavir:_2001,datta_crystal_2004}, originally marketed as solid capsules to treat HIV. 
This compound has at least two
polymorphs: the marketed and thoroughly tested polymorph (P$_{I}$) and a second more stable crystalline
phase P$_{II}$ which appeared after P$_{I}$ went to market.
P$_{II}$ is basically non active as a drug because of a much lower solubility than P$_{I}$. As such, and
most importantly because of the fact that P$_{II}$ had never been properly tested, Ritonavir was withdrawn
from the market in favor of a much safer alternative in the form of gel capsules.  Many other
examples~\cite{lee_practical_2014} could be listed, as various environmental factors (such as the temperature, the
degree of supersaturation, the kind of solvent or the presence of impurities) can play a role in determining the
final polymorph of many classes of molecular crystals. Thus, it is highly desirable to pinpoint
\textit{a priori} the conditions leading to the formation of a specific polymorph possessing the optimal
physical/chemical properties for the application of interest.

The term {\em nucleation from solution} encompasses a whole range of systems, from small molecules in aqueous or organic
solvents to proteins, peptides or other macromolecular systems in their natural environment.  These systems are very
diverse and an universal nucleation framework is probably not applicable to all these cases. The 
review by Dadey {\em et al.}\cite{davey_nucleation_2013} discusses the role of the solvent
in determining the final crystal. 
Many aspects of the nucleation of solute precipitates from solution have been
recently reviewed by Agarwal and Peters\cite{agarwal_solute_2014}.
In this section we limit the discussion to
small molecules in solution.

A central issue with MD simulations of nucleation from solution is the choice of order parameters able to distinguish
different polymorphs. Many of these \textit{collective variables} have been used in enhanced sampling simulations (see
Sec.~\ref{ESM}).  Several examples can be found in e.g.
Refs.~\citenum{santiso_general_2011,yu_order-parameter-aided_2014,duff_polymorph_2011,peters_competing_2009,shetty_novel_2002}.
MD simulations of nucleation from solution are particularly challenging because of finite size effects due the nature of
the solute/solvent system\cite{wedekind_finite-size_2006,grossier_reaching_2009}.  In the NVT and NPT ensembles, where
MD simulations of nucleation are usually performed, the total number of solute molecules is constant. However, the ratio
between the number of solute molecules in the crystalline phase and those within the solution varies during the
nucleation events, leading to a change in the chemical potential of the system.  This occurrence has negligible effects
in the thermodynamic limit~\cite{fnote7}, but it can affect substantially the outcomes of e.g. free energy based
enhanced sampling simulations.  Simulations of models containing a large number (10$^3$-10$^5$) of molecules can
alleviate the problem~\cite{zimmermann_nucleation_2015}, although this is not always the
case~\cite{duff_nucleation_2009,agarwal_solute_2014,agarwal_nucleation_2014}. An analytic correction to the free energy
for NPT simulations of nucleation of molecules from solution has been proposed in
Refs.~\citenum{agarwal_nucleation_2014,agarwal_solute_2014} on the basis of a number of previous works (see e.g.
Refs.~\citenum{reguera_phase_2003,wedekind_finite-size_2006,grossier_reaching_2009})and applied later on in
Ref.~\cite{salvalaglio_molecular-dynamics_2015} as well.  Alternative approaches include seeded MD
simulations~\cite{knott_homogeneous_2012,zimmermann_nucleation_2015} (see Sec.~\ref{Brute_force}) and simulations
mimicking the grand canonical ensemble $\mu$VT~\cite{agarwal_chemical_2014,perego_molecular_2015}, where the number of
constituents is not a constant  and the number of molecules in - in this case - the solution is allowed to evolve in
time.  It is worth noticing that nucleation of molecules in solution is a challenging playground for experiments as
well. For instance, quantitative data about nucleation of ionic solutions are amazingly hard to find within the current
literature. This is in stark contrast with the vast amount of data covering e.g. ice nucleation (as illustrated in
Sec.~\ref{WAH}).

\begin{figure}[t!]
\begin{center}
\includegraphics[width=7cm]{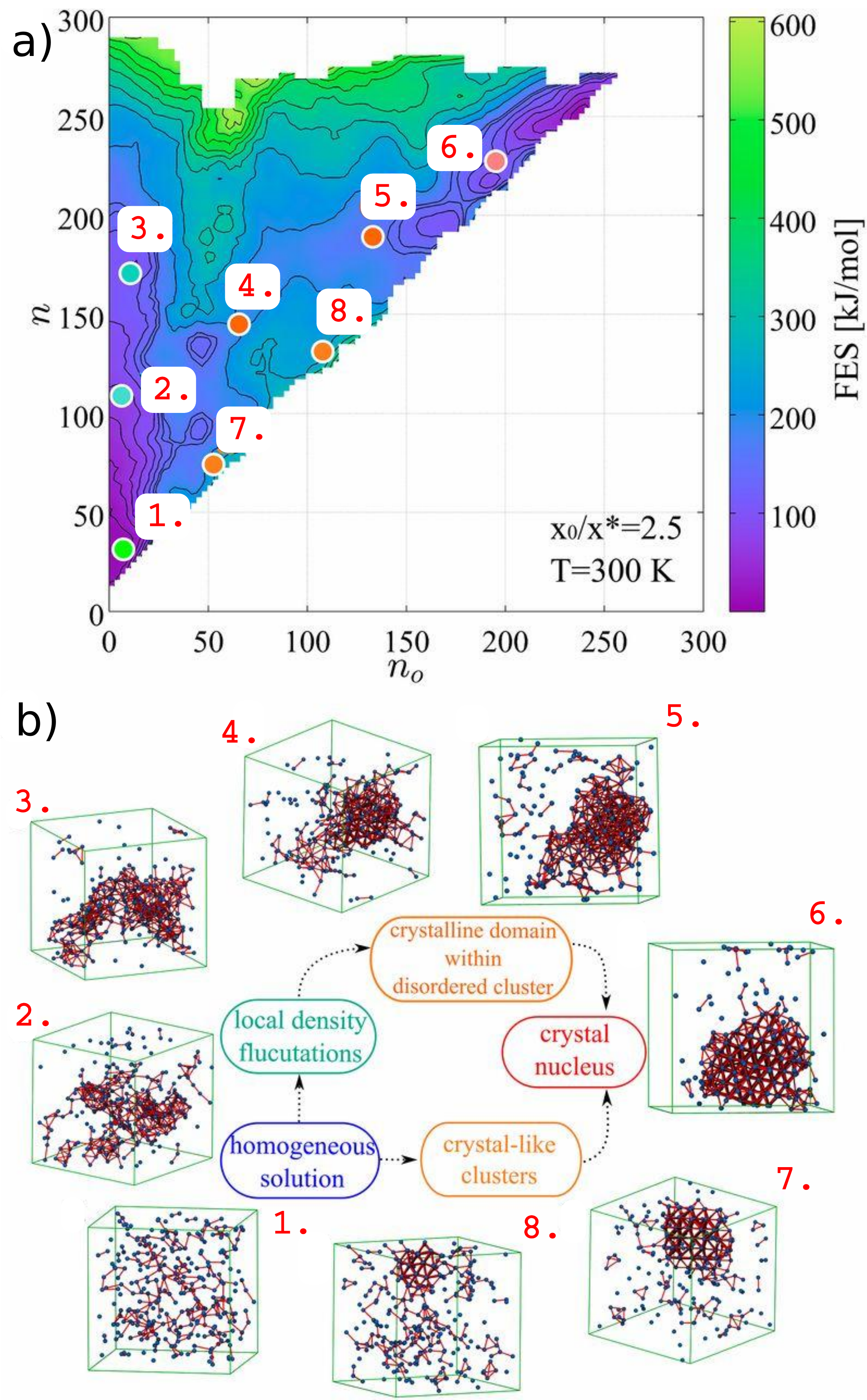}
\end{center}
\caption{ a) Free-energy surface (FES) associated with the early stages of nucleation of urea 
in aqueous solution. This has been obtained by Salvalaglio {\em et al.}\cite{salvalaglio_molecular-dynamics_2015} 
from a well tempered metadynamics simulation of 300 urea molecules and 3173 water molecules, within an 
isothermal-isobaric ensemble at $p=1$~bar and $T=300$~K  (simulation S2 in 
Ref.\cite{salvalaglio_molecular-dynamics_2015}, with correction term to the free-energy 
included in order to represent the case of a constant supersaturation of 2.5).
The contour plot of the FES is reported as a function of 
the number of molecules belonging to the largest connected cluster ($n$, in ordinate) and 
the number of molecules in a crystal-like configuration within the largest cluster ($n_o$, in abscissa). 
Note that $n \ge n_o$ by definition, and that CNT would prescribe that the evolution of the largest 
cluster in the simulation box is such that $n=n_o$ (i.e., only the diagonal of the contour plot is populated).
The presence of an off-diagonal basin provides evidence of a two-step nucleation of urea crystals from 
aqueous solutions. This is further supported by the representative states sampled during the 
nucleation process, shown in panel b). Urea molecules are represented as blue spheres, and red connections are drawn 
between urea molecules falling within a cutoff distance of 0.6~nm of each other.
Reprinted with permission from Ref.~\citenum{salvalaglio_molecular-dynamics_2015}. Copyright 2015, National Academy of Sciences.}
\label{fig:urea-water}
\end{figure}

\vspace{0.5cm}
\noindent \textbf{\textit{Nucleation of Organic Crystals}} \\
Among the countless organic compounds, urea molecules can be regarded as a benchmark for MD simulation of nucleation
from solution.  This is because urea is a system of great practical importance which: (i) displays fast nucleation kinetics;
and (ii) has only one experimentally characterized polymorph.  Early studies by Piana \textit{et
al.}~\cite{piana_understanding_2005,piana_simulating_2005} focused on the growth rate of urea crystals, which turned out
to be consistent with experimental results.  Years later, the inhibition of urea crystal growth by
additives was investigated by Salvalaglio {\em et al.}\cite{salvalaglio_uncovering_2012,salvalaglio_controlling_2013}.
The investigation of the early stages of nucleation has been tackled only recently by Salvalaglio \textit{et
al.}~\cite{salvalaglio_urea_2015} for urea molecules in aqueous or organic (ethanol, methanol and acetonitrile)
solvents. In these studies the authors employed metadynamics along with the generalized Amber force
field~\cite{cornell_second_1995,wang_development_2004}.  The resulting free energies, modified for finite size effects 
related to the solvent~\cite{salvalaglio_urea_2015}, suggested that different solvents led to different nucleation mechanisms.
While a single-step nucleation process is favored in methanol and ethanol, a two-step
mechanism (see Sec.~\ref{THEOF_TS}) emerges for urea molecules in acetonitrile and water, as depicted in Fig.~\ref{fig:urea-water}.
In this case, the initial formation of an amorphous - albeit
dense - cluster is followed by the evolution into a crystalline nucleus.
Note that according to the free energy surface reported in Fig.~\ref{fig:urea-water}a, the amorphous clusters
(configurations n.2 and n.3 in Fig.~\ref{fig:urea-water}a and Fig.~\ref{fig:urea-water}b) are unstable with respect to
the liquid phase, i.e. they are not metastable states having their own free energy bases but rather they originate from fluctuations within
the liquid phase. This evidence, together with the fact that the transition state
(configuration n.4 in Fig.~\ref{fig:urea-water}a and Fig.~\ref{fig:urea-water}b) displays a fully crystalline core,
prompts the following, long-standing question: \textit{if the critical nucleus is mostly crystalline and the amorphous
precursors are unstable with respect to the liquid phase, can we truly talk about a two-steps mechanism}? 
Ref.~\citenum{agarwal_solute_2014} suggests the term \textit{ripening regime two-step} when dealing with stable amorphous precursors and
\textit{crystallization-limited two-step} when the amorphous clusters are unstable and the limiting step is the formation of a crystalline core
within the clusters. 
Salvalaglio \textit{et
al.}~\cite{salvalaglio_urea_2015} also observed two polymorphs
(p$_{I}$ and p$_{II}$) in the early stages of the nucleation process. p$_{I}$ corresponds to the experimental crystal
structure and is the most stable structure in the limit of an infinite crystal~\cite{giberti_insight_2015}. p$_{II}$,
however, is more stable for small crystalline clusters.  In agreement with Ostwald rule (see Sec.~\ref{LJL}), the small
crystalline clusters that initially form in solution are of p$_{II}$ type, and the subsequent conversion from p$_{II}$
to p$_{I}$ seems to be an almost barrierless process.

A similar approach to Ref.~\cite{salvalaglio_urea_2015} was used to investigate crystal nucleation of 1,3,5-Tris(4-bromophenyl)benzene
molecules in water and methanol. These simulations showed the emergence of prenucleation clusters, consistent with 
recent experimental results~\cite{harano_heterogeneous_2012} based on single-molecule real-time transmission electron
microscopy (SMRT-TEM, see Sec.~\ref{Experimental_Methods}).
The formation of prenucleation clusters in the early stages of nucleation from solution has been observed in several
other
cases~\cite{dey_role_2010,harano_heterogeneous_2012,davey_nucleation_2013,gebauer_pre-nucleation_2014,zahn_thermodynamics_2015}.
This is of great relevance as CNT is not able to account for two (or multi) step nucleation.
MD simulations have been of help in several cases, validating or supporting a
particular mechanism.  For instance, MD simulations have provided evidence for two-step nucleation in
aqueous solutions of $\alpha$-Glycine~\cite{yani_glycine_2012} and $n$-octane (or $n$-octanol) solutions of D-/L-norleucine~\cite{ectors_oligomers_2015}.

\vspace{0.5cm}
\noindent \textbf{\textit{Nucleation of Sodium Chloride}} \\
Sodium chloride (NaCl) nucleation from supersaturated brines represents an interesting challenge for simulations, as the
system is relatively easy to model and experimental nucleation rates are available.

The first simulations of NaCl nucleation date back to Ohtaki et
Fukushima\cite{ohtaki_nucleation_1991}, who in the early 90's performed brute force MD simulations using 
very small systems (448 molecules including water molecules and ions) and exceedingly short simulation times ($\sim$10 ps).
Thus, the formation of small crystalline clusters they observed was most likely a consequence of finite size effects.
More recently, the TPS simulations of Zahn\cite{zahn_atomistic_2004} suggested that the centers of stability for
NaCl aggregates consists of non-hydrated Na$^+$ ions octahedrally coordinated with Cl$^-$ ions, albeit the results related to very small
simulations boxes (containing 310 molecules in total). 

\begin{figure}[b!]
\begin{center}
\includegraphics[width=7cm]{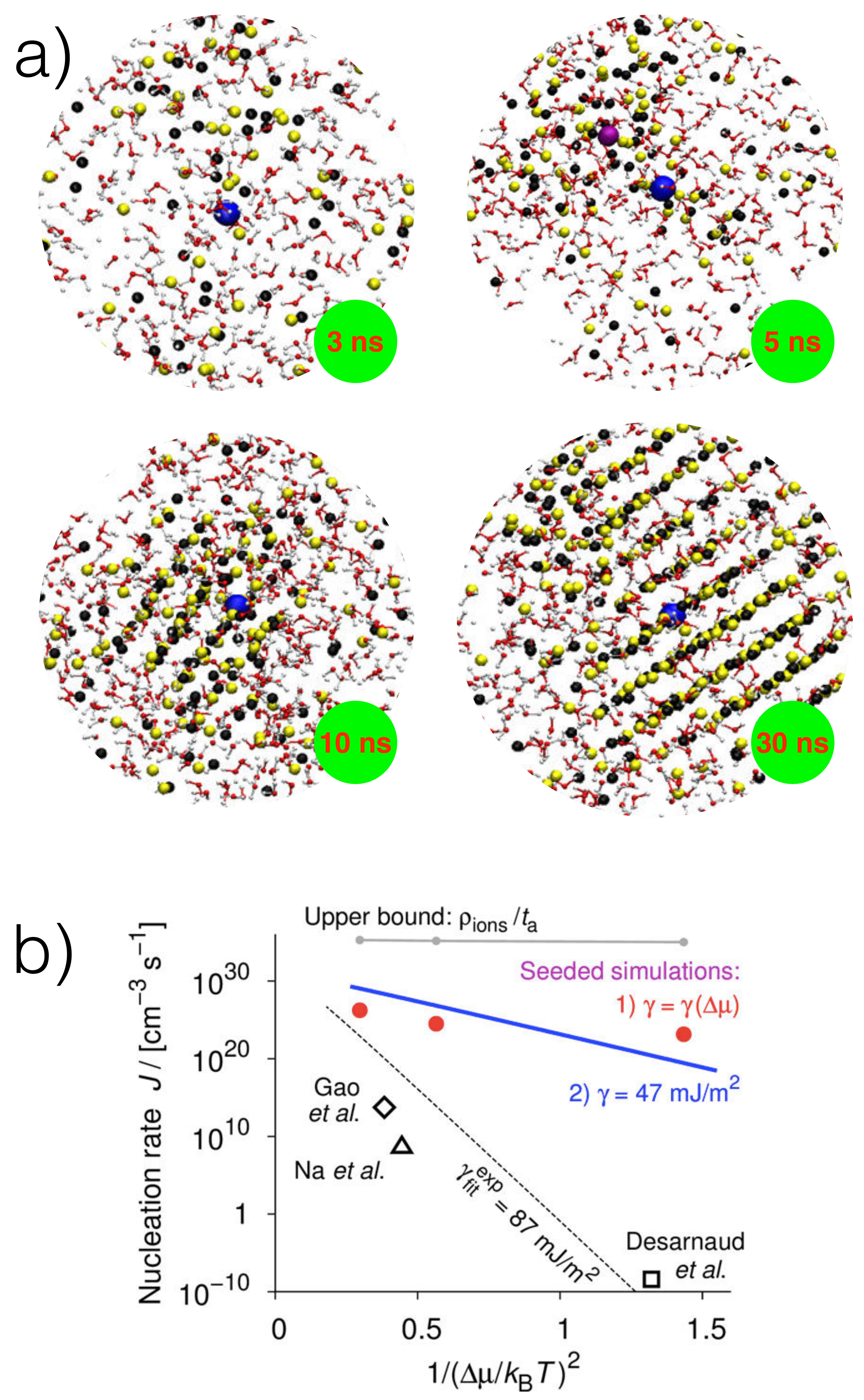} 
\end{center}
\caption{a) Snapshots from an MD simulation of crystal nucleation of NaCl from aqueous solution. The simulations,
carried out by Chakraborty and Patey\cite{chakraborty_how_2013}, involved 56,000 water molecules and 4000 ion pairs
(concentration is 3.97~$m$) in the NPT ensemble. All Na$^+$ (black) and Cl$^-$ (yellow) ions within 2 nm of a reference
Na$^+$ ion (larger and blue) are
shown together with water molecules (oxygens and hydrogens in red and white respectively) within 0.4 nm from each ion. 
From the relatively homogeneous solution (3 ns) an amorphous cluster of ions emerges (5 ns). This fluctuation in the
concentration of the ions leads to a subsequent ordering of the disordered cluster (10 ns) in a crystalline fashion (30
ns), consistently with a two-step nucleation mechanism. Reprinted with permission from Ref.~\cite{chakraborty_how_2013}. Copyright 2013, American Chemical Society.
b) Comparison of NaCl nucleation rates, $\mathcal{J}$, as a function of the driving force for nucleation, reported as $1/(\Delta\mu/k_B T)^2$. 
Red dots as well as blue and gray (continuous) lines have been
estimated via three different approaches by the simulations of Zimmermann {\em et al.}\cite{zimmermann_nucleation_2015}.
Experimental data obtained employing an electrodynamic levitator trap (Na {\em et
al.}\cite{na_cluster_1994}), an efflorescence chamber (Gao {\em et al.}\cite{gao_efflorescence_2007}), and
microcapillaries (Desarnaud {\em et al.}\cite{desarnaud_metastability_2014}) are also reported together with
a tentative fit ($\gamma_{fit}^{exp}$, dotted line). Note the substantial (up to about 30 orders of magnitude) discrepancy between
experiments and simulations. Reprinted with permission from Ref.~\cite{zimmermann_nucleation_2015}. Copyright 2015, American Chemical Society.} 
\label{fig:NaCl-water} 
\end{figure}

Tentative insight into the structure of the crystalline clusters came with the work
of Nahtigal {\em et al.}\cite{nahtigal_nucleation_2008}, featuring simulations of 4132 molecules (4000 water molecules and 132 ions) in the
673-1073 K range for supercritical water at different densities (0.17-0.34~g/cm$^3$). 
They reported a strong dependence of the crystalline cluster size distribution on the system
density, with larger clusters formed at lower densities. Moreover, the clusters appeared to be amorphous.
The emergence of amorphous precursors has been also reported in the work of Chakraborty and Patey\cite{chakraborty_how_2013,chakraborty_evidence_2013}, reporting
large scale MD simulations featuring 56,000 water molecules and 4000 ion pairs in the NPT ensemble. The SPC/E model\cite{ff.water.SPC-E} was
used for water, and the ion parameters were those used in the
OPLS\cite{chandrasekhar_energy_1984,aaqvist_ion-water_1990} force field. Their findings
provided strong evidence for a two-step mechanism of nucleation, where a dense but unstructured NaCl nucleus is formed first,
followed by a rearrangement into the rock salt structure, as depicted in Fig.~\ref{fig:NaCl-water}a.
On a similar note, metadynamics simulations performed by Giberti {\em et al.}\cite{giberti_transient_2013} using
the GROMOS\cite{ff.GROMOS53} force field for the ions and the SPC/E\cite{ff.water.SPC-E} model for water suggested the 
emergence of a wurtzite-like polymorph in the early stages of nucleation. This precursor could be an intermediate state along the
path from brine to the NaCl crystal. However, Alejandre and Hansen\cite{alejandre_ions_2007} pointed out a strong sensitivity of the nucleation mechanism
on the choice of the force field. 

In fact, very recent simulations by Zimmermann {\em et al.}\cite{zimmermann_nucleation_2015} demonstrated that the
GROMOS force field overestimates the stability of the wurtzite-like polymorph.  The authors employed a seeding approach
within a NVT setup for which the absence of depletion effects was explicitly verified~\cite{wedekind_finite-size_2006}.
The force fields used are  those developed by Joung and Cheatham\cite{joung_molecular_2009} for Na$^+$ and Cl$^-$, and
SPC/E\cite{ff.water.SPC-E} for water, which provide reliable solubilities and accurate chemical potential driving
force~\cite{aragones_solubility_2012}.  By using a methodology introduced in Ref.\cite{knott_homogeneous_2012}, the
interfacial free energy and the attachment frequency $\delta_n$ were deduced.  A thorough investigation of the latter
demonstrated that the limiting factor for $\delta_n$, which in turn strongly affects the kinetics of nucleation (see
Sec.~\ref{THEOF_1}), is not the diffusion of the ions within the solution, but instead the desolvation process needed
for the ions to get rid of the solvent and join the crystalline clusters.  Moreover, Zimmermann {\em et
al.}\cite{zimmermann_nucleation_2015} evaluated the free energy barrier to nucleation as well as the nucleation rate as
a function of supersaturation, providing three estimates by using different approaches. The results are compared with
experiments in Fig.~\ref{fig:NaCl-water}b, showing a substantial discrepancy as large as 30 orders of magnitude.
Interestingly, experimental nucleation rates are much smaller than what is observed in simulations, contrary to what has
been observed for e.g. colloids (see Sec.~\ref{COLL}). We stress that the work of Zimmermann {\em et al.} employed state
of the art computational techniques and explored NaCl nucleation in different conditions using a variety of approaches.
The fact that these \textit{tour de force} simulations yielded nucleation rates that differed significantly from
experiments casts yet another doubt on the possibility to compare effectively experiments and simulations. However,
it must be noted that Zimmermann {\em et al.}\cite{zimmermann_nucleation_2015} assumed a value of about 5.0 mol$_{NaCl}$/kg$_{H_2O}$ for the
NaCl solubility in water, as proposed in Ref.~\citenum{aragones_solubility_2012}. This differs substantially from
the values independently obtained by Moucka \textit{et al.}~\cite{moucka_molecular_2011} (3.64 mol$_{NaCl}$/kg$_{H_2O}$)
and more recently by Mester and Panagiotopoulos~\cite{mester_temperature-dependent_2015} (3.71
mol$_{NaCl}$/kg$_{H_2O}$). This discrepancy can explain the enormous mismatch reported by Zimmermann {\em et
al.}\cite{zimmermann_nucleation_2015}, once again demonstrating the severe sensitivity of nucleation rates with respect
to any of the ingredients involved in their calculations.

On a final note, we stress that many other examples of molecular dynamics simulations looking at specific aspects of
crystal nucleation from solution exist in the literature. For instance, a recent study by Anwar \textit{et
al.}~\cite{anwar_secondary_2015} describes
secondary crystal nucleation, where crystalline seeds are already present within the solution. The authors suggest, for
a generic solution represented by Lennard-Jones particles, a (secondary) nucleation mechanism enhanced by the existence
of PNCs (see Sec.~\ref{THEOF_TS}). Kawska \textit{et al.}~\cite{kawska_atomistic_2008} underlined instead 
the importance of proton transfer within the early stages of
nucleation of zinc oxide nano clusters from an ethanol solution. The emergence of similar ripening processes, selecting
specific crystalline polymorphs according to e.g. the effect of different solvents is still fairly unexplored but
bound to be of great relevance in the future.
Finally, several computational studies have dealt with the crystallization of calcium carbonate, which has
recently been reviewed extensively in Ref.~\cite{gebauer_pre-nucleation_2014} and thus, together with the broad topic of
crystal nucleation of biominerals, is not discussed in here.

\subsection{Natural Gas Hydrates}
\label{sec:gas-hydrates}

Natural gas hydrates are crystalline compounds in which small gas molecules are caged (or \textit{enclathrated}) in a
host framework of water molecules. Natural gas molecules (e.g. methane, ethane, propane) are hydrophobic. They are also
favored by conditions of high pressure and low temperature, and are found to occur naturally in the ocean bed and in
permafrost regions \cite{clathrates}. With exceptionally high gas storage capabilities and the fact that it is believed
that gas hydrates exceed conventional gas reserves by at least an order of magnitude \cite{KlaudaSandler2005sjc}, there
is interest in trying to exploit gas hydrates as a future energy resource. While gas hydrates may potentially play a
positive role in the energy industry's future, they are currently considered a hindrance: if mixed phases of water and
natural gas are allowed to cool in an oil pipeline, then a hydrate may form and block the line, causing production to
stall. Understanding the mechanism(s) by which gas hydrates nucleate is likely to play an important role in the rational
design of more effective hydrate inhibitors.

\vspace{0.5cm}
\noindent \textbf{\textit{Hydrate Structures}} \\
There are two main types of natural gas hydrates: structure I \textit{sI}, which has a cubic structure (space group
$Pm\bar{3}n$); and structure II \textit{sII}, which also has a cubic structure (space group $Fd\bar{3}m$). (There is
also a third, less common type \textit{sH}, which has an hexagonal crystal structure, but we do not discuss this any
further here.) Structurally, the water frameworks of both sI and sII hydrate are similar to Ice I$_h$, with each water
molecule finding itself in an approximately tetrahedral environment with its nearest neighbors. Unlike ice I$_h$, however,
the water framework consists of cages, with cavities large enough to accommodate a gas molecule.

\begin{figure}[t!]
  \centering
    \includegraphics[width=7cm]{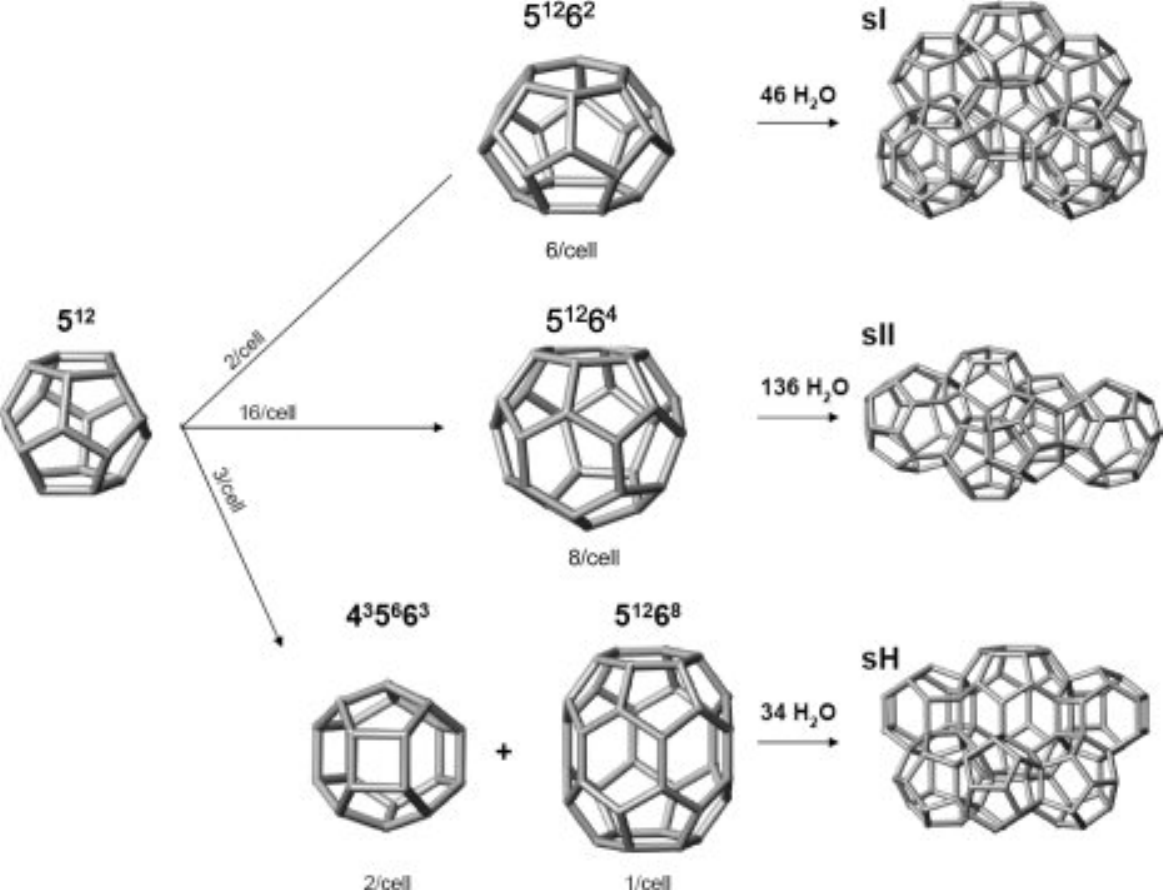}
    \caption{Crystal structure of the sI, sII and sH gas hydrates,
      along with the corresponding cage structures. Only the water molecule positions are shown as spheres connected by lines. Reprinted with permission from
      Ref.~\citenum{koh_natural_2007}. Copyright 2007, John Wiley and Sons.}
  \label{hydrates_F1}
\end{figure}

Between the sI and sII hydrate there exist three types of cages, which are denoted $5^{p}6^{h}$ depending upon the
number of five- and six-sided faces that make up the cage. For example, common to both the sI and sII hydrate is the
$5^{12}$ cage, where the water molecules sit on the vertices of a pentagonal dodecahedron. Along with the $5^{12}$ cage
the sI hydrate also consists of a $5^{12}6^{2}$ cage, which has two six-sided faces and twelve five-sided faces: there
are two $5^{12}$ cages and six $5^{12}6^{2}$ cages in the unit cell. The sII hydrate, on the other hand, has a unit cell
made up of sixteen $5^{12}$ cages and eight $5^{12}6^{4}$ cages. Due to the larger size of the $5^{12}6^{4}$ cage, the
sII structure forms in the presence of larger guest molecules such as propane, whereas small guest molecules such as
methane favor the sI hydrate. (This is not to say that small guest molecules are not present in sII, just that the
presence of larger guest molecules is necessary to stabilize the larger cavities.) The sI, sII and sH crystals
structures are shown in Fig.~\ref{hydrates_F1}, along with the individual cage structures. Further details regarding the
crystal structures of natural gas hydrates can be found in Ref.~\citenum{clathrates}.

\vspace{0.5cm}
\noindent \textbf{\textit{Homogeneous Nucleation}} \\
Historically, two main molecular mechanisms for hydrate nucleation have been proposed. First, Sloan and co-workers
\cite{MullerSloan1992sjc, SloanFleyfel1991sjc} proposed the \textit{labile cluster hypothesis} (LCH), which essentially
describes the nucleation process as the formation of isolated hydrate cages which then agglomerate to form a critical
hydrate nucleus. Second, the \textit{local structure hypothesis} (LSH) was proposed after umbrella sampling simulations
by Radhakrishnan and Trout \cite{RadhakrishnanTrout2002sjc} suggested that the guest molecules first arrange themselves
in a structure similar to the hydrate phase, which is accompanied by a perturbation (relative to the bulk mixture) of
the water molecules around the locally ordered guest molecules. For the same reasons already outlined elsewhere (see Sec.~\ref{Experimental_Methods}), 
it is experimentally challenging the verify which, if either, of these two nucleation mechanisms is correct.
What we will see in this section is how computer simulations of gas hydrate nucleation have been used to help shed light
on this process.

\begin{figure*}[t!]
  \centering
  \includegraphics[width=14cm]{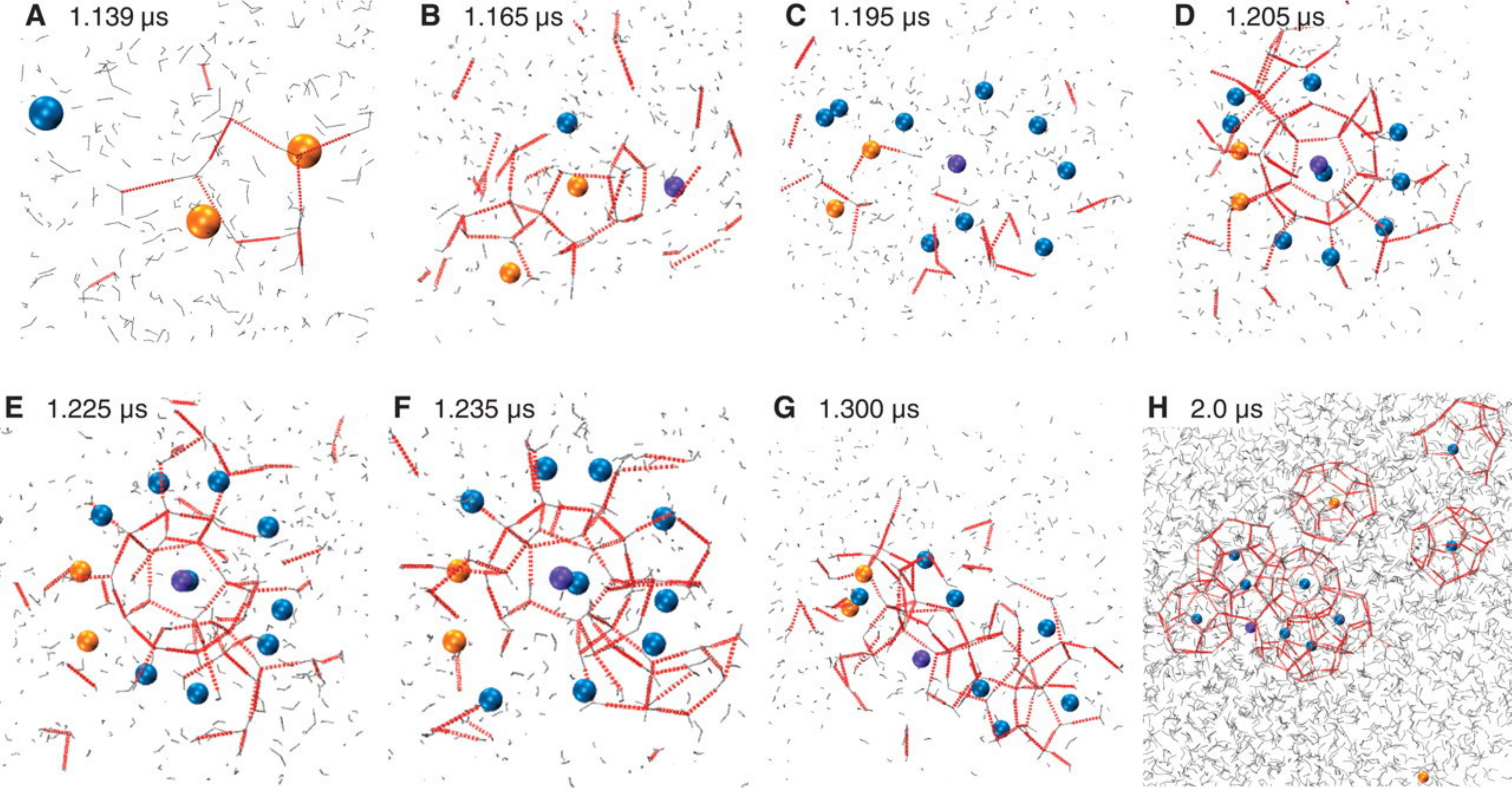}
  \caption{Early stages of hydrate nucleation observed by Walsh \textit{et al.} (A to C) A pair of methane molecules is
adsorbed on either  side of a single pentagonal face of water molecules. Partial cages form around this pair, near the
eventual central violet methane molecule, only to dissociate over several nanoseconds. (D and E) A small cage forms
around the violet methane and other methane molecules adsorb at 11 of the 12 pentagonal faces of the cage, creating the
bowl-like pattern shown. (F and G) The initial central cage opens on the end opposite to the formation of a network of
face-sharing cages, and rapid hydrate growth follows. (H) A snapshot of the system after hydrate growth, showing the
fates of those methane molecules that make up the initial bowl-like structure (other cages not shown).  Reprinted with permission from
Ref.~\citenum{walsh:science}. Copyright 2009, The American Association for the Advancement of Science.} 
\label{fig:HYWALSH} 
\end{figure*}

Although not the first computer simulation study of natural gas hydrate formation (see e.g.
Refs.~\cite{RadhakrishnanTrout2002sjc, HawtinRodger2008sjc, MoonRodger2007sjc, MoonRodger2003sjc}), one the most
influential simulation works on gas hydrate formation is that of Walsh \textit{et al.} \cite{walsh:science}, in which
methane hydrate formation was directly simulated under conditions of 250\,K and 500\,bar. It was found that nucleation
proceeded \emph{via} two methane and five water molecules cooperatively organizing into a stable structure, with the
methane molecules adsorbed on opposite sides of a pentagonal ring of water molecules. This initial structure allowed the
growth of more water faces and adsorbed methane, until a $5^{12}$ cage formed. This process took on the order of
50--100\,ns to complete. After persisting for {${\sim}$30\,ns}, this $5^{12}$ cage opened when two new water molecules
were inserted into the only face without an adsorbed methane molecule, on the opposite side to that where several new
full cages were completed. This opening of the original $5^{12}$ cage was then followed by the relatively fast growth of
methane hydrate. The early stages of hydrate nucleation are shown in Fig.~\ref{fig:HYWALSH}. After ${\sim}$240\,ns, the
original $5^{12}$ cage transformed into a $5^{12}6^{3}$ cage, a structure not found in any equilibrium hydrate
structure. Walsh \textit{et al.} also found that $5^{12}$ cages dominate, in terms of abundance, during the early stages
of nucleation. The formation of $5^{12}6^{2}$ cages (which along with the $5^{12}$ cages comprise the sI hydrate) are
the second most abundant, although their formation occurred approximately 100\,ns after the initial $5^{12}$ cages. A
significant amount of the larger $5^{12}6^{4}$ cages that are found in the sII hydrate was also observed, which was
rationalized by the large number of face-sharing $5^{12}$ cages providing an appropriate pattern. The $5^{12}6^{3}$
cages were also observed in an abundance close to that of the $5^{12}6^{2}$ cages. The final structure can be summarized
as a mixture of sI and sII motifs, linked by $5^{12}6^{3}$ cages. A similar structure had been previously reported as a
result of hydrate growth simulations \cite{VatamanuKusalik2006sjc, JacobsonMolinero2009sjc}.

\begin{figure*}[t!]
\begin{center}
\includegraphics[width=14cm]{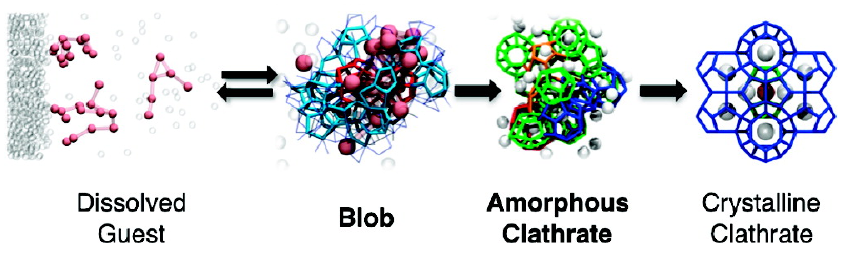}
\end{center}
\caption{Sketch of the nucleation mechanism of methane hydrates proposed in Ref.~\citenum{molinero:blobs}. Clusters of guest molecules aggregate in \textit{blobs}, which 
transform in to amorphous clathrates as soon as the water molecules arrange themselves in the cages characteristic of crystalline clathrate, which
eventually for upon the re-ordering of the guest molecules - and thus of the cages - in a crystalline fashion. Note that the
  difference between the \textit{blob} and \textit{amorphous
    clathrate} is that the water molecules have yet to be locked into
  clathrate hydrate cages in the former. Reprinted with permission from
  Ref.~\citenum{molinero:blobs}. Copyright 2010, American Chemical Society.}
\label{HY2}
\end{figure*}

Even though the work of Walsh \textit{et al.} \cite{walsh:science} provided useful insight into the hydrate nucleation
mechanism, the conclusions were based upon only two independent nucleation trajectories. Soon after the Walsh paper,
Jacobsen and Molinero \cite{molinero:blobs} reported a set of twelve simulations using a methane-water
model~\cite{molinero:mW-M} based on mW water under conditions of 210\,K and 500\,atm (the melting point of the model is
approximately 300\,K). Owing to the reduced computational cost of the coarse-grained model, Jacobsen and Molinero were
also able to study a much larger system size than Walsh \textit{et al.} (8000 water and 1153 guest molecules
\cite{molinero:blobs} vs 2944 water and 512 guest molecules \cite{walsh:science}). In agreement with Walsh \textit{et
al.}, the initial stages of the nucleation mechanism were also dominated by $5^{12}$ cages, and a mixture of sI and sII
motifs connected by $5^{12}6^{3}$ cages were observed. It was also observed that solvent-separated-pairs of guest
molecules were stabilized by greater numbers of guest molecules in the cluster. As gas hydrates are comprised of
solvent-separated-pairs of guest molecules as opposed to contact-pairs, this suggests that a resemblance to the LSH,
where the local ordering of guest molecules drives the nucleation of the hydrate. Jacobsen and Molinero, however, also
found a likeness to the LCH; clusters of guest molecules and their surrounding water molecules formed long-lived
\textit{blobs} that slowly diffuse in solution. These blobs could be considered large analogues of the labile clusters
proposed in the LCH. Through analysis of their simulation data, Jacobsen and Molinero concluded that the blob is a guest
rich precursor in the nucleation pathway of gas hydrates with small guest molecules (such as methane). Note that the
distinction between \textit{blobs} and \textit{amorphous} is that the water molecules have yet to be locked into the
clathrate hydrate cages in the former.  The overall nucleation mechanism is depicted in Fig.~\ref{HY2}.

Both the work of Walsh \textit{et al.} and Jacobson and Molinero suggest that amorphous hydrate structures are involved
in the nucleation mechanism, although both studies were carried out under high driving forces. In
Ref.~\citenum{molinero:amorphous-xtal}, Jacobsen and Molinero addressed the following two questions raised by the above
studies: \textit{how could amorphous nuclei grow into a crystalline form?}; and \textit{are amorphous nuclei precursors
intermediates for clathrate hydrates under less forcing conditions?}. By considering the size-dependent melting
temperature of spherical particles using the Gibbs-Thompson equation, Jacobson and Molinero found for all temperatures
that the size of the crystalline critical nucleus was always smaller than the amorphous critical nucleus, with the two
becoming virtually indistinguishable in terms of stability for very small nuclei of ${\sim}15$ guest molecules (i.e.
under very forcing conditions). From a thermodynamic perspective, this would suggest that nucleation would always
proceed \emph{via} a crystalline nucleus. The observation of amorphous
nuclei~\cite{walsh:science,molinero:blobs,HawtinRodger2008sjc, LiangKusalik2011sjc,SarupriaDebenedetti2012sjc} , even at
temperatures as high as 20\% supercooling, hints that their formation may be favored for kinetic reasons. Employing the
CNT expression for the free energy barrier suggested that the amorphous nuclei could be kinetically favored up to 17\%
supercooling if $\gamma_{a}/\gamma_{x} = 0.5$, where $\gamma_{a}$ and $\gamma_{x}$ are the liquid-amorphous and
liquid-crystal surface tensions, respectively. Jacobson and Molinero estimate $\gamma_{x} \approx 36$\,mJ/m$^{2}$ and
$16 < \gamma_{a} < 32$\,mJ/m$^{2}$, so it is certainly plausible that amorphous precursors are intermediates for
clathrate hydrates under certain conditions. The growth of clathrate hydrates from amorphous and crystalline seeds was
also studied, where it was found that crystalline clathrate can grow from amorphous nuclei. As the simulation led to
fast mass-transport, the growth of post critical nuclei was relatively quick, and the amorphous seed became encapsulated
by a (poly)crystalline shell. Under conditions where an amorphous nucleus forms first due to a smaller free energy
barrier, but where diffusion of the guest species becomes a limiting factor, it is likely that small nuclei would have
long enough to anneal to structures of greater crystallinity before growing to the macroscopic crystal phase.

It thus appears that gas hydrates may exhibit a multi-step nucleation process involving amorphous precursors for
reasonably forcing conditions, but for temperatures close to coexistence, it seems that nucleation should proceed
\emph{via} a single crystalline nucleus. By assuming a CNT expression for the free energy (as well as the total rate),
Knott \textit{et al.} \cite{homogeneous-unrealistic} used the \emph{seeding technique} (See Sec.~\ref{Brute_force}) to
compute the nucleation rate for sI methane hydrate with relatively mild supersaturation of methane, in a similar manner
to Espinosa \emph{et al.}~\cite{espinosa_homogeneous_2014} for homogeneous ice nucleation as discussed in
Sec.~\ref{IceHON}. They found vanishingly small homogeneous nucleation rates of $10^{-111}$ nuclei cm$^{-3}$ s$^{-1}$,
meaning that even with all of Earth's ocean waters, the induction time to form one crystal nucleus homogeneously would
be ${\sim}10^{80}$ years! Knott \textit{et al.}  therefore conclude that under mild conditions, hydrate nucleation must
occur heterogeneously.

\vspace{0.5cm}
\noindent \textbf{\textit{Heterogeneous Nucleation}} \\
Compared to homogeneous nucleation, the heterogeneous nucleation of gas hydrates is little studied. Liang \textit{et
al.} \cite{LiangKusalik2011sjc} investigated the steady state growth of a hydrate crystal in the presence of silica
surfaces, finding that the crystal preferentially grew in the bulk solution rather than at the interface with the solid.
It was also observed that in one simulation, local methane density fluctuations led to the spontaneous formation of a
methane bubble from solution, which was located at the silica interface. This had two effects on the observed growth:
(i) the methane bubble depleted most of the methane from solution, leading to an overall slowing down of the crystal
growth rate; and (ii) due to the location of the methane bubble, the silica surface effectively acts like a source of
methane, promoting growth of the crystal closer to the interface relative to the bulk.

Bai \textit{et al.} investigated the heterogeneous nucleation of CO$_2$ hydrate in the presence of a fully
hydroxylated silica surface, first in a two-phase system where the water and CO$_2$ are well mixed
\cite{BaiWang2011sjc}, and then in a three-phase system where the CO$_2$ and water are initially phase separated
\cite{BaiWang2012sjc}.  In the two-phase system, the authors report the formation of an ice-like layer at the silica
surface, above which a layer comprised of \textit{semi-5$^{12}$ cage-like} structures mediates the structural mismatch
between the ice-like contact layer and the sI hydrate structure above. In the three-phase system, nucleation is observed
at the three-phase contact line, along which the crystal nucleus also grows. This is attributed to the stabilizing
effect of the silica on the hydrate cages, plus the requirement for the availability of both water and CO$_2$. In a
later paper, Bai \textit{et al.} \cite{BaiWang2015sjc} investigated the effect of surface hydrophilicity (by decreasing
the percentage of surface hydroxyl groups) and crystallinity on the nucleation of CO$_2$ hydrate. They found that in
the case of decreased hydrophilicity, the ice-like layer at the crystalline surface vanishes, replaced instead by a
single liquid-like layer upon which the hydrate directly nucleates. Whereas shorter induction times to nucleation at the
less hydrophilic surfaces are reported, little dependence upon the crystallinity of the surface is observed. While
certainly an interesting observation, as only a single trajectory was performed for each system, studies in which
multiple trajectories are used to obtain a distribution of induction times would be desirable and, as the hydrate
actually appears to form away from the surface in all cases, a full comparison of the heterogeneous and homogeneous
rates would also be a worthwhile pursuit.

There have also been a number of studies investigating the potential role of ice in the nucleation of gas hydrates.
Pirzadeh and Kusalik~\cite{PirzadehKusalik2013sjc} performed MD simulations of methane hydrate nucleation in the
presence of ice surfaces,
 and reported that an increased density of methane at the interface induced structural
defects (coupled 5-8 rings) in the ice that facilitated the formation of hydrate cages. Nguyen \textit{et
al.}~\cite{nguyen2015structure} used MD
simulations to directly investigate the interface between a gas hydrate and ice, and found the existence of an
\textit{interfacial transition layer} (ITL) between the two crystal structures. The water
molecules in the ITL, which was found to be disordered and 2-3 layers of water in thickness, have a tetrahedrality and
potential energy that is intermediate between that of either of the crystal structures and liquid water. The authors
suggest that the ITL could assist the heterogeneous nucleation of gas hydrates from ice by providing a lower surface
free energy than either of the ice-liquid and hydrate-liquid interfaces. Differential scanning calorimetry experiments
by Zhang \textit{et al.}~\cite{zhang2004differential} found that ice and hydrate formation to occur simultaneously (on the experimental timescale),
which was attributed to the heterogeneous nucleation of ice, which in turn facilitated hydrate formation.
Poon and Peters~\cite{poon2013stochastic} provide a possible explanation for ice acting as
a heterogeneous nucleating agent for gas hydrates, aside from the structural considerations of
Refs.~\citenum{PirzadehKusalik2013sjc} and~\citenum{zhang2004differential}: At a growing ice front, the local
supersaturation of methane can be dramatically increased, to the extent that induction times to nucleation are reduced
by as much as a factor $10^{100}$.

Computer simulations of hydrate nucleation have certainly contributed to our understanding of the underlying mechanisms,
especially in the case of homogeneous nucleation. One fairly consistent observation across many simulation studies
\cite{RadhakrishnanTrout2002sjc, walsh:science, molinero:blobs, molinero:amorphous-xtal, HawtinRodger2008sjc,
MoonRodger2003sjc, BiLi2014sjc} suggests that some kind of ordering of dissolved guest molecules precedes the formation
of hydrate cages. Another is that amorphous nuclei, consisting of structural elements of both sI and sII hydrate form
when conditions are forcing enough. Nevertheless, open questions still remain. In particular, the prediction that
homogeneous nucleation rates are vanishingly small under mild conditions \cite{homogeneous-unrealistic} emphasizes the
need to better understand heterogeneous nucleation. To this end, enhanced sampling techniques such as FFS, which has
recently been applied to methane hydrate nucleation at 220\,K and 500\,bar \cite{BiLi2014sjc}, are likely to be useful,
although directly simulating nucleation under mild conditions is still likely to be a daunting task. Another
complicating factor is that, aside from the presence of solid particles, the conditions from which natural gas hydrates
form are often highly complex; for example in an oil or gas line there is a fluid flow and understanding how this
effects the methane distribution in water, is likely to be an important factor in determining how fast gas hydrates form
\cite{walsh:2011}. In this respect, the formation of natural gas hydrates is a truly multi-scale phenomenon.

\section{Future Perspectives}
\label{Discussion}

We have described only a fraction of the many computer simulation studies of
crystal nucleation in supercooled liquids and solutions. Still, we have learned that MD simulations have dramatically improved our fundamental
understanding of nucleation.  For instance, several studies on colloidal particles (see
Sec.~\ref{COLL}) provided evidence for two-step nucleation mechanisms, and the
investigation of LJ liquids has yielded valuable insight into the effect of confinement (see Sec.~\ref{LJL}). In
addition, the investigation of more realistic systems has delivered outcomes directly related to problems of 
great relevance. For example, the influence of different solvents on the early stages of urea crystallization (see
Sec.~\ref{sec.MIS}) has important consequences in fine chemistry and in the fertilizer industry, and the molecular
details of clathrate nucleation (see Sec.~\ref{sec.MIS}) could help to rationalize and prevent hydrate formation in oil or
natural gas pipelines.  Thus, it is fair to say that MD simulations have been and will remain
a powerful complement to experiments.

However, simulations are presently affected by several shortcomings, which hinder a reliable comparison with experimental nucleation rates and
limit nucleation studies to systems and/or conditions often far away from those investigated experimentally.
These weaknesses can be classified in to two main categories: (i) limitations related
to the accuracy of the computational model used to represent the system; and (ii) shortcomings due to the computational techniques employed to simulate
nucleation events. 

(i) In an ideal world, \textit{ab initio} calculations would be the tool of the trade. Unfortunately, in all but a handful of
cases such as the phase change materials presented in Sec.~\ref{BLJP}, the timescale problem makes \textit{ab initio}
simulations of crystal nucleation unfeasible (see Fig.~\ref{FIG_Brute_force_2}). As this will be the \textit{status quo} for the
next few decades, we are forced to focus our efforts on improving the current classical force fields
and on developing novel classical interatomic potentials.  This is a fundamental issue that affects computer simulations of materials as
a whole. While for nucleation of simple systems like colloids (Sec.~\ref{COLL}) this is not really
an issue, things start to fall apart when dealing with more realistic systems (see e.g. Sec.~\ref{sec.MIS} and
~\ref{sec:gas-hydrates}), and become even worse in the case of heterogeneous
nucleation (see e.g. Sec.~\ref{HIN}), as the description of the interface requires extremely transferable and
reliable force fields. Machine learning techniques~\cite{handley_next_2014} such as neural network potentials (see
Sec.~\ref{BLJP} and Refs.~\citenum{behler_representing_2014,bartok_gaussian_2010}) are emerging as possible candidates to allow for classical MD
simulations with an accuracy closer to first principles calculations, but the field is constantly looking for other
options that are capable of bringing simulations closer to reality.

(ii) The limitations of the computational techniques currently employed to study crystal nucleation are those
characteristic of rare events sampling. Brute force MD simulations (see Sec.~\ref{Brute_force}) allow for an  
unbiased investigation of nucleation events, but the timescale problem limits this approach to very few systems, typically 
very distant from realistic materials (see e.g. Sec.~\ref{COLL} and Sec.~\ref{LJL}) - although notable exceptions exist
(see Sec.~\ref{BLJP}). It is also worth noticing that
while brute force MD is not able to provide a full characterization of the nucleation process, 
useful insight can still be gained into e.g. pre-nucleation
events~\cite{tribello_molecular_2009,gebauer_pre-nucleation_2014}.
Enhanced sampling techniques (see Sec.~\ref{ESM}) are rapidly evolving and have the potential 
to take the field to the next level. However, free energy methods as they are do not give access to nucleation
kinetics and in the case of complex systems (see e.g. Sec.~\ref{IceHON} and Sec.~\ref{sec.MIS}) are strongly dependent
on the choice of the order parameter. On the other hand, in light of the body of work reviewed it seems that path sampling methods
can provide a more comprehensive picture of crystal nucleation. However,
at the moment these techniques are computationally expensive and a general implementation is not available yet, albeit
consistent efforts have been recently put in place.
We believe that the development of efficient enhanced sampling methods specific to crystal nucleation is one of the crucial
challenges ahead.  

At the moment, simulations of crystal nucleation of complex liquids are restricted to small (10$^2$-10$^4$ particles)
systems, most often in idealized conditions. For instance, it is presently very difficult to take into account
impurities or, in the case of heterogeneous nucleation, defects of the substrate.  
Indeed defects seem to be ubiquitous in many different systems, such as ice, hard spheres crystals, LJ crystals and organic crystals as well.
Defects are also often associated with polymorphism, but possibly because of the inherent difficulties in modeling them (or in 
characterizing them experimentally), they are under represented within the current literature.
These are important aspects that
almost always impact experimental measurements, and that should thus be included in simulations as well.  In general,
simulations of nucleation should allow us not only to provide microscopic insight, but also to make useful predictions
and/or to provide a general understanding to be applied to a variety of systems. These two ambitious goals are
particularly challenging for simulations of heterogeneous nucleation. In light of the literature we have reviewed in
here, we believe that much of the effort has to be devoted in the future to i) enable atomistic simulations of
heterogeneous nucleation dealing with increasingly realistic interfaces and to ii) obtain general, maybe non
material-specific trends able to point the community into the right direction, even at the cost of sacrificing  accuracy
to a certain extent.  On the other hand, we hope that the body of work reviewed here will inspire future experiments
targeting \textit{cleaner}, well defined systems by means of novel techniques, possibly characterized by better temporal
and spatial resolution.  Improving on the current limitations of the computational models and techniques would enable
simulations of much larger systems over much longer timescales, with a degree of accuracy that would allow a much more
fruitful comparison with experiments.  We think this should be the long term objective for the field.  Up to now, the
only way to connect simulations and experiments has been through the comparison of crystal nucleation rates, which even
now still exhibit substantial discrepancies for every single class of systems we have reviewed.  This is true not only
for complex liquids like water (see Sec.~\ref{IceHON}), but even for model systems such as colloids (Sec.~\ref{COLL}).
This, together with the fact that in some cases even experimental data are scattered across several orders of magnitude,
suggests that we are dealing with crystal nucleation in liquids within a flawed theoretical framework. 

As a matter of fact, CNT is now 90 years old. It is thus no wonder that every aspect of this battered theory has been criticized at some
point.  However, some aspects have been questioned more frequently than others. For instance, the emergence of two-step
(or even multi-step) mechanisms for nucleation, has been reported for many different systems (see
Sec.~\ref{COLL},~\ref{LJL},~\ref{sec.MIS} and \ref{sec:gas-hydrates}) and cannot be easily embedded in CNT as it is, albeit several improvement upon the original CNT formulation appeared within the last decade (see Sec.~\ref{THEOF_TS}).
Nonetheless, CNT is basically the only theory invoked by both experiments and simulations when dealing with crystal nucleation from the liquid phase. 
CNT is widely used because it offers a simple and unified picture for nucleation and it is often very useful. However, as demonstrated by both experiment and
simulations, even the basic rules governing the formation of the critical nucleus can change
drastically from one system to another.
Thus, we believe that any sort of theoretical universal approach, a brand new
CNT, so to say, will be unlikely to significantly further the field. Indeed we fear that the same reasoning will hold
for the computational methods required. We cannot think of a single enhanced sampling technique capable of tackling the
complexity of crystal nucleation as a whole. The interesting but uncomfortable truth is that each class of supercooled liquids often
possess a unique behavior, which in turn results into specific features ruling the crystal nucleation process.
Thus, it is very much possible that different systems within different conditions could require different, \textit{ad hoc}
flavors of CNT. While the latter has been evolving for decades, we believe that a sizable fraction of the new
developments in the field should aim at
producing particular flavors of CNT, specifically tailored to the problem at hand.

In conclusion, it is clear that MD simulations have proven themselves to be of the utmost importance in
unraveling the microscopic details of crystal nucleation in liquids. We have reviewed important advances which
provided valuable insight into 
fundamental issues and diverse nucleation scenarios, complementing experiments and
furthering our understanding of nucleation as a whole. Whether CNT can be effectively improved in a universal fashion is
unclear. We feel that the ultimate goal for simulations would be to get substantially closer
to the reality probed by experiments, and that in order to do so we have to sharpen our computational and possibly
theoretical tools. In particular, we believe that the community should invest in improving the classical interatomic
potentials available as well as the enhanced sampling techniques currently used, enabling accurate simulations of 
crystal nucleation for systems of practical relevance.

\section*{Acknowledgement}
This work was supported by the European Research Council under the European Union's Seventh Framework Programme
(FP/2007-2013)/ERC Grant Agreement number 616121 (HeteroIce project). A.M. is also supported by the Royal Society
through a Royal Society Wolfson Research Merit Award. We gratefully acknowledge Dr. Matteo Salvalaglio, Dr. Gareth Tribello, Dr. Richard Sear and
Prof. Daan Frenkel for insightful discussions and for reading an earlier version of the manuscript.

\clearpage


\begin{thebibliography}{475}
\bibitem [{fno({\natexlab{b}})}]{fnote_R1}%
  \BibitemOpen
  \href@noop {} {} ({\natexlab{b}}),\ \bibinfo {note} {in most cases,
  simulations of crystal nucleation do not embrace Classical Nucleation Theory
  as a whole, taking into account instead only selected elements of the theory.
  Common examples include i) assuming the size of the largest crystalline
  cluster as the only relevant order parameter to describe the nucleation
  process, ii) assuming the critical nuclei to be of spherical shape and iii)
  assuming a perfectly sharp interface between the crystalline and the liquid
  phase.}\BibitemShut {Stop}%
\bibitem [{fno({\natexlab{f}})}]{fnote10}%
  \BibitemOpen
  \href@noop {} {} ({\natexlab{f}}),\ \bibinfo {note} {to be clear, it is
  perfectly legitimate to measure the contact angle of a water droplet of
  several mm on a macroscopically planar surface. In contrast, defining a
  contact angle for a critical crystalline nucleus containing just a few
  hundred molecules on a surface characterized by a roughness of the same
  extent, is clearly much more difficult, if meaningful at all.}\BibitemShut
  {Stop}%
\bibitem [{fno({\natexlab{g}})}]{fnote11}%
  \BibitemOpen
  \href@noop {} {} ({\natexlab{g}}),\ \bibinfo {note} {cNT is still expected to
  capture nucleation on very short timescales at strong supercooling when the
  free energy barrier for nucleation, albeit being non-zero, is of the order of
  $k_B\cdot T$.}\BibitemShut {Stop}%
\bibitem [{fno({\natexlab{k}})}]{fnote14}%
  \BibitemOpen
  \href@noop {} {} ({\natexlab{k}}),\ \bibinfo {note} {brownian
  dynamics~\cite{Brownian_dynamics} differs from molecular dynamics mainly with
  respect to the dynamics of the system at short times. This difference is thus
  not expected to play a role in the estimate of nucleation rates, which depend
  chiefly on the long-time dynamics of the system.}\BibitemShut {Stop}%
\bibitem [{fno({\natexlab{m}})}]{fnote5}%
  \BibitemOpen
  \href@noop {} {} ({\natexlab{m}}),\ \bibinfo {note} {in this case the
  equilibrium crystalline phase that forms upon homogeneous nucleation is a
  RHCP made of stacked of hexagonal layers.}\BibitemShut {Stop}%
\bibitem [{fno({\natexlab{n}})}]{fnote6}%
  \BibitemOpen
  \href@noop {} {} ({\natexlab{n}}),\ \bibinfo {note} {we are referring to the
  average population of critical nuclei at a given temperature and pressure. In
  other words, the number of critical nuclei expected per unit volume, which
  should not be confused with the density of the crystalline phase within the
  nuclei.}\BibitemShut {Stop}%
\bibitem [{fno({\natexlab{o}})}]{fnote7}%
  \BibitemOpen
  \href@noop {} {} ({\natexlab{o}}),\ \bibinfo {note} {statistical mechanics
  tells us that macroscopic thermodynamic variables are well behaved for a
  system with a number of molecules of the order of $N_{A}$, with $N_{A}$ being
  Avogadro's number.}\BibitemShut {Stop}%
\end{thebibliography}
\end{document}